\newif{\ifarxiv}
\newif{\ifdraft}
\newif{\ifremarks}
\newif{\ifjhep}

\arxivtrue 
\drafttrue 
\remarkstrue 

\ifdraft\else\fi

\ifjhep
\documentclass[11pt,reqno,preprint,a4paper]{article}
\usepackage{jheppub}
\else
\documentclass[12pt,a4paper]{article}
\fi

\usepackage[utf8]{inputenc}
\usepackage[dvipsnames,table]{xcolor}
\usepackage{soul}

\ifjhep\else
\usepackage[left=2.3cm,right=2.3cm,top=2cm,bottom=2.5cm]{geometry}
\usepackage{cite}
\usepackage[bookmarks=true]{hyperref}
\fi

\usepackage{parskip}
\usepackage{booktabs}
\ifremarks\newcommand{\remarktb}[1]{%
  {\renewcommand{\bfdefault}{b}\color[RGB]{0,150,0}{\textbf{T:~#1}}}}\fi
\providecommand{\remarktb}[1]{\ignorespaces}
\ifremarks\newcommand{\remarkfc}[1]{%
  {\renewcommand{\bfdefault}{b}\color[RGB]{0,0,150}{\textbf{F:~#1}}}}\fi
\providecommand{\remarkfc}[1]{\ignorespaces}
\ifremarks\newcommand{\remarkcb}[1]{%
  {\renewcommand{\bfdefault}{b}\color[RGB]{150,0,150}{\textbf{C:~#1}}}}\fi
\providecommand{\remarkcb}[1]{\ignorespaces}
\ifremarks\newcommand{\remarkab}[1]{%
  {\renewcommand{\bfdefault}{b}\color[RGB]{0,150,150}{\textbf{A:~#1}}}}\fi
\providecommand{\remarkab}[1]{\ignorespaces}

\usepackage[T1]{fontenc}
\usepackage{lmodern}
\usepackage{microtype}

\usepackage{xspace}
\def\etal.{et\penalty50\ al.}
\newcommand*{\eg}{e.\,g.\@\xspace}
\newcommand*{\ie}{i.\,e.\@\xspace}

\providecommand{\hypersetup}[1]{}
\providecommand{\hyperref}[2][]{#2}
\providecommand{\texorpdfstring}[2]{#1}
\providecommand{\pdfbookmark}[3][]{}
\newcommand{\email}[1]{\href{mailto:#1}{\nolinkurl{#1}}}

\hypersetup{plainpages=false}
\hypersetup{pdfpagemode=UseOutlines}
\hypersetup{bookmarksnumbered=true}
\hypersetup{bookmarksopen=true}
\hypersetup{pdfstartview=FitH}
\hypersetup{colorlinks=true}
\hypersetup{citecolor=[rgb]{0 .3 0}}
\hypersetup{urlcolor=[rgb]{.35 0 0}}
\hypersetup{linkcolor=[rgb]{0 0 .5}}
\hypersetup{citebordercolor={.6 .9 .6}}
\hypersetup{urlbordercolor={.7 .8 1}}
\hypersetup{linkbordercolor={1 .7 .7}}
\hypersetup{pdfborder={0 0 .5}}

\makeatletter
\let\@myabstract\@empty
\let\@keywords\@empty
\let\@subject\@empty
\providecommand{\affiliation}[1]{\gdef\@affiliation{#1}}
\providecommand{\myabstract}[1]{\gdef\@myabstract{#1}}
\providecommand{\keywords}[1]{\gdef\@keywords{#1}}
\providecommand{\subject}[1]{\gdef\@subject{#1}}
\def\thetitle{\@title}
\def\theauthor{\@author}
\def\theaffiliation{\@affiliation}
\def\theabstract{\@myabstract}
\def\thesubject{\@subject}
\def\thedate{\@date}
\def\thekeywords{\@keywords}
\makeatother
\AtBeginDocument{
\hypersetup{pdftitle={\thetitle}}%
\hypersetup{pdfauthor={\theauthor}}%
\hypersetup{pdfsubject={\thesubject}}%
\hypersetup{pdfkeywords={\thekeywords}}%
}

\usepackage{graphbox}

\usepackage[font=small,labelfont=bf,margin=0.05\textwidth]{caption}

\usepackage{amsmath,amssymb,mathtools}
\usepackage{amsbsy}
\usepackage[mathbold,autobold,greekcaps,greeklower]{mathfixs}
\providecommand{\mathbold}{\mathbf}
\newcommand{\nn}{\nonumber}
\allowdisplaybreaks
\numberwithin{equation}{section}

\newcommand{\namedref}[2]{\hyperref[#2]{#1~\ref*{#2}}}
\newcommand{\secref}[1]{\namedref{Section}{#1}}
\newcommand{\appref}[1]{\namedref{Appendix}{#1}}
\newcommand{\tabref}[1]{\namedref{Table}{#1}}
\newcommand{\figref}[1]{\namedref{Figure}{#1}}

\makeatletter
\def\mr@ignsp#1 {\ifx\:#1\@empty\else #1\expandafter\mr@ignsp\fi}%
\newcommand{\multiref}[1]{\begingroup
\xdef\mr@no@sparg{\expandafter\mr@ignsp#1 \: }%
\def\mr@comma{}%
\@for\mr@refs:=\mr@no@sparg\do{\mr@comma\def\mr@comma{,}\ref{\mr@refs}}%
\endgroup}
\makeatother
\renewcommand{\eqref}[1]{(\multiref{#1})}

\newcommand{\sfrac}[2]{{\textstyle\frac{#1}{#2}}}
\newcommand{\half}{\sfrac{1}{2}}

\newcommand{\suprm}[1]{^{\text{#1}}}
\newcommand{\subrm}[1]{_{\text{#1}}}
\newcommand{\alg}[1]{\mathfrak{#1}}
\newcommand{\grp}[1]{\mathrm{#1}}
\newcommand{\mathematica}{\textsc{Mathematica}\@\xspace}

\newcommand{\filename}[1]{\texttt{#1}}
\newcommand{\ancfile}[1]{\href{https://arxiv.org/src/2509.14332/anc/#1}{\filename{#1}}}
\colorlet{c1}{Black}
\colorlet{c2}{Black}
\colorlet{c3}{Black}
\colorlet{c4}{Black}
\colorlet{c5}{Black}
\newcommand{\col}[1]{\textcolor{c#1}{#1}}
\definecolor{red1}{RGB}{180,0,0}
\definecolor{green1}{RGB}{0,100,0}
\def\NMHV{{\scriptscriptstyle\text{NMHV}}}
\def\NMHVf#1{{\scriptscriptstyle\text{N$^#1$MHV}}}
\def\NkMHV{\NMHVf{k}}
\def\NlMHV{\NMHVf{\ell}}

\RequirePackage[extdef]{delimset}
\makeatletter
\providecommand{\brkleft}[1][r]{\begingroup\def\dlm@use{\delim(.}%
\if r#1 \def\dlm@use{\delim(.}\fi%
\if s#1 \def\dlm@use{\delim[.}\fi%
\if c#1 \def\dlm@use{\delim\{.}\fi%
\if a#1 \def\dlm@use{\delim<.}\fi%
\expandafter\endgroup\dlm@use}
\providecommand{\brkright}[1][r]{\begingroup\def\dlm@use{\delim.)}%
\if r#1 \def\dlm@use{\delim.)}\fi%
\if s#1 \def\dlm@use{\delim.]}\fi%
\if c#1 \def\dlm@use{\delim.\}}\fi%
\if a#1 \def\dlm@use{\delim.>}\fi%
\expandafter\endgroup\dlm@use}
\makeatother

\DeclareMathOperator{\tr}{tr}

\DeclareMathOperator{\Li}{Li}

\DeclareMathOperator*{\res}{Res}
\DeclareMathOperator{\sgn}{sign}

\newcommand{\order}[1]{\mathcal{O}(#1)}

\newcommand{\beq}{\begin{equation}}
\newcommand{\eeq}{\end{equation}}
\newcommand{\bba}{\begin{align}}
\newcommand{\eea}{\end{align}}

\newcommand{\ee}{\mathrm{e}}
\newcommand{\ii}{i}
\newcommand{\eps}{\varepsilon}

\newcommand{\superN}{\mathcal{N}}
\newcommand{\Nc}{N\subrm{c}}

\newcommand{\dd}[2][]{\mathinner{\mathrm{d}\ifx#1\empty\else{^#1}\fi#2}}

\newcommand{\Lint}{L\subrm{int}}
\newcommand{\superO}{\mathbb{O}}
\newcommand{\superG}{\mathbb{G}}
\newcommand{\Zref}{Z_\star}
\newcommand{\twentyprime}{\mathbf{20'}}

\usepackage{diagbox}

\title{Higher-Point Correlators in \texorpdfstring{$\mathcal{N}=4$}{N=4}
SYM:\texorpdfstring{\\}{ }Generating Functions}

\author{%
Till Bargheer\texorpdfstring{$^1$}{},
Albert Bekov\texorpdfstring{$^1$}{},
Carlos Bercini\texorpdfstring{$^1$}{},
Frank Coronado\texorpdfstring{$^2$}{}}

\myabstract{We construct generating functions of five- and
six-point correlators up to two loops at weak 't~Hooft coupling in planar
$\mathcal{N}=4$ SYM.
These generating functions unify the correlators of the lightest scalar operator in the
stress-tensor multiplet with those of all higher R-charge single-trace half-BPS scalar operators, thereby extending previous results for four-point loop integrands. At the integrated level,  they are represented as sums of conformal integrals with coefficients exhibiting ten-dimensional poles  that combine  spacetime and R-charge distances.
Our results show that higher-order poles are captured by
products of lower-point generating functions.
We also extract new OPE data on spinning structure constants, and
compare these to integrability-based computations, finding good
agreement.}

\subject{Mathematical Physics}


\begin{document}

\pdfbookmark[1]{Title Page}{title}

\thispagestyle{empty}
\setcounter{page}{0}

\renewcommand{\thefootnote}{\fnsymbol{footnote}}
\setcounter{footnote}{0}

\hfill
\texttt{DESY-25-126}

\mbox{}
\vfill

\begin{center}

{\Large\textbf{\mathversion{bold}\thetitle}\par}

\vspace{1cm}

\textsc{\theauthor}

\bigskip

\begingroup
\footnotesize\itshape

${}^1$ Deutsches Elektronen-Synchrotron DESY,
Notkestr.~85, 22607 Hamburg, Germany

${}^2$ Institut f\"ur Theoretische Physik, ETH Zurich, CH-8093 Z\"urich, Switzerland

\endgroup

\bigskip

\begingroup
\small\ttfamily
\email{till.bargheer@desy.de},
\email{albert.bekov@desy.de},\\
\email{carlosbercini@gmail.com},
\email{fcidrogo@gmail.com}
\endgroup
\par

\vspace{1cm}

\textbf{Abstract}
\vspace{5mm}

\begin{minipage}{12cm}
\theabstract
\end{minipage}

\end{center}

\vfill
\vfill
\newpage

\renewcommand{\thefootnote}{\arabic{footnote}}

\hrule
\pdfbookmark[1]{\contentsname}{contents}
\setcounter{tocdepth}{3}
\microtypesetup{protrusion=false}
\tableofcontents
\microtypesetup{protrusion=true}
\vspace{3ex}
\hrule

\newpage
\section{Introduction}

Computing observables with many scales at higher loop orders in
four-dimensional gauge theory remains an outstanding challenge.
The deepest probes, both qualitatively and
quantitatively, are possible in the
highly symmetric setting of $\superN=4$ super Yang--Mills theory
(SYM), which is moreover connected to quantum gravity through holography.
As in any conformal field theory, correlation functions of local
operators are among its most fundamental observables.
The most extensively studied correlation functions in $\superN=4$ SYM
are those of the lightest scalar protected operators in the stress-tensor multiplet.
In the 't~Hooft planar limit, their four-point correlator is known analytically to various orders in both the weak-coupling~\cite{Eden:2011we,Drummond:2013nda} and strong-coupling series~\cite{Arutyunov:2000py,Goncalves:2014ffa,Alday:2023mvu}. At finite coupling, powerful tools such as localization and the numerical conformal bootstrap have also been applied~\cite{Beem:2016wfs,Chester:2023ehi,Caron-Huot:2022sdy,Caron-Huot:2024tzr}.
More recently, higher-point correlators have been computed at weak~\cite{Bargheer:2022sfd} and strong coupling~\cite{Goncalves:2019znr, Goncalves:2025jcg}, further broadening the scope of applications. These observables provide access to non-protected structure constants through OPE limits, connect to Wilson loops~\cite{Alday:2010zy} and scattering amplitudes~\cite{Alday:2007hr, Berkovits:2008ic, Beisert:2008iq,Bern:2008ap, Drummond:2008aq, Mason:2010yk,Caron-Huot:2010ryg,Eden:2010zz} via polygonal light-like limits and T-duality, and capture graviton scattering in $\grp{AdS}_5\times \grp{S}^5$. They also yield detector correlators through light-ray transforms~\cite{Belitsky:2013ofa,Moult:2025nhu}.

The stress-tensor supermultiplet is the lightest in an infinite family
of half-BPS single-trace operators, whose scaling dimensions are
protected and determined by their $\grp{SO}(6)$ R-charge.
Holographically, these are dual to the graviton supermultiplet and its
infinite tower of Kaluza–Klein (KK) modes ($\grp{S}^5$ harmonics) in
$\grp{AdS}_5\times \grp{S}^5$. On the CFT side, they encode the
dynamics and symmetries of the internal $\grp{S}^5$ manifold of the bulk dual. Furthermore,  besides sharing many of the applications of lightest operators, the correlators of higher KK modes allow us to explore other interesting limits.  These include the BMN limit and  large R-charge limits, where integrability-based techniques, such as hexagonalization~\cite{Basso:2015zoa,Fleury:2016ykk}, become more effective. In particular,   four-point correlators with large R-charge and special polarizations (the ``octagon'') are known analytically at finite coupling~\cite{Coronado:2018cxj,Kostov:2019stn,Belitsky:2020qrm,Bajnok:2024epf}.

A particularly striking development has been the identification of a hidden ten-dimensional symmetry.  More
specifically, the planar four-point correlators of half-BPS single-trace
operators with arbitrary R-charge can be resummed into a single
generating function, which enjoys a ten-dimensional conformal
invariance. This $\grp{SO}(10,2)$ symmetry unifies the
$\grp{SO}(4,2)$ spacetime and $\grp{SO}(6)$ R symmetries, acting on
the ten-dimensional vector $X=(x,y)$ composed of the 4d spacetime
coordinates $x$ and 6d R-polarizations $y$. This hidden symmetry has been identified in two regimes of the
planar 't~Hooft coupling: At strong coupling, in the tree-level supergravity
regime of the bulk dual~\cite{Caron-Huot:2018kta}, and at weak coupling in the perturbative
series of the loop integrand~\cite{Caron-Huot:2021usw}, defined via
the Lagrangian insertion method~\cite{Intriligator:1998ig,Eden:2011we}. However, this 10d symmetry is known to be broken away from these two
regimes, by stringy $\alpha'$ corrections in the supergravity side, and
when performing the 4d spacetime integrals over the Lagrangian insertions on the
perturbative CFT side. To date, there is no
first-principle derivation of this ten-dimensional symmetry, and thus its
fundamental origin as well as the mechanism of its breaking remain unclear.

These observations naturally raise two questions. The first is whether
this symmetry extends beyond four points to higher-point
correlators, a problem that remains open. The second is whether the
computation of these correlators can be streamlined by considering
instead the correlators of the ``master operator'', which encapsulates
the full tower of KK modes. This idea was investigated in the
weak-coupling regime at the loop-integrand level
in~\cite{Caron-Huot:2023wdh}. In that context, the ``master operator'' was
identified with the logarithm of a half-BPS superdeterminant operator, which generates all scalar KK modes and their supersymmetric partners,
including the chiral Lagrangian operator.
This identification, in turn, provided a set of effective Feynman rules for computing the corresponding ``master'' supercorrelators in the theory's self-dual sector, where the ``master'' loop integrands emerge naturally as specific supercomponents.
The rules employ an
effective scalar propagator with a ten-dimensional pole, combining spacetime and R-charge distances,
and additional interaction vertices that encode Grassmann
supercomponent data. These rules exhibit the ten-dimensional pole structure of the loop integrands at arbitrary multiplicity and loop order, but they do not yet render the hidden 10d conformal symmetry manifest.  For instance, establishing this symmetry for four-point loop integrands still requires summing all  Feynman graphs and invoking supersymmetry, while the higher-point case remains even less understood.

In the absence of a fundamental explanation or derivation of the
ten-dimensional symmetry, we can only resort to further exploration.
In particular, in the four-point function, the ten-dimensional
$\grp{SO}(10,2)$ conformal symmetry only emerges
after passing to the reduced correlator, which means factoring out a
universal superconformal invariant that arises due to a partial
non-renormalization theorem for the stress-tensor correlator~\cite{Eden:2000bk}.
It is therefore not clear if and how the symmetry will be
realized at higher points. There could be multiple higher-point
invariants, perhaps related to different (super)polarizations of the
dual closed-string amplitudes.%
\footnote{See~\cite{Fernandes:2025eqe} for a recently observed
decomposition of the five-point Mellin-space function at strong coupling.}
In this paper, we approach this uncertainty by considering the
next-simplest examples beyond four points, namely the five-point and
six-point correlators, and explore to what extent we can organize the
results in terms of ten-dimensional objects.

The AdS$_5$/CFT$_4$ system in the planar limit is an integrable
model~\cite{Beisert:2010jr}. Thus, with the
right methods, all observables should in principle be computable at
any value of the 't~Hooft coupling. These methods are best developed
for two-point functions, alias scaling dimensions of local
operators~\cite{Gromov:2013pga}.
For dynamical observables in the form of higher-point
functions of local operators, the state-of-the-art integrability-based method
is hexagonalization~\cite{Basso:2015zoa,Fleury:2016ykk,Eden:2016xvg},
which effectively is an expansion around the
limit of infinite operator charges. Therefore, its predictions are so far mostly
limited to relatively specific settings
(the simplest is the ``octagon'' limit mentioned above)
and/or low loop orders.%
\footnote{Notably, it also applies to the computation
of higher-genus terms in the $1/\Nc^2$ 't~Hooft
expansion~\cite{Bargheer:2017nne,Bargheer:2018jvq}.}
To further develop the method towards finite charge, acquiring more perturbative data
is essential, especially for operators whose charges can range
from large to small. This is exactly what is provided by the
correlators that we consider, which lends further motivation for their
perturbative computation.

\medskip

In this paper, we make some progress in the explicit computation of
higher-point correlators of BPS operators with arbitrary R-charge, by
following the methods that were used in~\cite{Chicherin:2014uca,Fleury:2019ydf,Bargheer:2022sfd} for
$\twentyprime$ operators, and were generalized to KK towers (and determinant
operators) in~\cite{Caron-Huot:2023wdh}. Specifically, we compute
the loop-integrand generating function for five-point correlators of BPS
operators at two-loop order, as well as the corresponding six-point function at one
loop. While we cannot make completely conclusive statements about their
ten-dimensional symmetry, we do make some observations about
their general structure. In particular, we
observe a nesting structure of higher-order ten-dimensional poles, controlled by
lower-point generating functions. We also recast the generating
function as a linear combination of a few correlators with low and
fixed R-charges whose coefficients are simple ten-dimensional poles.

\begin{figure}[t]
\centering
\includegraphics[align=c]{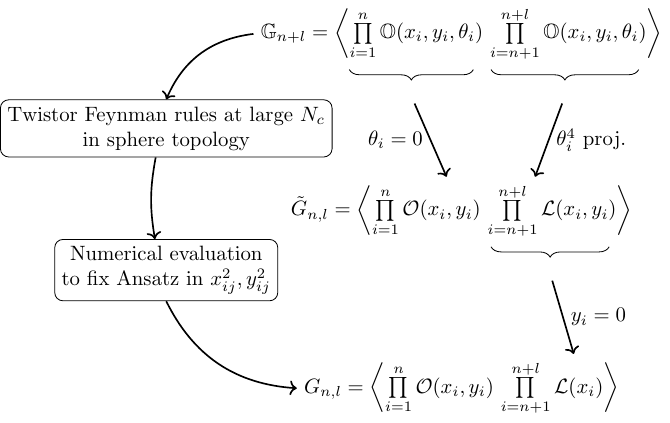}
\caption{Correlators in self-dual super Yang--Mills theory and the methods to
compute them. The super-correlator is computed by using twistor
Feynman rules, as reviewed in \secref{sec:twistor-rules}. And the
loop integrands $G_{n,l}$ are extracted by performing Grassmann
$\theta$-projections and R-charge $y$-projections.  We obtain explicit results in terms of
conformal integrands by matching a numerical evaluation of the
supercorrelator with an ansatz in $x$-$y$ kinematic space.}
\label{fig:CorrelatorsMethods}
\end{figure}

The paper is organized as follows: \secref{sec:super-corr-loop}
introduces the correlators and generating functions that we compute.
\secref{sec:twistors} describes and reviews in detail the twistor
methods that we use to obtain concrete expressions, see \figref{fig:CorrelatorsMethods}. In
\secref{sec:Gn0treelevel} to \secref{sec:six-point-integrand}, we
present our various results. \secref{sec:higher-poles} describes the
systematics of higher-order ten-dimensional poles. In
\secref{sec:OPEData}, we pass to the coupling-dependent $\superN=4$
SYM correlators, where the Lagrangian operators are integrated over 4d
spacetime. This results in our main formula~\eqref{eq:G5SYM} for the
five-point correlator at two loops.
Furthermore, we expand our new correlators in OPE limits,
match the result against available data from integrability, and extract some
new CFT data. We conclude with an outlook in
\secref{sec:conclusions-outlook}. A number of appendices contain further
details.

\section{Super-Correlator and Loop Integrands}
\label{sec:super-corr-loop}

We study correlation functions of the
half-BPS operator
\begin{equation}\label{eq:Oscalar}
\mathcal{O}(x,y)= \sum_{k=2}^\infty\, \mathcal{O}_k(x,y)
\,,\qquad
\mathcal{O}_k(x,y)=\frac{1}{k}\tr\left[y\cdot\phi(x)\right]^k + \text{multi-traces}
\,.
\end{equation}
This operator resums the tower of scalar protected operators
$\mathcal{O}_k$ with integer scaling dimension~$k$. In its definition,
the trace is taken on the adjoint $\grp{U}(\Nc)$ matrices.
Besides the single-trace term, we include additional multi-trace terms
to make these  operators orthogonal to all other pure multi-trace
operators. This is the single-particle basis introduced
in~\cite{Aprile:2020uxk}.%
\footnote{The conceptual motivation for these single-particle
operators is that they are natural duals to
single-particle scalar supergravity states on
$\grp{AdS}_5\times\grp{S}^5$. See~\cite{Arutyunov:1999en} for a first
study that fixed the double-trace admixture.}

The half-BPS condition requires the null condition of the
six-dimensional R-charge polarization vector:%
\footnote{In \secref{sec:integrated-correlators} we work with the equivalent $\grp{SO}(6)$ vector basis: $y^{I}y_I=0$ with $I=1,\dots,6$.}
\beq
y\cdot y \equiv y^{AB} y_{AB} = 0 \quad \text{with}\quad A,B = 1,2,3,4
\,.
\eeq
The two- and three-point correlators of these operators are tree-level
exact due to supersymmetry~\cite{Intriligator:1998ig,Intriligator:1999ff}. On the other hand, the four- and higher-point correlators receive
quantum corrections in general kinematics. These perturbative
corrections can be computed by means of the Lagrangian insertion
method as~\cite{Intriligator:1998ig,Eden:2011we}:
\begin{equation}
\left\langle\prod_{i=1}^{n}\mathcal{O}(x_i,y_i)\right\rangle_{\!\text{SYM}}=
\sum_{\ell=0}^\infty\frac{(-g^{2})^{\ell}}{\ell!}
\int\brk[s]3{\,\prod_{k=1}^\ell\frac{\dd[4]{x_{n+k}}}{\pi^2}}
G_{n,\ell}\,,
\label{eq:integratedcorrelator}
\end{equation}
where the left-hand-side label $\text{SYM}$ indicates that the correlator
is computed in the full interacting theory. In contrast, the $\ell$-loop integrand
$G_{n,\ell}$ is a  correlator of the $n$ original
operators with $\ell$ extra insertions of the chiral Lagrangian
$\Lint$ computed in the self-dual sector of the theory
($\text{SDYM}$):
\begin{equation}
\label{eq:Gnl-integrand}
G_{n,\ell} \equiv \left\langle \prod_{i=1}^{n}\mathcal{O}(x_i,y_i)  \prod_{i=1}^{\ell}\Lint(x_i)  \right\rangle\subrm{\!SDYM}
\,.
\end{equation}
More specifically, the interaction term $g^{2}\Lint$ completes the self-dual sector to the full SYM Lagrangian in the Chalmers--Siegel
formulation~\cite{Chalmers:1996rq}. In this paper, we focus on the planar limit. At large $\Nc$, the connected part of the correlator scales as:
\beq\label{eq:Gnlscaling}
G_{n,\ell} \sim \Nc^{2-n}
\,.
\eeq
The correlator $G_{n,\ell}$ is in general a rational function of the
spacetime and R-charge distances: $x_{ij}^2\equiv (x_{i}-x_{j})^2$ and
$y_{ij}^2\equiv (y_{i}-y_{j})^2$, and it can also depend on other
Lorentz and R-charge invariants that appear for $n\geq 5$ (see for
instance the odd part of the integrand in~\eqref{eq:R6b}, and the
R-structure in~\eqref{eq:R6a}). The loop-integrand $G_{n,\ell}$ of $\mathcal{O}$
serves as a generating function for the loop-integrands of operators
$\mathcal{O}_{k}$ with fixed but arbitrary R-charge.

The best studied are the four-point loop integrands $G_{4,\ell}$, which are rational functions of only  $x_{ij}^2$ and $y_{ij}^2$. Furthermore, superconformal Ward identities fix these
integrands to take the form~\cite{Eden:2000bk}:
\begin{equation}
\label{eq:susywardG4}
G_{4,\ell} = \frac{2\mathcal{R}_{1234}}{\Nc^{2}}\prod_{i<j}^{4}x_{ij}^2\times \mathcal{H}_{4,\ell}(X_{ij}^2)
\,,
\end{equation}
where $\mathcal{R}_{1234}$ is a simple polynomial, defined
in~\eqref{eq:R1234first}, which vanishes for special kinematics
$(x,y)$ with enhanced preserved supersymmetry~\cite{Drukker:2009sf}.
The factor
$\mathcal{H}_{4,\ell}$ is known as the reduced (integrand) correlator, and
it enjoys a ten-dimensional symmetry that combines spacetime and
R-charge kinematics on the same footing as $X_{ij}^2\equiv x_{ij}^2+y_{ij}^2$. This
symmetry was observed in~\cite{Caron-Huot:2021usw}, checked up to five
loops by using the bootstrap results of~\cite{Chicherin:2015edu,
Chicherin:2018avq}, and is conjectured to hold at all loops. We review
explicit three-loop results in \secref{sec:4pointreview}.

On the other hand, higher-point integrands are less constrained by supersymmetry. In particular, there is no known generalization of the factorization in~\eqref{eq:susywardG4} for five- and higher-point integrands, and hence also the generalization of the ten-dimensional symmetry is unclear. In this paper, we explore the structure of the simplest non-trivial five- and six-point integrands:
\beq\label{eq:Gtargets}
G_{5,1}\text{ in \secref{sec:Oneloop5point}}
,\quad
G_{5,2}\text{ in \secref{sec:Twoloop5point}}
,\quad\text{and}\quad
G_{6,1}\text{ in \secref{sec:Oneloop6point}}
.
\eeq
In order to compute these correlators, we follow the method put forward in~\cite{Caron-Huot:2023wdh}. In that paper, the loop integrand in~\eqref{eq:Gnl-integrand} was generalized in two ways, by adding all (chiral) superdescendants, and the higher R-charge partners of the chiral Lagrangian. This allows to study the correlators of the (chiral) supersymmetric extension of the operator $\mathcal{O}$:
\begin{align}
\superO(x,y,\theta)
&\equiv
\ee^{\theta^{A\alpha}\,Q_{A\alpha}}\mathcal{O}(x,y)
=\mathcal{O}(x,y) +\cdots -\frac{1}{2}(\theta.y.\theta)^2\,\frac{\Lint(x,y)}{\Nc}
\,.
\label{eq:Osuper}
\end{align}
The truncation at order $\theta^{4}$ is due to the half-BPS condition.
The top component, proportional to $\theta^4$, is a generalization of
the chiral Lagrangian $\Lint(x)$, which can be extracted
in the limit $y\to 0$ as:%
\footnote{An explicit definition of
the chiral Lagrangian can be found in eq.~(2.3)
of~\cite{Caron-Huot:2023wdh}. This definition includes an overall~$\Nc$
factor, which explains why the $\ell$-loop integrand has the large
$\Nc$ scaling in eq.~\eqref{eq:Gnlscaling}. In this normalization, the
top component of $\superO$ gives $\Lint/\Nc$ as in~\eqref{eq:Osuper}.}
\beq
\Lint(x,y)
\xrightarrow{\;y\to 0\;}
\Lint(x)
\,.
\eeq
We define the supercorrelator in the self-dual sector as:
\beq\label{eq:Gsuper}
\superG_{n} \equiv \left\langle \prod_{i=1}^{n}\superO(x_i,y_i,\theta_i) \right\rangle\subrm{\!SDYM}
\,.
\eeq
It is then evident that this supersymmetric correlator contains the
loop integrand $G_{n,\ell}$ as a component that can be obtained by
performing projections in the Grassmann ($\theta$) and R-charge ($y$)
variables. We state these projections in detail at the end of this
section in~\eqref{eq:integrandComponent}. Before, we review some of
the basic properties of the supercorrelator.

The free scalar
propagator is normalized to:
\begin{equation}
\avg!{\brk{y_1\cdot\phi(x_1)}\brk{y_2\cdot\phi(x_2)}}\subrm{free}
=\frac{d_{12}}{\Nc}
\qquad \text{with} \qquad
d_{ij}
\equiv\frac{2\,y_i\cdot{y_j}}{x_{ij}^2}
\equiv\frac{-y_{ij}^2}{x_{ij}^2}
\,.
\label{eq:dijdef}
\end{equation}
Under this normalization, we have the large-$\Nc$ scaling:
\beq
\superG_{n} \sim \Nc^{2-n}
\,.
\eeq
Thanks to non-renormalization theorems, the two-, three- and four-point supercorrelators just evaluate to their scalar bottom components:
\beq
\superG_{n} = G_{n,0}\quad\text{for }n=1,2,3,4
\,.
\eeq
These correlators can be computed by Wick contractions using the scalar propagator~\eqref{eq:dijdef}:
\begin{align}
\superG_2&= \log(1+D_{12})+\order{\Nc^{-1}} \overset{y\to0}{=} \left(d_{12} + \frac{d_{12}^2}{2} +\frac{d_{12}^3}{3} +\cdots\right)+\order{\Nc^{-1}} \,,\nonumber \\
\superG_3&= \frac{1}{\Nc}\,D_{12}D_{23}D_{31} +\order{\Nc^{-2}} \,,\nonumber\\
\superG_4&=\frac{1}{\Nc^2} \big{[}D_{12}D_{23}D_{34}D_{14}(1 + 2 D_{13}+ D_{13}^{2} + 2 D_{24} + D_{24}^2) + (1\leftrightarrow 2) + (1\leftrightarrow 4) \nonumber\\
&\qquad\quad + 2 D_{12}D_{13}D_{14}D_{23}D_{24}D_{34}\big{]}  + \order{\Nc^{-3}}
\,,
\label{eq:GG4}
\end{align}
where we introduce a notation for the effective propagator $D_{ij}\equiv{d_{ij}}/\brk{1-d_{ij}}$ as well as the four-, six- and ten-dimensional distances:
\beq
D_{ij} \equiv \frac{-y_{ij}^2}{X_{ij}^2}
\,,\quad
x_{ij}^2\equiv (x_{i}-x_{j})^2 \,,\quad y_{ij}^2\equiv (y_{i}-y_{j})^2 = -2\, y_{i}\cdot y_{j}\,,\quad X_{ij}^2 \equiv x_{ij}^2 +y_{ij}^2
\label{eq:DijDef}
\eeq
The three-point function is given by a single term, thanks to our choice of basis. In the single-particle basis, extremal correlators vanish.

Starting at five points, we have a non-trivial Grassmann dependence on the supercorrelator, which can be organized as:
\beq
\superG_{n} = G_{n,0}+ \sum_{k=1}^{n-4} \superG_{n}^{\NkMHV}\quad\text{for }n\geq 5
\,.
\eeq
The ``MHV'' or bottom component $G_{n,0}$ is Grassmann independent (we collect further results on $G_{n,0}$ for $n\geq 5$ in \secref{sec:Gn0treelevel}).
The $\text{N}^{k}\text{MHV}$ component has Grassmann degree $4k$, and should be decomposable into a basis of superconformal invariants.\footnote{This $\text{N}^{k}\text{MHV}$-nomenclature was used in~\cite{Caron-Huot:2023wdh} in the same supercorrelator context, and is inspired by the Grassmann decomposition of the gluon super-amplitude~\cite{Drummond:2008vq}. Furthermore, from the correlator-amplitude duality,  the light-like limit of the stress-tensor supercorrelator is identical to the (square of) the gluon superamplitude~\cite{Eden:2011yp,Eden:2011ku}.} For the top component  $\superG^{\NkMHV}_{k+4}$, there is a single superconformal invariant~\cite{Eden:2000bk},
whose coefficient is proportional to the reduced correlator $H_{4,k}$ in~\eqref{eq:susywardG4}. However, other, lower components are expected to have a larger basis of susy invariants, see for  instance the three invariants for $\superG^{\NMHV}_6$ in eq.~\eqref{eq:G6NMHV}. There is no classification of these susy invariants in general, see however~\cite{Caron-Huot:2023wdh} for the $\text{NMHV}$ case. In this paper, our main focus is on  the NMHV and $\text{N}^{2}\text{MHV}$ components of seven-point supercorrelator, which contain the loop integrands we are studying.

The supersymmetric correlator contains the loop integrand as a
component that is obtained by performing Grassmann ($\theta$) and
R-charge ($y$) projections as:
\begin{align}
\superG_{n+\ell} &\equiv \left\langle \prod_{i=1}^{n+\ell}\superO(x_i,y_i,\theta_i) \right\rangle\subrm{\!SDYM}
\,,
\label{eq:Gsuper2}
\\
\tilde{G}_{n,\ell} &\equiv
\Nc^{\ell}\,\eval*{
\int\dd[4]{\theta_{n+1}} \dots \dd[4]{\theta_{n+\ell}}
\superG_{n+\ell}^{\NlMHV}
}_{\theta_i\to0}
= \brk[a]*{\prod_{i=1}^{n}\mathcal{O}(x_i,y_i) \prod_{i=1}^{\ell}\Lint(x_i,y_i)}\subrm{\!SDYM}
\label{eq:thetaprojection}
\,,\\
G_{n,\ell} &\equiv \eval*{\tilde{G}_{n,\ell}}_{y_{n+1},\dots,y_{n+\ell}\to0}
\,.
\label{eq:integrandComponent}
\end{align}
These projections are also summarized in \figref{fig:CorrelatorsMethods}. For our targets in~\eqref{eq:Gtargets}, we need the supercorrelators
\beq\label{eq:G7projections}
\superG_6^{\NMHV}\rightarrow G_{5,1}
\,,\quad
\superG_7^{\NMHVf{2}}\rightarrow G_{5,2}
\quad\text{and}\quad
\superG_7^{\NMHV}\rightarrow G_{6,1}
\,.
\eeq
Finally, following the definition~\eqref{eq:Oscalar}
of the operator $\mathcal{O}$, its
correlators $G_{n,\ell}$ serve as generating functions of correlators
of fixed-charge operators $\mathcal{O}_{k}$ at each point:
\begin{equation}
G_{n,\ell}=
\sum_{\mathclap{k_1,\dots,k_n=2}}^{\infty}
\,\avg{k_1\dots k_n}_\ell
\,.
\label{eq:GnlFromFixedCharges}
\end{equation}
These correlators can be extracted by performing the rescaling $y_{i}\to t_{i} y_{i}$ and picking the coefficient of $t_{1}^{k_1}t_{2}^{k_{2}}\cdots t_{n}^{k_n}$ in the series of $t_{i}\to 0$. This  can be effectively performed by using differential operations accompanied by some factorials:
\begin{align}
\brk[a]*{k_1k_2\cdots k_n}_\ell
&\equiv
\brk[a]*{\prod_{i=1}^{n}\mathcal{O}_{k_i}(x_i,y_i) \prod_{i=1}^{\ell}\Lint(x_i)}\subrm{\!SDYM}
\nn\\ &=
\eval*{
\frac{1}{k_1!}
\frac{\partial^{k_1}}{\partial t_1^{k_1}}
\cdots
\frac{1}{k_n!}
\frac{\partial^{k_n}}{\partial t_n^{k_n}}
\,G_{n,\ell}(x_i, t_i\,y_{i})
}_{t_i\to0}
\,.
\label{eq:Rweightcomponent}
\end{align}
%

\section{From Twistors Rules to Loop Integrands}
\label{sec:twistors}

In this section, we explain how we obtain our expressions for the
correlators~\eqref{eq:integrandComponent}
from the twistor rules that were derived in~\cite{Caron-Huot:2023wdh},
which generalize those for the lightest operators of~\cite{Chicherin:2014uca}.
The reader mostly
interested in the results can skip ahead to the subsequent sections.

The super-correlator $\superG_{n+\ell}$ of
super-multiplets~\eqref{eq:Osuper} can
be computed using the twistor Feynman rules developed
in~\cite{Caron-Huot:2023wdh}. These rules efficiently integrate out
the self-dual quantum corrections to the correlators
$\superG_{n+\ell}$. However, the usual caveat inherent to the twistor
formulation of $\superN=4$ sYM theory applies: All intermediate
expressions depend on the choice of an arbitrary but fixed reference
twistor~${\Zref}$. This gauge dependency
makes it virtually impossible to analytically convert the result
generated by the twistor rules to a manifestly invariant expression.
As was done in past
computations~\cite{Caron-Huot:2023wdh,Bargheer:2022sfd,Fleury:2019ydf},
we therefore resort to creating a suitable ansatz in terms of
spacetime and internal invariants $x_{ij}^2$, $y_{ij}^2$, and fix all
freedom in the ansatz by numerical comparison to the twistor result.
To make this process feasible, we restrict the super-correlator
$\superG_{n+\ell}$ to its
bosonic integrand component $G_{n,\ell}$ by performing a projection in
the variables $\theta_i$ and $y_i$ on the twistor side. We can
construct a finite rational ansatz for the generating function
$G_{n,\ell}$, based on its pole structure.
However, due to the large size of this ansatz, we found it practical
to further extract fixed-charge component integrands
$\avg{k_1k_2\dots k_n}_\ell$~\eqref{eq:Rweightcomponent} from the
twistor expression for $G_{n,\ell}$. For these, we can
construct relatively small ansätze
that are easily fixed by numerical comparison to the twistor expression.
After obtaining sufficiently many fixed-charge integrands, we
can finally reconstruct the full generating function $G_{n,\ell}$.

Below, we explain all of the above steps in more detail, starting
with a summary of the twistor rules (\secref{sec:twistor-rules}),
continuing with the projection to the bosonic component $G_{n,\ell}$
(\secref{sec:proj-loop-integr}), observations on the structure of the
integrand and the construction of the ansatz
(\secref{sec:structure-integrand}), and finishing with the extraction
of fixed-charge integrands (\secref{sec:fixed-charge-proj}) and a note
on the separation of parity-even and odd parts (\secref{sec:parity}).

\subsection{Twistor Rules for Supercorrelators}
\label{sec:twistor-rules}

\begin{figure}
\centering
\includegraphics[align=c,width=9cm]{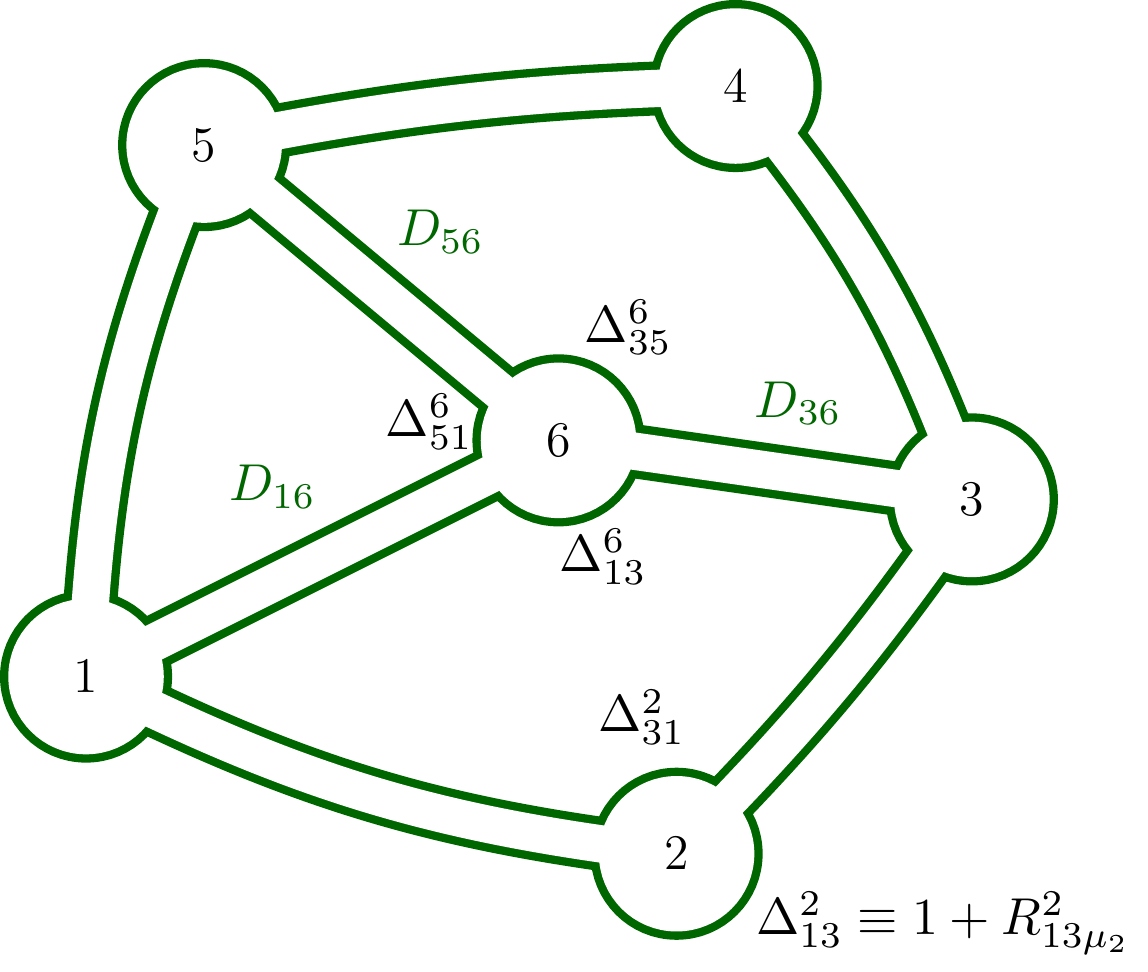}
\caption{Feynman Rules for the planar supersymmetric correlator of operator $\superO$ in the self-dual sector. We show a planar graph in the double-line notation representing the  matrix structure of our composite operators under the gauge group. For each labeled planar graph, we associate the effective propagator $D_{ij}$ to each edge connecting vertices $i$ and $j$.  The weight $\Delta^i_{jk} =1 + R^{i}_{jk\mu_{i}}$  dresses a section of the color line in vertex $i$ delimited by edges $[ij]$ and $[ik]$. We are only depicting the dressings of operators $2$ and $6$. In the single-particle basis, the super-correlator $\superG_{n}$
is obtained by summing over all planar graphs with $n$ vertices of
valency at least~2 and faces with~3 or more sides. }
\label{fig:FatGraphRules}
\end{figure}

We briefly summarize the twistor rules of~\cite{Caron-Huot:2023wdh},
illustrated in \figref{fig:FatGraphRules}.

\paragraph{Supertwistor Space.}

Each supermultiplet $\superO_i=\superO(x_i,y_i,\theta_i)$ is characterized by a subspace $\mathbb{CP}^{1\vert 2}$ in supertwistor space, that can be parameterized by $(\lambda,\psi)$ in the following way
\begin{align}
	Z_i(\lambda)&=\lambda^1 Z_{i,1}+\lambda^2 Z_{i,2}=\lambda^\beta Z_{i,\beta}\,,\\
	\eta_i(\lambda,\psi)&=\lambda^\alpha (\theta_i)^a_\alpha W_{i,a}+\psi^{a'}Y_{i,a'}\,.
\end{align}
Here, $Z_{i,\beta}$ are two bosonic twistors located on the
respective subspace, and $W_{i,a}$ and $Y_{i,a'}$ split the chiral
superspace into two parts, such that the supermultiplets only depend
on the four Grassmann variables $(\theta_i)^a_\alpha$. We can specify
the two bosonic twistors and the decomposition by (here, $B=(b,b')$)
\begin{equation} \label{eq:paramdecompose}
(Z_{i,\beta})_{\alpha\dot\alpha}
=\left(\epsilon_{\alpha\beta},x_{i,\beta\dot\alpha}\right)
\,,\quad
W^B_{i,a}=\left(\delta^b_a,\,0\right)
\,,\quad\text{and}\quad
Y^B_{i,a'}=\left(y^b_{i,a'},\,\delta^{b'}_{a'}\right)\,,
\end{equation}
such that the following $4\times 4$ determinants (denoted by angled
brackets) compute the basic Lorentz and R-charge invariants with unit proportionality factors
\begin{align}
	\langle Z_{i,1}Z_{i,2} Z_{j,1}Z_{j,2}\rangle &\equiv \det (x_i-x_j)^{\dot\alpha \beta}=x_{ij}^2\,,\\
	\langle Y_{i,1}Y_{i,2} Y_{j,1}Y_{j,2}\rangle &\equiv \det (y_i-y_j)^{a b'}=y_{ij}^2\,.
\end{align}

\paragraph{Edges and Vertices.}

The super-correlator $\superG_{n+\ell}$ is expressed as a sum over graphs, where each
vertex represents one of the $n+\ell$ operators $\superO$.
The rules for graph evaluation are as follows:
An edge connecting two vertices $i$ and $j$ represents a bundle of
arbitrarily many ordinary propagators $d_{ij}$ that is re-summed into
a single effective propagator
\begin{equation}
\label{eq:fatedge}
D_{ij}=\frac{d_{ij}}{1-d_{ij}}=d_{ij}+(d_{ij})^2+(d_{ij})^3+\dots\,.
\end{equation}
Additionally, each operator $i$ comes with a vertex factor
$V^i_{j_1\dots j_n}$, of which the lower indices label the operators it
connects with. The order of the lower indices reflects the cyclic
ordering of the edges around the operator $i$. This vertex factor is given by:
\beq\label{eq:Vcyclic}
V^i_{j_1 j_2\dots j_n} \equiv  \Delta^{i}_{j_1j_2}\,\Delta^{i}_{j_2j_3}\cdots \Delta^{i}_{j_nj_1}
\,.
\eeq
At the graph level, see \figref{fig:FatGraphRules}, the weight
$\Delta^{i}_{jk}$ is inserted on the $i$-cycle segment that is
delimited by the edges $(ij)$ and $(ik)$. It is given by:
\beq
\Delta^{i}_{jk}\,=\, 1 + R^{i}_{jk\mu}
\,,
\eeq
where $R$ is a fermionic delta function that carries all the
dependence on the Grassmann superspace coordinate $\theta$. It is
defined more explicitly in eq.~\eqref{eq:Rinv} below, with the replacements
$\lambda_{il}\to \mu$ and $\psi_{il}\to \psi_{\mu}$. Here, $\mu$ is an
arbitrary spinor common to all weights in the same $i$-cycle.
Sometimes it is advantageous to make specific choices of $\mu$ on each
cycle in order to simplify the fermionic dependence. However, in the
present paper, we prefer to get rid of this reference spinor by using
the $\Delta$-algebra:
\beq
\Delta^{i}_{jk}\Delta^{i}_{kj}=1
\quad \text{ and }\quad
\Delta^{i}_{jkl}\equiv \Delta^i_{jk}\Delta^i_{kl}\Delta^i_{lj}
\,,
\eeq
where the weight $\Delta$ with three lower indices is now independent
of the reference spinor~\cite{Caron-Huot:2023wdh}. We can make
repeated use of this $\Delta$-algebra to rewrite the vertex factors~\eqref{eq:Vcyclic} as:
\begin{equation}\label{eq:Vnocyclic}
V^i_{j_1 j_2\dots j_n}
=\Delta^i_{j_1 j_2 j_3}\Delta^i_{j_1 j_3 j_4}\dots\Delta^i_{j_1j_{n-1}j_n}
\,.
\end{equation}
This representation is now manifestly independent of reference spinor
$\mu_i$, however it obscures its cyclic symmetry since it picks $j_1$
as a reference point. Nevertheless, the $\Delta$-algebra implies that
all choices are equivalent. In this work, we
exclusively use the representation~\eqref{eq:Vnocyclic} for the
vertex factor.

Finally, vertices with one or two edges receive trivial weights:
\begin{equation}
V^i_{j}=1 \quad\text{and}\quad V^i_{jk}= 1\,.
\end{equation}

\paragraph{The R-Invariant.}

The weights $\Delta^i_{jkl}$ depend on the four points $i,j,k,l$, and are given by
\begin{equation}\label{eq:Rinv}	\Delta^i_{jkl}=1+R^i_{jkl}\quad\text{with}\quad R^i_{jkl}\equiv\frac{\delta^{0\vert 2}\left(\langle\lambda_{ij}\lambda_{ik}\rangle\psi_{il}+\langle\lambda_{il}\lambda_{ij}\rangle\psi_{ik}+\langle\lambda_{ik}\lambda_{il}\rangle\psi_{ij}\right)}{\langle\lambda_{ij}\lambda_{ik}\rangle\langle\lambda_{il}\lambda_{ij}\rangle\langle\lambda_{ik}\lambda_{il}\rangle}\,,
\end{equation}
where $\delta^{0\vert 2}$ denotes a two-dimensional fermionic delta
function, and the angled brackets are $2\times 2$ determinants:
$\brk[a]{\lambda\mu}\equiv\lambda^\alpha\varepsilon_{\alpha\beta}\mu^\beta$.
The functions $R^i_{jkl}$ are four-point super-invariants, and
are off-shell generalizations of the on-shell invariants that define
tree-level scattering amplitudes~\cite{Chicherin:2014uca}, and have
more recently appeared in super form
factors of local operators~\cite{Basso:2023bwv}. They are completely
antisymmetric in their lower indices. The spinors $\lambda$ are fixed
by the following conditions:
\begin{equation} \label{eq:onshell}
	\lambda_{ij}^\alpha=\epsilon^{\alpha\beta}\frac{\langle Z_{i,\beta}\Zref Z_{j,1}Z_{j,2}\rangle}{\langle Z_{i,1}Z_{i,2} Z_{j,1}Z_{j,2}\rangle}\quad\text{and}\quad \psi_{ij}^{a'}=\epsilon^{a'b'}\frac{\langle Y_{i,b'}E_{ij} Y_{j,1}Y_{j,2}\rangle}{\langle Y_{i,1}Y_{i,2} Y_{j,1}Y_{j,2}\rangle}\,,
\end{equation}
where $\Zref$ denotes a fixed but arbitrary reference twistor, and
$E_{ij}$ is given by
\begin{equation}
	E_{ij}^A=\lambda^\alpha_{ij}(\theta_i)^a_\alpha W_{i,a}^A +\lambda^\alpha_{ji}(\theta_j)^a_\alpha W_{j,a}^A\,.
\end{equation}

\paragraph{Example.}

Let us exemplify these rules by considering a graph that contributes
to the super-correlator $\superG_6$. Applying the twistor Feynman
rules gives
\begin{align}
\includegraphics[align=c]{FigF60TreeGraph}
&= D_{12}D_{13}D_{23}D_{24}D_{25}D_{35}D_{36}D_{46}D_{56}
\,
V^1_{23}V^2_{4135}V^3_{1652}V^4_{26}V^5_{236}V^6_{345}
\nonumber\\[-2ex]
&= D_{12}D_{13}D_{23}D_{24}D_{25}D_{35}D_{36}D_{46}D_{56}
\,
\Delta^2_{413}\Delta^2_{435}\Delta^3_{165}\Delta^3_{152}\Delta^5_{236}\Delta^6_{345}
\,.
\label{eq:example42}
\end{align}
In the last step, we use the representation~\eqref{eq:Vnocyclic} for the vertex factors.

\paragraph{Graphs.}

The last ingredient that we have to specify for the computation of the
super-correlator is the set of
graphs that one needs to sum over. Since we work in the large-$\Nc$
limit, we have to sum over all inequivalent ribbon graphs (also called
fat graphs). Besides the collection of vertices and edges, these
ribbon graphs specify a definite ordering of the edges around each
vertex. This accounts for the color structure of the underlying
Feynman graphs in the large-$\Nc$ limit. The ordering of
edges implies that the faces of each graph become well-defined disks,
and the collection of faces, vertices, and edges define
a punctured compact surface on which the graph is drawn. These are
nothing else than the surfaces in 't~Hooft's genus
expansion~\cite{tHooft:1974jz}. In the sum over graphs, we must
include graphs with multiple edges that connect the same pair of
vertices, as long as these edges are homotopically distinct.
Homotopically equivalent edges must be identified. In order to compute the planar,
connected super-correlator $\superG_{n+\ell}$, all
planar, connected ribbon graphs with $(n + \ell)$ vertices
and any number of edges must be considered. In the single-particle
basis, only graphs whose vertices are at least of valency two
contribute (see \tabref{tab:graph-counting}).

There are various ways to generate the complete list of ribbon graphs
for given genus and number of vertices in practice. One algorithm that
we employed is described in Appendix~B of~\cite{Bargheer:2018jvq}: It
generates all inequivalent graphs by adding bridges to a given set of
ribbon graphs in all possible ways, starting with the ``empty'' graph
that has no edges, but only vertices. Graphs whose genus exceeds the
target genus are discarded. In this construction, the ordering of
edges around each vertex is prescribed from the very beginning. A
different and perhaps more efficient algorithm is explained near
Table~1 of~\cite{Caron-Huot:2023wdh}: Start with all ordinary graphs
(collection of vertices and edges) with the wanted number of vertices.
Promote each graph to a set of ribbon graphs by decorating it with a
prescribed ordering of edges at each vertex in all possible ways.
Finally, ``split'' each edge into homotopically distinct edges in all
possible ways, without increasing the genus.

\begin{table}
\centering
\begin{tabular}{crrrrrrr}
\toprule
genus$\backslash$$n$ & 3       & 4      & 5     & 6      & 7        & 8     & 9 \\
\midrule
0                   & 1       & 4      & 21    & 216    & 3318     & 62767 & 1313096 \\
1                   & 21      & 584    & 20186 & 712862 & 24870531 &       & \\
2                   & 3910    & 542735 &       &        &          &       & \\
3                   & 2902406 &        &       &        &          &       & \\
\bottomrule
\end{tabular}
\caption{Numbers of graphs that contribute to the planar, connected
super-correlator $\superG_n$ for various numbers of points $n$ and
different genera, in the single-particle basis. These are all planar,
connected ribbon graphs with $n$ vertices, where all vertices have
valency at least two. The numbers were obtained by explicit graph
construction.}
\label{tab:graph-counting}
\end{table}
%

\subsection{Projection to Loop Integrands}
\label{sec:proj-loop-integr}

Using the twistor rules explained above, the
super-correlator $\superG_{n+\ell}$ is written as a sum of products of propagators $D_{ij}$
and vertex factors $\Delta^i_{jkl}$. The resulting expression however
has several drawbacks:
\begin{itemize}
\item When expanded in terms of R-invariants~\eqref{eq:Rinv}, the
number of terms grows prohibitively large.
\item The super-correlator is not expressed in terms of basic
spacetime and internal invariants $x_{ij}^2$, $y_{ij}^2$.
\item Each R-invariant depends on the reference twistor $\Zref$,
even though the full correlator $\superG_{n+\ell}$ is independent of
$\Zref$.
\end{itemize}
It would be great to resolve these points for the full
super-correlator $\superG_{n+\ell}$. In this work, we restrict ourselves to
the bosonic integrand component
$G_{n,\ell}$~\eqref{eq:integrandComponent}, \ie the generating
function of loop integrands for scalar half-BPS operators
$\mathcal{O}(x,y)$, which allows us to find a manifestly invariant
closed-form expression.
To obtain $G_{n,\ell}$ from $\superG_{n+\ell}$, we must perform
projections
in the Grassmann variables $\theta$
and
in the internal variables~$y$.

\paragraph{Grassmann Projection.}

A partial projection in the Grassmann variables can be easily done:
The scalar half-BPS operator are
the lowest component in the supermultiplet and thus come with Grassmann
degree zero. In contrast, the chiral Lagrangians are the top component
and thus come with Grassmann degree four. Hence we want to extract the
component of $\theta_{n+1}^4\dots\theta_{n+\ell}^4$.
Since each R-invariant~\eqref{eq:Rinv} is homogeneous of degree two,
we can expand all vertex factors $V$ in terms of $R^i_{jkl}$, and keep
only products of exactly $2\ell$ R~factors. Moreover, only a subset of
R-products will contribute to the component
$\theta_{n+1}^4\dots\theta_{n+\ell}^4$. All others can be set to zero.
For example, one has
\begin{equation}
R^{i}_{jkl}=0 \quad\text{if}\quad i,j,k,l\le n\,.
\label{eq:Rzeros}
\end{equation}
Concretely, we keep only products where each R~factor contains at
least one of the indices $n+1,\dots,n+\ell$, and the product contains
each of those indices at least twice. Among these products, some still
vanish although not apparent from their symbolic expression. We thus
probe all products numerically and remove any that evaluate to zero.

\paragraph{Lagrangian Projection.}

After the $\theta$-projection, the super-correlator is projected to
$\tilde{G}_{n,\ell}$~\eqref{eq:thetaprojection}, which still contains
all higher KK siblings of the interaction Lagrangian $\Lint(x)$. To
project out the higher KK modes, we set $y_i\rightarrow 0$ for $i>n$,
and thus obtain $G_{n,\ell}$. This projection is slightly more subtle, since
individual R-invariants can diverge when taking any~$y_i$ to zero.
Only the complete products of R-invariants and propagators $D_{ij}$
remain finite or vanish. In order to take the limit,
it is necessary to know how products of R-invariants scale with the
polarization vectors~$y_i$. We can measure these scalings by dressing
the polarization vectors $y_i$ with scalar factors $t_i$:
\begin{equation}
\label{eq:ytscaling}
y^b_{i,a'}\rightarrow t_i^{1/2}\, y^b_{i,a'}
\qquad \text{for }i>n
\,.
\end{equation}
The R-invariants depend on $y_i$
through~$Y$ and~$W$, which scale as%
\footnote{While the scaling of $Y_{i}$ can be directly inferred
from~\eqref{eq:paramdecompose}, the scaling of $W_{i}$ might seem
surprising. $\grp{SU}(4)$ unitarity conditions impose that
\begin{equation*}
W^B_{i,a} \overline{Y}^c_{i,B}=\delta^c_a
\quad\text{with}\quad
\overline{Y}^c_{i,B}=(\delta^c_b,\,-y_{i,b'}^c)
\quad\text{such that}\quad
Y^B_{i,a'}\overline{Y}^a_{i,B}=0
\,.
\end{equation*}
Since $\overline{Y}_i$ scales with the same power as $Y_i$, the above
condition implies that $W_i$ scales with the inverse power of the
respective polarization vector.}
\begin{equation}
Y^B_{i,a'}\rightarrow t_i^{1/2}\,Y^B_{i,a'} \quad\text{and}\quad W^B_{i,a}\rightarrow t_i^{-1/2}\,W^B_{i,a} \,.
\label{eq:YWscaling}
\end{equation}
The scaling of a given R-product can in principle be found analytically by explicit
expansion.
However, we found it much easier to probe the scaling numerically by inserting
random numerical values for the spacetime coordinates and polarization
vectors, while keeping the scalar factors~$t_i$ symbolic.%
\footnote{Although not immediately apparent from the definition~\eqref{eq:Rinv},
after the projection to the $\theta_{n+1}^4\dots\theta_{n+\ell}^4$
component, each product of R-invariants is a homogeneous function of
all $t_i$. Heuristically, we find that for every $i>n$, every product of R-invariants
scales as $t_i^{-2-\#i}$, where $\#i$ is the number of occurrences of $i$ in
the upper labels of the R-invariants in the product. See \eg the
examples in~\eqref{eq:exampleterms}.}

Moreover, the fat propagators $D_{ij}$ can be expanded in
ordinary propagators $d_{ij}$, which scale as $d_{ij}\rightarrow t_i
t_j d_{ij}$. Once the scaling behavior of all factors in a given
product is known, one can safely take the $t_i\rightarrow0$ limit,
thus projecting to the loop integrand $G_{n,\ell}$.

\paragraph{Example.}

Let us again exemplify this procedure by
considering~\eqref{eq:example42} and computing which terms survive
when projecting
to the four-point two-loop integrand $G_{4,2}$. First, we set $\theta_i=0$
for $i\le n$ and obtain
\begin{align}
\Delta^2_{413}\Delta^2_{435}\Delta^3_{165}\Delta^3_{152}\Delta^5_{236}\Delta^6_{345}
&=1\,\brk{1+R^2_{435}}\,\brk{1+R^3_{165}}\,\brk{1+R^3_{152}}\,\brk{1+R^5_{236}}\,\brk{1+R^6_{345}}
\nonumber\\
&=R^2_{435}R^3_{165}R^3_{152}R^5_{236}+R^2_{435}R^3_{165}R^3_{152}R^6_{345}+R^2_{435}R^3_{165}R^5_{236}R^6_{345}
\nonumber\\
&\qquad+R^2_{435}R^3_{152}R^5_{236}R^6_{345}+R^3_{165}R^3_{152}R^5_{236}R^6_{345}+\dots\,.
\end{align}
The dots indicate terms with fewer or more R~factors,
which are omitted, since they cannot contribute to the two-loop
integrand. We compute the R-charge weights of the five four-products by
probing them numerically and obtain
\begin{align}
R^2_{435}R^3_{165}R^3_{152}R^5_{236}
&\sim
t_5^{-3}\times t_6^{-2}
\,,\nn\\
R^2_{435}R^3_{165}R^3_{152}R^6_{345}
&\sim
t_5^{-2}\times t_6^{-3}
\,,\nn\\
R^2_{435}R^3_{165}R^5_{236}R^6_{345}
\,,
R^2_{435}R^3_{152}R^5_{236}R^6_{345}
\,,
R^3_{165}R^3_{152}R^5_{236}R^6_{345}
&\sim
t_5^{-3}\times t_6^{-3}
\,.
\label{eq:exampleterms}
\end{align}
On the other side, the product of propagators scales as
\begin{equation}
D_{12}D_{13}D_{23}D_{24}D_{25}D_{35}D_{36}D_{46}D_{56}
\sim
t_5^{3}\times t_6^{3}+\text{higher powers in }t_5,t_6
\,.
\end{equation}
We conclude that in the limit $t_5,t_6\rightarrow 0$, only the terms
in the last line of~\eqref{eq:exampleterms} survive, while the others only contribute to
higher KK siblings of the Lagrangian, and are thus set to zero.
Furthermore, in the fat propagators $D_{ij}$ that
involve one of the Lagrangian insertion points, only the leading terms
contribute. All in all, we obtain
\begin{multline}
\eval{\eqref{eq:example42}}_{G_{4,2}}
=D_{12}D_{13}D_{23}D_{24}\times d_{25}d_{35}d_{36}d_{46}d_{56}
\\
\times\left(R^2_{435}R^3_{165}R^5_{236}R^6_{345}+R^2_{435}R^3_{152}R^5_{236}R^6_{345}+R^3_{165}R^3_{152}R^5_{236}R^6_{345}\right)
\label{eq:exampleG42}
\,.
\end{multline}

\subsection{Structure of the Integrand}
\label{sec:structure-integrand}

We have explained how to extract the integrand generating function
$G_{n,\ell}$ from the super-correlator $\superG_{n+\ell}$, constructed
from the twistor rules. What remains is to translate the twistor
expression to coordinate space, \ie to an expression in terms of basic
Lorentz and internal invariants $x_{ij}^2$, $y_{ij}^2$. This proves
quite difficult, in particular due to the dependence of all
subexpressions on the reference twistor $\Zref$.

\paragraph{Singularities.}

To obtain an expression in coordinate space, we therefore opt to construct
an ansatz, whose unknown coefficients are determined by a
numerical fit to the twistor expression, as was done in
earlier computations~\cite{Fleury:2019ydf,Bargheer:2022sfd}.
The form of the ansatz for $G_{n,\ell}$ follows from
the twistor rules together with
the fact that $G_{n,\ell}$ is a
generating function for fixed-charge loop
integrands~\eqref{eq:Rweightcomponent}. Namely, it can be written as a
finite polynomial $P_{n,\ell}$ in $d_{ij}$ and $x_{ij}^2$, multiplied by an overall
monomial that absorbs all singularities:
\begin{equation}
G_{n,\ell}=
\brk[s]3{\mspace{2mu}
  \prod_{1\leq i<j\leq n}
  \brk2{
    \frac{x_{ij}^2}{X_{ij}^2}
  }^{\mspace{-5mu}n-3+\delta_{\ell,0}}
}
\brk[s]3{\mspace{2mu}
  \prod_{\substack{1\leq i<j\leq n+\ell\\n<j}}
  \frac{1}{x_{ij}^2}
}
P_{n,\ell}
\,.
\label{eq:GFansatzRefined}
\end{equation}
The fact that $G_{n,\ell}$ is a polynomial in the fat
propagators $D_{ij}\sim1/X_{ij}^2$ with $i,j\leq n$ immediately follows
from the twistor rules. All $D_{ij}$ with $i>n$ or $j>n$ are reduced
to powers of ordinary propagators $d_{ij}$, because we projected the
Lagrangian points to their lowest-charge component ($y_i\to0$ for
$i>n$). Moreover, because we
restrict to graphs of sphere topology (the leading term in the planar limit),
the maximal power of any $D_{ij}$ is $n-2$.%
\footnote{This is the
maximal number of homotopically distinct lines between any two
punctures on the $n$-punctured sphere. The bound makes use of the fact
that the fat propagators do not get re-normalized, \ie two identical
propagators $D_{ij}$ cannot be separated only by Lagrangian points.}
In fact, the maximal exponent $n-2$ on $D_{ij}$ only occurs for
$\ell=0$. The reason is that the only $n$-point graph with $n-2$
propagators $D_{ij}$ (whose vertices have valency two or more) has the
following topology (here $n=5$):
\begin{equation}
\includegraphics[align=c]{Fig5ptGraph3IdProp}
\label{eq:maxPoleGraph}
\end{equation}
However, this graph receives no loop corrections, \ie inserting any
number of Lagrangian insertions into the faces gives a zero
contribution, because all external operators are BPS, and
all faces are triangular and therefore protected.%
\footnote{If this was not the case, three-point functions of BPS
operators, which are products of two triangular faces (at leading order
in $1/\Nc^2$), would receive loop corrections. But three-point
functions of BPS operators are tree-level exact, hence also triangular faces
are free of loop
corrections~\cite{Intriligator:1998ig,Intriligator:1999ff}. Strictly
speaking, this
only holds at the integrated level, but we find that it is
also true for the integrand.}
Loop corrections (correlators with Lagrangian insertions) are only
non-trivial if the graph of $D_{ij}$ propagators has at least one face
that touches four or more vertices. For such graphs, the maximal
exponent on any $D_{ij}$ is $n-3$, which explains the
$\delta_{\ell,0}$ in the exponent of~\eqref{eq:GFansatzRefined}.
Using $D_{ij}=d_{ij}/(1-d_{ij})=d_{ij}\,x_{ij}^2/X_{ij}^2$,
it follows that the first factor in~\eqref{eq:GFansatzRefined} absorbs
all singularities in $X_{ij}^2$, while maintaining polynomiality in $d_{ij}$.

The inverse powers of $x_{ij}^2$ in the second factor
of~\eqref{eq:GFansatzRefined} absorb all divergences in
$x_{ij}^2\to0$ limits, and their exponents follow from the OPEs of the
operators $\mathcal{O}$ and $\Lint$, as well as the fact that we chose
to write $P_{n,\ell}$ as a polynomial in $d_{ij}$ as opposed to $y_{ij}^2$.
The conformal weights of the BPS and Lagrangian
operators in $G_{n,\ell}$~\eqref{eq:Gnl-integrand} fix the scaling weights in $x_i$
of the polynomial $P_{n,\ell}$ to $+\ell$ for the external
points $x_{i\leq n}$,
and $\ell+n-5$ for the Lagrangian points $x_{n<i}$.
The degree in $d_{ij}$ is also bounded, and grows with the loop order $\ell$.

\paragraph{Divide and Conquer.}

In principle, all freedom in the polynomial $P_{n,\ell}$ can now
be determined by numerical comparison to the twistor expression on
sufficiently many different kinematic points. However, the ansatz
for $P_{n,\ell}$ quickly becomes very large, unless $n$ and $\ell$
are very small.%
\footnote{The limiting factor of this approach is the size of the
dense linear system that must be solved, which is determined by the number of
unknowns in the ansatz. In practice, we use \mathematica’s
\texttt{NSolve}, which handles such systems feasibly up to about
10,000 unknowns. A full ansatz for the generating function would far
exceed this limit, so the problem must be approached in smaller
steps.}
This is alleviated by noting that~\eqref{eq:GFansatzRefined} can be
significantly refined: Since
not all powers of $D_{ij}$ can appear at once (the maximal number of
edges of a sphere graph with $n$ vertices is $3n-6$), the
freedom in the polynomial $P_{n,\ell}$ can be
substantially reduced by splitting the ansatz into various terms that
come with different combinations of powers of the various $D_{ij}$.
In other words, $P_{n,\ell}$ can be obtained by grouping $P_{n,\ell}$
into different $D_{ij}$ monomials, and fixing the (much smaller) ansätze
for each of these numerically.

In practice, we employ a slightly different but equivalent ``divide
and conquer'' strategy: We resort to extracting fixed-charge
component correlators $\avg{k_1\dots k_n}_\ell$ from the twistor
expression for~$G_{n,\ell}$.
For these fixed-charge integrands, equally compact ansätze can be constructed
and numerically matched against the twistor answer, which is
independent of the reference twistor.
Once sufficiently many
fixed-charge correlators are obtained,
matching~\eqref{eq:GFansatzRefined}
against the expansion~\eqref{eq:GnlFromFixedCharges} determines all degrees of freedom in the
polynomial ansatz~\eqref{eq:GFansatzRefined},
and thus the full generating function $G_{n,\ell}$ is
recovered.
This procedure divides the problem of
fitting $G_{n,\ell}$ to smaller
computations that can be done successively.

\subsection{Fixed-R-Charge Projection}
\label{sec:fixed-charge-proj}

\paragraph{Twistor Projection.}

To extract fixed-charge components $\avg{k_1\dots k_n}_\ell$ from the
twistor expression for $G_{n,\ell}$, we
use~\eqref{eq:Rweightcomponent} by applying the
scaling~\eqref{eq:ytscaling}, but this time for the first $n$ points:
\begin{equation}
\label{eq:ytscaling2}
y^b_{i,a'}\rightarrow t_i^{1/2}\, y^b_{i,a'}
\qquad \text{for }i\leq n
\,.
\end{equation}
Expanding both the products of R-invariants and the fat
propagators $D_{ij}$ to sufficiently high powers
in the scaling parameters $t_i$, we can collect all terms proportional
to $t_1^{k_1}\dots t_n^{k_n}$ that contribute to $\avg{k_1\dots k_n}_\ell$.

\paragraph{Example.}

Let us again turn to the example in~\eqref{eq:exampleG42} and derive
what terms contribute to the two-loop integrand $\brk[a]*{2442}_2$ of
operators with charges $2$, $4$, $4$, and $2$,
see~\eqref{eq:Rweightcomponent} for the relation to $G_{4,2}$.
Introducing scale factors $t_i$ via~\eqref{eq:ytscaling2}, using~\eqref{eq:YWscaling}, and
numerically probing the remaining three R-products in~\eqref{eq:exampleG42} yields
\begin{align}
R^2_{435}R^3_{165}R^5_{236}R^6_{345}
\,,\,
R^2_{435}R^3_{152}R^5_{236}R^6_{345}
\,&\sim\,
t_2^{-1}\times t_3^{-1}
\,,
\label{eq:2442component}
\\
R^3_{165}R^3_{152}R^5_{236}R^6_{345}
\,&\sim\,
t_3^{-2}
\,,
\end{align}
while the propagator factors scale as
\begin{equation}
D_{12}D_{13}D_{23}D_{24}\times d_{25}d_{35}d_{36}d_{46}d_{56}
\,\sim\,
t_1^{2}\times t_2^{4}\times t_3^{4}\times t_4^{2}
+\text{higher powers in }t_i
\,.
\end{equation}
One can see that the only terms proportional to $t_1^2t_2^4t_3^4t_4^2$
are the two products in~\eqref{eq:2442component} dressed with ordinary
propagators as follows:
\begin{align}
\eval{\eqref{eq:exampleG42}}_{\brk[a]*{2442}_2}
=
d_{12}d_{13}d^2_{23}d_{24}
\,d_{25}d_{35}d_{36}d_{46}d_{56}
\brk*{
R^2_{435}R^3_{165}R^5_{236}R^6_{345}+R^2_{435}R^3_{152}R^5_{236}R^6_{345}
}
\,.
\end{align}
%

\paragraph{Fixed-Charge Ansatz.}

Finally, let us briefly explain how to construct the coordinate-space
ansatz for the fixed-charge correlators $\avg{k_1\dots k_n}_\ell$,
following~\cite{Bargheer:2022sfd}. Due to Lorentz and R-symmetry
invariance, it must be a function of the basic invariants
$x_{ij}^2=(x_i-x_j)^2$ and $y_{ij}^2=(y_i-y_j)^2=-2\,y_i\cdot y_j$.
Considering the various OPE limits among the fixed-charge BPS operators and
Lagrangian insertions, the correlator can be written as
\begin{equation}
\brk[a]*{k_1\dots k_n}_\ell=
\frac{C}{\prod_{i=1}^{n+l}
\prod_{i< j=n+1}^{n+l}x_{ij}^2}
\sum_{\mathbf{a}}
\brk3{\,\prod_{i<j=1}^{n}d_{ij}^{a_{ij}}}
P_\mathbf{a}^\ell(x_{ij}^2)\,,
\label{eq:ansatz}
\end{equation}
where $C$ is a constant that depends on the number of colors $N$
and the charges $k_i$ of the external operators. The sum runs over all
possible propagator $y$-structures $\mathbf{a}=\{a_{ij}\}$ that are
compatible with the operator charges, \ie that satisfy
\begin{equation}
k_i=\sum_{j\neq i}^{n} a_{ij}\,\quad \text{for all}\quad i=1,\dots, n\,.
\end{equation}
Finally, $P_\mathbf{a}^\ell(x_{ij}^2)$ is a polynomial in $x_{ij}^2$
whose conformal weights in $x_i$ is fixed by the conformal weights of the operators: The
BPS operators have weights $k_i$, whereas the
Lagrangian insertions have weight four. Taking into account the
powers of $d_{ij}$ as well as the denominator in~\eqref{eq:ansatz},
the total degree of $P_\mathbf{a}^\ell$ in the external points $i\leq n$
must be $+\ell$, and $(n+\ell-5)$ in the internal points $i>n$. Moreover,
the polynomial is constrained by the permutation symmetries of the
correlator $\avg{k_1\dots k_n}_\ell$ and the respective $y$-structure.

The coefficients of the polynomials $P_\mathbf{a}^\ell$ are
numerically fitted against the twistor expression. There are a
few further constraints one can
impose on the polynomials beforehand, which we discuss in
\appref{app:fixedChargeAnsatz}, where
we also illustrate the construction with a concrete example.

\subsection{A Note on Parity}
\label{sec:parity}

The correlator $G_{n,\ell}$ can be decomposed into parity-even and parity-odd
parts. The ansätze for the parity-even part
are~\eqref{eq:GFansatzRefined} and~\eqref{eq:ansatz}, where in both
cases the spacetime dependence enters only through parity-symmetric
distances~$x_{ij}^2$. The ansätze for the parity-odd part are
identical, only that now every term in the polynomials $P_{n,\ell}$,
$P_{\mathbf{a}}^\ell$ must contain a parity-odd factor
\begin{equation}
\eps_{\mu\nu\rho\sigma} x_{i}^\mu x_{j}^\nu x_{k}^\rho x_{l}^\sigma
\,.
\end{equation}
In order to properly build the ansätze, it is useful to write these
parity-odd factors into a conformally covariant object that has
well-defined conformal weights in each point. To this end, we
transform the four-dimensional Minkowski spacetime points to
six-dimensional projective vectors $X^I$ upon which the conformal
symmetry $\grp{SO}(4,2)$ acts linearly. For example, a suitable
representation for these vectors is%
\footnote{This $X^{I}$, with $I=1,\dots,6$, is a six-dimensional
vector, which should not be confused with the ten-dimensional vector
$X=(x,y)$ used ubiquitously in our results. We only use the 6d vector
$X^{I}$ to write the antisymmetric invariant $X^{\text{anti}}$ that
enters the odd part of the five-point function~\eqref{eq:R6b}.}
\begin{equation} \label{eq:vec6d}
X^I = \left(\frac{1-x^2}{2},\, x^\mu,\, \frac{1+x^2}{2}\right)
\,.
\end{equation}
Starting from six points, it is possible to write down a parity-odd covariant object by contracting all indices with the Levi-Civita tensor
\begin{equation} \label{eq:Xanti}
X^\mathrm{anti}_{123456} = \eps_{IJKLMP}X^I_1X^J_2X^K_3X^L_4X^M_5X^P_6
\,.
\end{equation}
With the representation~\eqref{eq:vec6d} at hand, one can readily relate this covariant to the four-dimensional notation by
\begin{align}
X^\mathrm{anti}_{123456} &=
\frac{1}{2}\frac{1}{4!}\sum_{\sigma \in \grp{S}_5}
\sgn(\sigma)
\,x_{\sigma_56}^2
\,\eps_{\mu\nu\rho\tau}
x_{\sigma_16}^\mu x_{\sigma_26}^\nu x_{\sigma_36}^\rho x_{\sigma_46}^\tau
\nn\\
&=
\frac{1}{2}
\,x_{56}^2
\,\eps_{\mu\nu\rho\tau}
x_{16}^\mu x_{26}^\nu x_{36}^\rho x_{46}^\tau
+(\grp{C}_5\text{ perms})
\,.
\label{eq:XantiRewritten}
\end{align}
Here, the point 6 is not special in any fundamental way. We simply choose to shift by $x_6^\mu$ in order to obtain a translationally invariant expression. For a more detailed discussion, we refer to Appendix B of~\cite{Ambrosio:2013pba}.

In the numerical twistor computation, even and odd parts are easily
separated: Since we work in Lorentzian signature, the parity-odd terms
appear as imaginary parts, whereas parity-even parts are real. Thus it
is easy to isolate even and odd parts, and they can be separately
matched against the respective ansatz.

\section{Tree-Level Correlators \texorpdfstring{$G_{n,0}$}{Gn0}}
\label{sec:Gn0treelevel}

In this section, we present the tree-level connected correlators at
leading order in large~$\Nc$. They are obtained from Wick contractions represented by planar graphs, with each edge weighted by the effective ten-dimensional propagator $D_{ij}$~\eqref{eq:DijDef}. As discussed in \secref{sec:twistor-rules}, the graphs of $G_{n,0}$
serve as seeds for the twistor construction of the supercorrelator
$\mathbb{G}_{n}$, from which one extracts the loop integrands
$G_{n-\ell,\ell}$ with
loop orders $\ell=1,\dots, n-4$. The ten-dimensional poles of these integrands can in part be traced back to those of $G_{n,0}$.

\paragraph{Four Points.}

The four-point generating function~\eqref{eq:GG4} reads
\begin{align}
G_{4,0}&=\frac{1}{\Nc^2}
\brk3{\,\prod_{i<j=1}^4 \!\! D_{ij}}
\brk[s]*{
\frac{D_{12}}{4 D_{34}}
+
\frac{1}{12}
+
\frac{1}{2 D_{24}}
+
\frac{1}{8 D_{13}D_{24}}
+
(\grp{S}_4\text{ perms})
}
\,,
\end{align}
where $+(\grp{S}_4\text{ perms})$ stands for $23$ repetitions of all
previous terms with $\grp{S}_4$-permuted point labels. We will use
this notation everywhere below.

\paragraph{Five Points.}

Summing the $21$ different single-particle five-point graphs (see
figure 6 in~\cite{Caron-Huot:2021usw}) and
summing over inequivalent permutations, we find
the five-point tree-level generating function
\begin{align}
\label{eq:G50}
G_{5,0}
&=\frac{1}{\Nc^3}\brk3{\,\prod_{i<j=1}^5 \!\! D_{ij}}
\bigg[
\frac{1}{6D_{34}D_{35}D_{45}}{D_{12}^2}
+\frac{1}{2D_{24}D_{35}D_{45}}\brk*{1+\frac{1}{D_{23}}}\,{D_{12}D_{13}}
\\ & \mspace{60mu}
+\frac{1}{2D_{35}D_{45}}\brk*{1+\brk*{4+\frac{1}{D_{13}}+\frac{2}{D_{23}}}\frac{1}{D_{24}}+\frac{1}{D_{34}}}{D_{12}}
\nn\\ & \mspace{60mu}
+\frac{1}{60D_{45}} \Big(
    10
    + 15 \brk*{\frac{1}{D_{12}} + \frac{4}{D_{14}}}
    + 30 \frac{1}{D_{14}} \brk*{\frac{1}{D_{15}} + \frac{1}{D_{23}} + \frac{4}{D_{25}}}
\nn\\ & \mspace{130mu}
    + \frac{10}{D_{14}} \brk*{\frac{1}{D_{15}D_{23}} + \frac{3}{D_{12}D_{25}} + \frac{6}{D_{13}D_{25}}}
    + \frac{6}{D_{13}D_{14}D_{23}D_{25}}
\Big)
\nn\\ & \mspace{480mu}
+ (\text{$\grp{S}_5$ perms})
\bigg]
\nn\,,
\end{align}
Note that there are only $18$ terms in the sum~\eqref{eq:G50}, whereas
there are $21$ different tree-level graphs. The reason is that some tree-level
graphs have different topologies, but lead to the same products of
$D_{ij}$ propagators (which thus get extra symmetry factors).
For example, the three ribbon graphs
\begin{equation}\label{eq:5pointGraphs}
\includegraphics[align=c]{FigF50TreeGraph1}
\quad
\includegraphics[align=c]{FigF50TreeGraph2}
\quad
\includegraphics[align=c]{FigF50TreeGraph3}
\end{equation}
represent non-identical contractions, but
all contribute the same term $1/D_{45}D_{14}D_{25}$ to the square
bracket in~\eqref{eq:G50}.  On the middle graph, the vertex labels can
be permuted in $120$ inequivalent ways, whereas the left and right
graphs have only $60$ permutations each. This leads to the overall
term $2/D_{45}D_{14}D_{25}$ in the square bracket.

\paragraph{Six Points.}

The six-point generating function at tree level can be written as a
sum of $127$ terms, plus $\grp{S}_6$ permutations. We only quote the
terms with the largest number of repeating propagators ($4$~in this case):
\begin{align}
\label{eq:G60}
G_{6,0}&=\frac{1}{\Nc^4}\brk3{\,\prod_{i<j=1}^6 \!\! D_{ij}}
\bigg[
\frac{D_{12}^3}{8 D_{34} D_{35} D_{36} D_{45} D_{46} D_{56}}
+ \brk*{1 + \frac{1}{D_{23}}}\frac{D_{12}^2 D_{13}}{D_{24} D_{35} D_{36} D_{45} D_{46} D_{56}}
\\ & \mspace{-10mu}
+ \frac{D_{12} D_{13} D_{14}}{2 D_{23} D_{25} D_{36} D_{45} D_{46} D_{56}}
+ \frac{D_{12} D_{13} D_{15}}{D_{23} D_{24} D_{35} D_{36} D_{45} D_{46} D_{56}}
+ \frac{D_{12} D_{13} D_{23}}{6 D_{14} D_{25} D_{36} D_{45} D_{46} D_{56}}
\nn\\ & \mspace{-10mu}
+ \brk*{1 + \frac{2}{D_{23}} + \frac{1}{D_{14} D_{23}}}
\frac{D_{12} D_{13}  D_{3 4}}{2 D_{15} D_{24} D_{25} D_{36} D_{46} D_{56}}
+ (\text{lower-order terms})
+ (\text{$\grp{S}_6$ perms})
\bigg]
\nn\,,
\end{align}
where ``lower-order terms'' stands for terms with fewer factors
$D_{ij}$ in the numerator. See the attached \mathematica file
\ancfile{G60.m} for the full expression.

\paragraph{Higher Points.}

At more than six points, the twistor computation becomes very
demanding due to the large number of contributing terms, and the
increasingly large ansatz. Beyond six points, we only computed the
leading-order generating function $G_{7,0}$. As expected, it displays
ten-dimensional poles of degree up to five. It is too long to display,
we supply it in the attached \mathematica file \ancfile{G70.m}.

\section{Review of Four-Point Integrands \texorpdfstring{$G_{4,\ell}$}{G4l}}
\label{sec:4pointreview}

The loop integrands of planar four-point correlators with arbitrary
R-charge were computed to three loops~\cite{Chicherin:2015edu}, and
later to five loops~\cite{Chicherin:2018avq}. The results show that
all correlators can be obtained from a finite basis of fixed-charge
correlators $\brk[a]{k_1,\dots,k_4}_{\ell}$ with small R-charges~($k_i<\ell+2$), and the
dimension of the basis grows with the loop order.

Subsequently, it was observed that
performing a resummation of all R-charge correlators results in a
compact generation function~\cite{Caron-Huot:2021usw}, which we presently denote as
$G_{4,\ell}$. This resummation also unveiled a hidden ten-dimensional
symmetry. A consequence of this symmetry is that the generating
function can be simply obtained from the stress-tensor (reduced)
integrand by performing the uplift:
$x_{ij}^2\to X_{ij}^2\equiv x_{ij}^2+y_{ij}^2$.

In the following, we review the known results for the four-point
generating function $G_{4,\ell}$ up to three loops. As shown below in
\secref{sec:five-point-integrand} and \secref{sec:six-point-integrand},
the correlators $G_{4,\ell}$ also enter our results on higher-point
integrands, as coefficients of their higher-order 10d poles.

The one-loop four-point integrand can be obtained from the
$\text{NMHV}$ component of the five-point supercorrelator, given
in~\cite{Caron-Huot:2023wdh} as:
\begin{equation}
\superG_5^{\NMHV}=
\frac{-2\,\mathcal{I}^{(5)}_{12345}}{\Nc^3}\prod_{i<j}^{5}\frac{y_{ij}^2}{X_{ij}^2}
\,,
\label{eq:G5super}
\end{equation}
where the superconformal invariant can be expressed as a degree-two polynomial on the fermionic delta function $R$, defined in~\eqref{eq:Rinv}, as:
\begin{equation}
\label{eq:I5superinv}
\mathcal{I}^{(5)}_{12345}
= \sum_{10}\frac{R^{1}_{345}R^{2}_{345}}{d_{12}d_{34}d_{45}d_{35}}
+ \sum_{60}\frac{R^{1}_{234}R^{2}_{135}}{d_{15}d_{24}d_{34}d_{35}}
+ \sum_{15}\frac{R^{1}_{234}R^{1}_{345}}{d_{23}d_{34}d_{45}d_{52}}
\end{equation}
This combination of $R^{i}_{jkl}$ with $d_{ij}$ coefficients has
permutation invariance on the five points. The numbers on the sums
count the $\grp{S}_5$ inequivalent permutations, where in the last sum we
have to take into account the identity $R^1_{234}R^1_{345}=R^1_{245}R^1_{345}$. Most importantly, it is
independent of the reference twistor $\Zref$, used in defining the
spinors of~\eqref{eq:onshell}, and hence superconformally invariant. For
an alternative representation of this same invariant see eq.~(5.22)
of~\cite{Chicherin:2014uca}. In order to extract the loop integrand
$G_{4,1}$, here we specialize to the Grassmann
projection~\eqref{eq:thetaprojection} of this invariant:
\begin{equation}
\label{eq:R5}
R^{(5)}_{1234;5}\equiv
\eval2{\mathcal{I}^{(5)}_{12345}}_{\theta_5^4}\times\prod_{i<j}^{5}d_{ij}
= -\frac{\mathcal{R}_{1234}}{x_{15}^2 x_{25}^2 x_{35}^2 x_{45}^2}
\end{equation}
with the four-point polynomial defined as:
\begin{equation}
\label{eq:R1234first}
\mathcal{R}_{1234}\equiv \sfrac{1}{8}
x_{12}^2x_{34}^2\,(d_{12}d_{34}-d_{13}d_{24})(d_{12}d_{34}-d_{14}d_{23})
+(\grp{S}_4\text{ perms})
\,.
\end{equation}
Finally, by doing the R-charge projection on the last point
$y_{5}\to 0$ to obtain the Lagrangian operator $\Lint(x_6)$, we can extract the
one-loop integrand $G_{4,1}$. By repeating this computation for other
supercorrelators of the form $\mathbb{G}_{4+\ell}^{\NlMHV}$, we
can extract the loop integrands $G_{4,\ell}$. Here we present the
first three loop orders:
\begin{align}
\superG_{5}^{\NMHV}\longrightarrow G_{4,1} &= \frac{2\mathcal{R}_{1234}}{\Nc^2}\prod_{i<j}^{4}x_{ij}^2\times\frac{1}{\prod\limits_{1\leq i<j\leq 5}X_{ij}^2}\,\bigg{|}_{y_5\to 0}
\label{eq:G41}
\\
\superG_{6}^{\NMHVf{2}}\longrightarrow G_{4,2} &= \frac{2\mathcal{R}_{1234}}{\Nc^2}\prod_{i<j}^{4}x_{ij}^2\times\frac{X_{12}^2 X_{34}^2 X_{56}^2 + \text{14 perm.} }{\prod\limits_{1\leq i<j \leq 6}X_{ij}^2}\,\bigg{|}_{y_5,y_6\to 0}
\label{eq:G42}
\\
\superG_{7}^{\NMHVf{3}}\longrightarrow G_{4,3} &= \frac{2\mathcal{R}_{1234}}{\Nc^2}\prod_{i<j}^{4}x_{ij}^2\times\frac{(X_{12}^2)^2(X_{34}^2 X_{45}^2 X_{56}^2 X_{67}^2 X_{73}^2)+ \text{251 perm.} }{\prod\limits_{1\leq i<j \leq 7}X_{ij}^2}\,\bigg{|}_{y_5,y_6,y_7\to 0}
\label{eq:G43}
\end{align}
They all have the polynomial $\mathcal{R}_{1234}$ as a prefactor,
as in~\eqref{eq:susywardG4}. The coefficient of $\mathcal{R}_{1234}$,
known as reduced correlator, displays ten-dimensional conformal
invariance, made explicit by its exclusive dependence on the 10d distances $X_{ij}^2$. Furthermore, the form of the reduced integrand is the same as for the lightest operator but with the  replacement $(x_{ij}^2)^{\text{4d}}\to (X_{ij}^2)^{\text{10d}}$. This means that, in principle, the current knowledge of the stress tensor integrand up to twelve loops, see~\cite{He:2024cej} and~\cite{Bourjaily:2025iad}, also gives the generating function for all KK modes by doing this 10d uplift.\footnote{This statement has a
caveat pertaining to the existence of 4d Gram identities. These are
polynomials in $X_{ij}^2$ which vanish when reduced to four-dimensional $x_{ij}^2$, due
to the lower dimensionality of the vector space. If present, such
terms will be missed by the uplift of the stress-tensor correlator.} There are other approaches based on positive geometry, see~\cite{Eden:2017fow,He:2024xed,He:2025rza}, which construct the loop integrand of the stress-tensor correlator and could be generalized to our 10d generating functions.

At the integrand level, this 10d symmetry   presents spacetime and R-charge kinematics on the same footing, however, this is explicitly broken at the integrated level, since we only integrate the Lagrangian positions on 4d spacetime.  Moreover, this integration can be expressed on a basis of conformal integrals, see eq.~\eqref{eq:G42} below, which depend on 4d conformal cross-ratios and are known analytically in general kinematics up to three-loop order~\cite{Drummond:2013nda}. At higher-loop orders, there is recent progress on computing the relevant basis of conformal integrals, see~\cite{He:2025vqt} and~\cite{He:2025lzd}.

\section{Five-Point Integrands}
\label{sec:five-point-integrand}

In this section we report on the five-point integrands $G_{5,1}$ and
$G_{5,2}$ that we obtained following the method of \secref{sec:twistors}.

\subsection{Some General Properties: Zeros and Poles}

The loop integrands are rational functions of the variables
$d_{ij}$ and $x^2_{ij}$, and as such are characterized by their zeros
and poles in these variables.

\paragraph{Zeros.}

Our integrands $G_{n,\ell}$ vanish on the topological twist introduced by
Drukker and Pfleka~\cite{Drukker:2009sf}, which effectively enforces the spacetime and R-charge alignment:
$y_{ij}^2 = x_{ij}^2$. Written in a
$\grp{SO}(4,2)\times\grp{SO}(6)_{R}$ covariant form, these zeroes
are located at the alignment of the spacetime and R-charge cross ratios:
\beq
\frac{x_{ij}^2 x_{kl}^2}{x_{ik}^2 x_{jl}^2}=\frac{y_{ij}^2 y_{kl}^2}{y_{ik}^2 y_{jl}^2} \quad \text{or} \quad \frac{d_{ij}d_{kl}}{d_{ik}d_{jl}}=1
\,.
\label{eq:susyzerocross}
\eeq
This property is made manifest in our results by the ubiquitous presence of a two-by-two determinant denoted as:
\begin{equation}
\label{eq:Vdet2}
V^{ij}_{kl} \equiv d_{ij}\,d_{kl}-d_{ik}\,d_{jl}
\,.
\end{equation}
See for instance the polynomial coefficients in~\eqref{eq:P31and2} for the use of $V$. Furthermore, our results also depend on higher-rank $d$-determinants, see for instance~\eqref{eq:danti5}.

\paragraph{Poles.}

Besides ordinary propagator poles, the generating functions
$G_{n,\ell}$ display ten-dimensional poles that combine spacetime and R-charge distances.
By construction, all poles of $G_{n,\ell}$ originate from ordinary
propagators $d_{ij}$~\eqref{eq:dijdef}, \ie are located at zeros of $x_{ij}^2$, or from
effective propagators $D_{ij}$~\eqref{eq:DijDef}~\eqref{eq:fatedge}, \ie are located at
zeros of ten-dimensional distances $X_{ij}^2\equiv \brk{x_{ij}^2+y_{ij}^2}$.
The ordinary propagator poles $d_{ij}$ arise since we project
$\superO$ to $L\subrm{int}$ at the points
$\brk{n+1}$ through $\brk{n+\ell}$, which in particular projects the
conformal weights at these points to four, and therefore truncates the
infinite series $D_{ij}=d_{ij}+d_{ij}^2+\dots$ to mere $d_{ij}$ factors. The resulting poles in $x_{ij}^2$
will be combined to integrands of conformally invariant integrals that
only depend on four-dimensional distances $x_{ij}^2$. The coefficients
of these integrals will only have ten-dimensional poles. For
convenience, we introduce the variables $w_{ij}$ which map to the other
variables $d, D, x, y, X$ in~\eqref{eq:DijDef} as:
\begin{equation}
w_{ij} \equiv 1- d_{ij} =\frac{1}{1+D_{ij}}=\frac{d_{ij}}{D_{ij}}
=\frac{x_{ij}^2+y_{ij}^2}{x_{ij}^2}
=\frac{X_{ij}^2}{x_{ij}^2}
\,.
\label{eq:wDef}
\end{equation}
Our results below look more compact by expressing the ten-dimensional poles as poles in $w_{ij}$.
Higher-point functions also display higher-order ten-dimensional
poles, whose coefficients are given by lower-point functions. We
will see this exemplified in~\eqref{eq:G51-Rbasis}
and~\eqref{eq:G52splitcharges} below, and provide a general analysis
of the higher-order poles in \secref{sec:higher-poles}.

\subsection{One-Loop Integrand \texorpdfstring{$G_{5,1}$}{G51}}
\label{sec:Oneloop5point}

The integrand $G_{5,1}$ can be extracted
from the  six-point super-correlator $\mathbb{G}_6^{\NMHV}$
computed in~\cite{Caron-Huot:2023wdh}.
This was given by a combination of two six-point superconformal
invariants~$\mathcal{I}^{(6a)}$ and~$\mathcal{I}^{(6b)}$, and the
five-point invariant $\mathcal{I}^{(5)}$ defined in~\eqref{eq:I5superinv}:
\begin{equation}\label{eq:G6NMHV}
\Nc^{4}\times\superG_6^{\NMHV}
=\left(2\, \mathcal{I}^{(6a)}_{123456} -2\,\mathcal{I}^{(6b)}_{123456}\right)\prod_{i<j}^6 D_{ij}
+\left[\tilde{C}^{(5)}_{12345,6}\, \mathcal{I}^{(5)}_{12345}\prod_{i<j}^5D_{ij}+ (\grp{C}_6\text{ perms})\right].
\end{equation}
Here, $\grp{C}_6$ is the cyclic group on six points, that is the last term
is summed over six cyclic permutations.

The coefficient of the five-point invariant $\mathcal{I}^{(5)}$,
defined in~\eqref{eq:I5superinv}, is a polynomial in the effective
10d propagator $D_{ij}$:
\begin{align}
\tilde{C}^{(5)}_{12345,6} &=
4 D_{16}D_{26}D_{36}D_{46}D_{56}
+2\sum_{5} D_{16}D_{26}D_{36}D_{46}
-2\sum_{10} D_{16}D_{26}(1+D_{12})
\,.
\end{align}
It has permutation invariance on the labels before the comma, and the
numbers on the sums count the $\grp{S}_5$ inequivalent permutations.

The simplest of the six-point invariants is given by:
\begin{align} \label{I6b}
\mathcal{I}^{(6b)}_{123456}
&=\sum_{90} \frac{R^1_{234}R^4_{561}}{d_{23}d_{56}}
 \,\frac{\det \left[d_{ij}\right]^{i=1,2,3}_{j=4,5,6}}{\prod_{\substack{i=1,2,3\\ j=4,5,6}} d_{ij}}
\nonumber\\& \quad
+\sum_{360} \frac{R^1_{234} R^2_{135}}{d_{34}d_{35}d_{36} d_{12}d_{24}d_{45}d_{51}}
\left[\frac{d_{12}}{d_{16}d_{26}} -\frac{d_{15}}{d_{16}d_{56}}-\frac{d_{24}}{d_{26}d_{46}} + \frac{d_{45}}{d_{46}d_{56}}\right]
\nonumber\\& \quad
+\sum_{90} \frac{R^1_{234}R^2_{134}}{d_{12}d_{34}d_{35}d_{36}d_{45}d_{46}}\left[\frac{1}{d_{15}d_{26}}+\frac{1}{d_{16}d_{25}}\right]
-\sum_{180} \frac{R^1_{234}R^1_{235}}{d_{16}d_{23}d_{24}d_{25}d_{34}d_{35}d_{46}d_{56}}\,.
\end{align}
It is independent of the reference twistor $\Zref$, and it is
proportional to the stress-tensor supercorrelator, see eq.~(3.20)
of~\cite{Chicherin:2015bza}. An expression for the invariant
$\mathcal{I}^{(6a)}$ can be found in eq.~(6.36)
of~\cite{Caron-Huot:2023wdh}. This invariant has a higher degree on
$y_6$, and it does not survive the R-charge projection that extracts the
Lagrangian operator.

In order to extract the integrand $G_{5,1}$ we consider the Grassmann and R-charge projections:

\paragraph{Grassmann Projection of Super-Invariants.}

By following the method described in \secref{sec:proj-loop-integr}, we
obtain the Grassmann projections of the six-point invariants:%
\footnote{The parity-odd parts that involve $X\suprm{anti}$ and
$Y\suprm{anti}$ suggest that the invariants $R^{(6b)}$ and $R^{(6a)}$
should combine symmetrically, such that $X\suprm{anti}$ and
$Y\suprm{anti}$ join into a larger ten-dimensional invariant. And in fact
they do combine symmetrically in~\eqref{eq:G6NMHV},~\eqref{eq:G51fullKK}.
A lesson to be drawn from this is that
higher-point correlators with ten-dimensional invariance may contain larger
invariants besides the simple ten-dimensional distances $X_{ij}^2$.}
\begin{align}
R^{(6b)}_{12345;6}&\equiv
\eval2{\mathcal{I}^{(6b)}_{123456}}_{\theta_6^4}
\times
\prod_{i<j=1}^{6}d_{ij}
\label{eq:R6b}
\\
&=
\frac{1}{4}\brk[s]*{\frac{x_{12}^2 x_{34}^2\,d_{15}\,V^{12}_{34}\,V^{23}_{54}}{\,x_{16}^2 x_{26}^2 x_{36}^2 x_{46}^2}
+(\grp{S}_5\text{ perms})}
+4\ii\, \frac{
X\suprm{anti}_{123456}\times d\suprm{anti}_{12345}
}{
x_{16}^2 x_{26}^2 x_{36}^2 x_{46}^2 x_{56}^2
}\nonumber
\,,\\
R^{(6a)}_{12345;6}
&\equiv
\mathcal{I}^{(6a)}_{123456}\bigg{|}_{\theta^{4}_{6}}
\times
\prod_{i<j=1}^{6}d_{ij}
\label{eq:R6a}
\\\nn
&=
\mspace{-1mu}
\frac{1}{4}
\brk[s]*{
\frac{d_{12}d_{23}d_{34}d_{45}d_{51}d_{56}}{x_{16}^2x_{26}^2x_{36}^2x_{46}^2}
\brk*{\mspace{-1mu}
x_{12}^2x_{34}^2V^{12}_{34}+x_{14}^2x_{23}^2V^{13}_{42}
\mspace{-1mu}}
\mspace{-0.5mu}
+\mspace{-0.5mu}\brk{\grp{S}_5\,\text{perms}}
}
\mspace{-1mu}
+\mspace{-1mu}4\ii\,\frac{
Y\suprm{anti}_{123456}\times d\suprm{anti}_{12345}
}{
x_{16}^2 x_{26}^2 x_{36}^2 x_{46}^2 x_{56}^2
},
\end{align}
where the $V$-factors are defined in~\eqref{eq:Vdet2} and  the function $d\suprm{anti}_{12345}$ is given by:
\begin{equation}
\label{eq:danti5}
d\suprm{anti}_{12345} = \frac{1}{10}\sum_{\sigma\in \grp{S}_5} \text{sign}(\sigma) d_{\sigma_{1}\sigma_{2}}d_{\sigma_{2}\sigma_{3}}d_{\sigma_{3}\sigma_{4}}d_{\sigma_{4}\sigma_{5}}d_{\sigma_{5}\sigma_{1}}
\,.
\end{equation}
In $R^{(6b)}_{12345;6}$, the unique parity-odd conformal covariant
$X\suprm{anti}_{123456}$ at six points appears, which was discussed
before around~\eqref{eq:Xanti}. Analogously, we define an
antisymmetric Y-structure using the projective six-dimensional
null-vectors $Y^I$ that transform under the fundamental representation
of the internal group $\grp{SO}(6)$:\footnote{
Starting from the twistor-like decomposition of the polarization vectors
\begin{equation*}
y^{AB}_{i} = \epsilon^{a b}Y_{i,a}^{A}Y_{i,b}^{B} \text{\quad and choosing:\quad} \left( \begin{matrix} Y_{i,1} \\ Y_{i,2}\end{matrix}\right) =\left( \begin{matrix} 1 & 0 & y_{i}^{11} & y_{i}^{12} \\
0 & 1 & y_{i}^{21} & y_{i}^{22} \end{matrix}\right)
\,,
\end{equation*}
one can obtain the six-dimensional projective polarization vectors by using appropriate sigma matrices $\sigma_{AB}^I$ that relate the antisymmetric representation of $\grp{SU}(4)$ to the fundamental of $\grp{SO}(6)$ by $Y^I_i = \sigma_{AB}^I\,y^{AB}_{i}$. As a representation for $\sigma_{AB}^I$ we choose
$2\, \sigma^I = \left(\sigma_{12},\,i\,\sigma_{23},\,-\sigma_{20},\,-i\,\sigma_{21},\,-\sigma_{32},\,i\,\sigma_{21}\right)$,
with $\sigma_{\mu\nu}=\sigma_\mu\,\otimes\,\sigma_\nu$, and $\sigma_\mu$ being the Pauli matrices.}
\begin{equation}
Y\suprm{anti}_{123456}=\eps_{IJKLMP}Y_1^IY_2^JY_3^KY_4^LY_5^MY_6^P
\,.
\end{equation}
Notice that, when choosing the polarization matrices $y^{ab'}$ to be
real, $Y\suprm{anti}_{123456}$ will in general be imaginary, yielding
a real contribution in~\eqref{eq:R6a}. This is consistent, since the
term is parity-even and thus should be real. In contrast, the
$X\suprm{anti}_{123456}$ term in~\eqref{eq:R6b} is parity-odd and thus
imaginary.
Putting these projections together, we obtain the correlator:
\begin{align}
\tilde{G}_{5,1}&=
\frac{1}{N_c^3}\frac{
2 R^{(6a)}_{12345;6}
-2 R^{(6b)}_{12345;6}
}{\prod_{i<j}^6 w_{ij}} +
\Bigg{[}\frac{1}{N_c^3}
\frac{\tilde{C}^{(5)}_{12346,5} \, R^{(5)}_{1234;6}}{\prod_{i<j}^4 w_{ij}
\prod_{i=1}^4 w_{i6}}
+(\grp{C}_5\text{ perms.}) \Bigg{]}
\,,
\label{eq:G51fullKK}
\end{align}
where $R^{(5)}_{1234;6}$ was defined in~\eqref{eq:R5}.
This SDYM correlator includes all the higher-R-charge partners of the
Lagrangian operator (KK modes in the bulk dual dictionary). In order
to obtain the loop integrand $G_{n,\ell}$, we project out these KK modes.

\paragraph{R-charge Projection Down to Loop Integrand.}

Taking the final projection $y_6\to 0$, we obtain the loop integrand $G_{5,1}$.
Notice that the susy invariant $\mathcal{I}^{(6a)}$ will only
contribute to correlators that include higher KK modes of the chiral
Lagrangian. The coefficient of the five-point invariant becomes
\begin{equation}
\label{eq:defC5}
C^{(5)}_{1234,5} \equiv
\prod_{i=1}^{4}w_{i5}\times \tilde{C}^{(5)}_{12346,5}
\bigg{|}_{y_{6}\to 0}
=\frac{1}{2}\brk*{
\frac{d_{15}d_{25}d_{35}d_{45}}{6}
-\frac{d_{15}d_{25}\,w_{35}w_{45}}{w_{12}}
+(\grp{S}_4\text{ perms})
}\,.
\end{equation}
Since $R^{(6b)}_{12345;6}$ and $R^{(5)}_{1234;6}$ do not depend on
$y_6$, they remain unchanged by this projection and the one-loop
integrand can be expressed as
\begin{equation}
G_{5,1}=\frac{1}{N_c^3}
\frac{
-2 R^{(6b)}_{12345;6}
+\brk[s]*{C^{(5)}_{1234,5} R^{(5)}_{1234;6}
+(\grp{C}_5\text{ perms.})}
}{\prod_{i<j}^5 w_{ij}}
\,.
\end{equation}
The one-loop integrand can also be expressed as a sum over one-loop
box integrals. To achieve this, we introduce the five-point
polynomial $\mathcal{R}_{1234,5}$, defined as the even part of the
numerator of~\eqref{eq:R6b}:%
\footnote{This polynomial appears again in
the two-loop integrand, see $f_5$ in eq.~\eqref{eq:fk} below.}
\begin{align}
\mathcal{R}_{1234,5}&\equiv
-\half\,x_{12}^2 x_{34}^2
\,d_{15}V^{12}_{34}V^{23}_{54}
+(\grp{S}_4\text{ perms})
\label{eq:R1234c5}
\\\nn
&=
x_{12}^2 x_{34}^2
\brk*{
 d_{45}V^{12}_{35}V^{12}_{43}
+d_{35}V^{12}_{34}V^{12}_{45}
-d_{15}V^{12}_{34}V^{23}_{54}
-d_{15}V^{12}_{43}V^{24}_{53}
}
+(1\leftrightarrow 3)+(1\leftrightarrow 4)
\,,
\end{align}
which is independent of the Lagrangian insertion point (6), symmetric
on the points $1,\dots,4$, and has $y$-weight $2$ in all five points $1,\dots,5$. It satisfies a simple OPE-like relation to the four-point polynomial~\eqref{eq:R1234first} as:
\beq
\lim_{y_{5}\to y_{4}} \frac{\mathcal{R}_{1234,5}}{2\,d_{45}}=\mathcal{R}_{1234}
\eeq
Finally, we can rewrite the one-loop integrand as a sum over box
diagrams, with numerators given by the polynomials
in~\eqref{eq:R1234first} and~\eqref{eq:R1234c5} as:
\begin{equation}
G_{5,1}=\frac{1}{\Nc^3}
\prod_{i<j=1}^5\frac{1}{w_{ij}}
\brk3{
\frac{\mathcal{R}_{1234,5}-C^{(5)}_{1234;5}\,\mathcal{R}_{1234}-4\ii\,d^\mathrm{anti}_{12345}\,\eps_{\kappa\lambda\mu\nu}
x_{16}^\kappa x_{26}^\lambda x_{36}^\mu x_{46}^\nu
}{x_{16}^2x_{26}^2x_{36}^2x_{46}^2}
+(\grp{C}_5\text{ perms})
}
\,,
\label{eq:G51}
\end{equation}
where we have separated off the parity-odd part and written it
using~\eqref{eq:XantiRewritten}.

\subsection{Two-Loop Integrand \texorpdfstring{$G_{5,2}$}{G52}}
\label{sec:Twoloop5point}

In this section we present results on the five-point two-loop
integrand $G_{5,2}$, which we extract as a Grassmann and R-charge
projection of the seven-point supercorrelator
$\mathbb{G}_{7}^{\NMHVf{2}}$. After performing this projection, we
resort to the ``divide and conquer'' strategy, described in
\secref{sec:structure-integrand}, to fix the numerator that multiplies the
ten-dimensional poles of the generating function $G_{5,2}$. In
\secref{sec:G52Rchargebasis}, we present this numerator in terms of a
finite basis of correlators with fixed R-charge, starting with the
stress-tensor correlator. Alternatively, in
\secref{sec:G52integrals}, we present $G_{5,2}$ in a basis of conformal
integrals, whose coefficients carry the 10d poles. This latter
representation is handy for the OPE limits we consider later in
\secref{sec:OPEData}.

\subsubsection{Generating Function: Basis of Fixed-Charge Correlators}
\label{sec:G52Rchargebasis}

We present the generating function in a finite basis
of correlators with fixed R-charge, with coefficients containing simple
and double poles on the 10d distances $X_{ij}^2$. For convenience, we use the variables
$w_{ij}$ introduced earlier~\eqref{eq:wDef} to express these 10d poles.

For the one-loop integrand, this representation follows
straightforwardly from~\eqref{eq:G51} and reads as:
\begin{equation}\label{eq:G51-Rbasis}
	G_{5,1}=
	\frac{\langle 22222\rangle_{1}}{\prod\limits_{1\leq i<j\leq 5} w_{ij}}\,
	-\frac{1}{2\Nc}\brk*{
	\frac{C^{(5)}_{1234;5}}{w_{15}w_{25}w_{35}w_{45}}\,G_{4,1}
		+(\grp{C}_5\text{ perms.})
	}\,,
\end{equation}
The first term only contains simple poles on each pair $w_{ij}$, and
the numerator is given by the $\twentyprime$ or stress-tensor
integrand,
see~\eqref{eq:Rweightcomponent} for the notation.
This correlator can be identified by the terms in~\eqref{eq:G51} which
depend on the polynomial $\mathcal{R}_{1234,5}$,
defined in~\eqref{eq:R1234c5}. More explicitly, this is given by:
\begin{equation}
\langle 22222 \rangle_{1}
=-\frac{2 R^{(6b)}_{12345;6}}{\Nc^3}
=\frac{1}{\Nc^3}\,
\frac{\mathcal{R}_{1234,5}-4i\,d^\mathrm{anti}_{12345}\,\eps_{\kappa\lambda\mu\nu}
x_{16}^\kappa x_{26}^\lambda x_{36}^\mu x_{46}^\nu}{x_{16}^2x_{26}^2x_{36}^2x_{46}^2}
+(\grp{C}_5\text{ perms.})
\,.
\label{eq:22222_1}
\end{equation}
The second term in~\eqref{eq:G51-Rbasis} is given by terms
in~\eqref{eq:G51} controlled by the polynomial $\mathcal{R}_{1234}$.
It can be identified with the four-point generating function $G_{4,1}$
($\sim \Nc^{-2}$) in~\eqref{eq:G41}, dressed with a coefficient which
contains all the double poles of the five-point integrand.
See~\eqref{eq:defC5} for the explicit definition of the coefficient.

Inspired by the one-loop results above, we searched for a similar
representation of the two-loop integrand. Indeed, we found that the
double-pole terms are accounted for by the same coefficient
in~\eqref{eq:G51-Rbasis}, now dressed by the two-loop four-point
integrand $G_{4,2}$ in~\eqref{eq:G42}. On the other hand, the simple-pole
numerator is slightly more complicated, and is given by a finite basis
of fixed-charge correlators with small R-charge, in addition to the $\twentyprime$
correlator. In detail, the parity-even part of the generating function reads as follows:
\begin{align}\label{eq:G52splitcharges}
	G_{5,2}=\mathord{}&
   \frac{\langle 22222\rangle_{2}
	+\langle 42222\rangle_{2}
	+\langle 33222\rangle_{2}^\star
	+\langle 33332\rangle_{2}^\star	+\langle 33334\rangle_{2}^\star
	+\langle 44444\rangle_{2}^\star
	+(\text{ineq perm.})}{\prod\limits_{1\leq i<j\leq 5} w_{ij}}\,
	\nonumber\\
	&
-\frac{1}{2\Nc}\brk*{
	\frac{C^{(5)}_{1234;5}}{w_{15}w_{25}w_{35}w_{45}}\,G_{4,2}
		+(\grp{C}_5\text{ perms})
	}\,.
\end{align}
Again, the angled brackets denote fixed-weight
correlators~\eqref{eq:Rweightcomponent}, and the star indicates a
modification to the respective correlator. In the first line, each
(modified) correlator in the numerator must be permuted over all
inequivalent configurations such that the expression becomes
$\grp{S}_5$-symmetric. In particular, the
contributions are defined as
\begin{align}
	\langle 33222\rangle_{2}^\star =\mathord{}& \langle 33222\rangle_{2}
	-d_{12}\langle 22222\rangle_{2}
	-\brk*{\frac{d_{15}d_{25}}{2}\langle 2222\rangle_{2}+(\grp{C}_3\text{ perms})}\,,
	\\
	\langle 33332\rangle_{2}^\star =\mathord{}& \langle 33332\rangle_{2}
	-\Bigl(\frac{d_{34}}{4}\langle 33222\rangle_{2}^\star+\frac{d_{12}d_{34}}{8}\langle 22222\rangle_{2}
    +\frac{d_{35}d_{45}}{4}\langle 3322\rangle_{2}
		+(\grp{C}_4\text{ perms})\Bigr)\,,
	\nonumber \\
	\langle 33334\rangle_{2}^\star =\mathord{}& \langle 33334\rangle_{2}
	-\Bigl(\frac{d_{15}}{6}\langle 23333\rangle_{2}^\star+\frac{d_{35}d_{45}}{4}\langle 33222\rangle_{2}^\star+\frac{d_{12}d_{35}}{2}\langle 22233\rangle_{2}^\star
    +\frac{d_{12}d_{34}}{8}\langle 22224\rangle_{2}
	\nonumber\\
	&
    \!\!\!+\frac{d_{15}d_{25}d_{34}}{4}\langle 22222\rangle_{2}
    +\frac{d_{35}d_{45}^2}{2}\langle 3322\rangle_{2}^{[1,2,3,5]}-\frac{d_{15}d_{25}d_{35}d_{45}}{24}\langle 2222\rangle_{2}
	+(\grp{C}_4\text{ perms})\Bigr),
	\nonumber \\
	\langle 44444\rangle_{2}^\star =\mathord{}&
    \frac{(d_{14}d_{23}x_{14}^2x_{23}^2+d_{13}d_{24}x_{13}^2x_{24}^2+d_{12}d_{34}x_{12}^2x_{34}^2)d_{15}d_{25}d_{35}d_{45}}{12 N_c^3\; x_{16}^2x_{26}^2x_{36}^2x_{46}^2x_{17}^2x_{27}^2x_{37}^2x_{47}^2}\mathcal{R}_{1234}+(\grp{C}_5\text{ perms})\,.\nonumber
\end{align}
The modifications are such that they preserve the permutation
symmetries of the unstarred correlators $\brk[a]{\dots}_2$.
Since, the contribution $\langle 44444\rangle_{2}^\star$ contains only
few terms, we refrained from formulating it in terms of the full
correlator, but simply state it directly. We provide this basis of
R-charge correlators in terms of conformal integrals in the ancillary
\mathematica file \ancfile{G52CorrelatorBasis.m}.
Our result for the lightest correlator $\brk[a]{22222}_2$ matches
with~\cite{Bargheer:2022sfd}, and the correlator $\brk[a]{44444}_2$,
extracted from~\eqref{eq:G52splitcharges}
via~\eqref{eq:Rweightcomponent}, is consistent with the decagon
correlator of~\cite{Fleury:2020ykw,Bercini:2024pya}.

This representation in a finite basis of R-charge correlators is
expected based on the ``saturation of bridges'' in the planar limit at
weak coupling~\cite{Chicherin:2015edu}, see for instance Figure~1
in~\cite{Caron-Huot:2021usw}. When the number of free propagators
between two single-trace BPS operators becomes larger than the loop
order, to stay in the planar topology, this bridge of propagators
becomes uncrossable by loop corrections. Based on this argument, we
can expect that at $\ell$-loop order the basis of R-charge
correlators, sufficient to get the full generating function, starts
with $\langle 22\cdots 2\rangle$ and ends with
$\langle pp\cdots p\rangle$ with $p=2\ell$ (or slightly larger). This
latter R-charge correlator corresponds to the ``simplest
correlators'', at $\ell$-loop order, such as the so-called
octagon~\cite{Coronado:2018cxj} and decagon~\cite{Fleury:2020ykw} in
the four- and five-point cases.%
\footnote{The same principle applies at subleading orders in $1/\Nc^2$
(\ie higher-genus corrections), and was used
in~\cite{Bargheer:2019kxb,Bargheer:2018jvq,Bargheer:2017nne} to
factorize higher-genus large-charge correlators into patches of disk
topology that can be computed individually from integrability/hexagonalization.}
While this representation can become
handy in the absence of another organizational principle, it can also
obscure the symmetries of the generating function. For instance the
generating functions $G_{4,\ell}$ have a ten-dimensional symmetry as
shown in eqs.~\eqref{eq:G41}-\eqref{eq:G43}, however this is not at all
evident in the R-charge basis representation
of~\cite{Chicherin:2015edu,Chicherin:2018avq}. Below, we show an
alternative representation of the five-point integrand in a basis of
conformal integrals and provide their coefficients explicitly.

The presence of the four-point correlator sitting on the double-poles
of the five-point correlator can be explained by the same principle,
combined with an analysis of the graphs
that contribute to the twistor computation.
Moreover, we expect this nesting structure of lower-point functions
controlling the higher-order poles to be more general. In
\secref{sec:Oneloop6point}, we present another example of this nesting
for the six-point correlator, which contains the five-point and
four-point correlators controlling the double and triple poles
respectively. A more detailed argument as well as concrete formulas
are presented in \secref{sec:higher-poles}.

\subsubsection{Generating Function: Basis of Conformal Integrals}
\label{sec:G52integrals}

We organize the generating function in terms of
integrands of seven independent two-loop conformal integrals
$\mathcal{I}_1,\dots,\mathcal{I}_7$:
\begin{align}
G_{5,2} &=\frac{1}{\Nc^3}
\frac{\sum_{i=1}^{7}f_i\times \mathcal{I}_i \, + (\grp{S}_5\,\text{perms.})}{\prod_{1\leq i<j\leq 5}w_{ij}}
-\frac{1}{2\Nc}\brk*{
		\frac{C^{(5)}_{1234;5}}{w_{15}w_{25}w_{35}w_{45}}\,G_{4,2}
		+(\grp{C}_5\text{ perms.})
	}\,.
\label{eq:G52}
\end{align}
Here, all dependence on the Lagrangian insertion coordinates $x_6$ and
$x_7$ is absorbed in the basis of seven integrands $\mathcal{I}_k$ of
conformal integrals, which we define as
\begin{align}
\mspace{50mu}\includegraphics[align=c,hsmash=c]{FigInt1}
&&\mspace{70mu}
\mathcal{I}_{1}^{[\col1,\col2,\col3]}
&=\frac{1}{(x_{\col16}^2 x_{\col26}^2 x_{\col36}^2)x_{67}^2(x_{\col17}^2 x_{\col27}^2 x_{\col37}^2)}
\,,\nonumber\\
\mspace{50mu}\includegraphics[align=c,hsmash=c]{FigInt2}
&&\mspace{70mu}
\mathcal{I}_{2}^{[\col1,\col2,\col3|\col4,\col5]}
&= \frac{1}{2}\left(\frac{x_{\col56}^2\,x_{\col47}^2}{(x_{\col16}^2 x_{\col26}^2 x_{\col36}^2 x_{\col46}^2 )x_{67}^2(x_{\col17}^2 x_{\col27}^2 x_{\col37}^2 x_{\col57}^2 )}+(6\leftrightarrow 7)\right)
\,,\nonumber\\
\mspace{50mu}\includegraphics[align=c,hsmash=c]{FigInt3}
&&\mspace{70mu}
\mathcal{I}_{3}^{[\col1,\col2,\col3|\col4,\col5]}
&= \frac{1}{2}\left(\frac{1}{(x_{\col16}^2 x_{\col26}^2 x_{\col36}^2 x_{\col46}^2)(x_{\col17}^2 x_{\col27}^2 x_{\col37}^2 x_{\col57}^2)}+(6\leftrightarrow 7)\right)
\,,\nonumber\\
\mspace{50mu}\includegraphics[align=c,hsmash=c]{FigInt4}
&&\mspace{70mu}
\mathcal{I}_{4}^{[\col1,\col2,\col3,\col4]}
&= \frac{1}{(x_{\col16}^2 x_{\col26}^2 x_{\col36}^2 x_{\col46}^2) (x_{\col17}^2 x_{\col27}^2 x_{\col37}^2 x_{\col47}^2)}
\,,\nonumber\\
\mspace{50mu}\includegraphics[align=c,hsmash=c]{FigInt5}
&&\mspace{70mu}
\mathcal{I}_{5}^{[\col1,\col2|\col3,\col4]}
&=\frac{1}{2}\left(\frac{1}{(x_{\col16}^2 x_{\col26}^2 x_{\col36}^2) x_{67}^2(x_{\col17}^2 x_{\col27}^2 x_{\col47}^2) }+ (6\leftrightarrow 7)\right)
\,,\nonumber\\
\mspace{50mu}\includegraphics[align=c,hsmash=c]{FigInt6}
&&\mspace{70mu}
\mathcal{I}_{6}^{[\col1,\col2|\col3,\col4|\col5]}
&= \frac{1}{2}\left(\frac{x_{\col56}^2}{(x_{\col16}^2 x_{\col26}^2 x_{\col36}^2 x_{\col46}^2 )x_{67}^2 (x_{\col17}^2 x_{\col27}^2 x_{\col57}^2 ) }+(6\leftrightarrow 7)\right)
\,,\nonumber\\
\mspace{50mu}\includegraphics[align=c,hsmash=c]{FigInt7}
&&\mspace{70mu}
\mathcal{I}_{7}^{[\col1,\col2|\col3,\col4|\col5]}
&= \frac{1}{2}\left(\frac{1}{(x_{\col36}^2 x_{\col46}^2 x_{\col56}^2 ) x_{67}^2 (x_{\col17}^2 x_{\col27}^2 x_{\col57}^2   ) } +(6\leftrightarrow 7)\right)
\,.
\label{eq:5pt2LoopInts}
\end{align}
We organize the integrals according to their symmetries under
permutations of their labels $\set{1,\dots,5}$.
The first elements of the basis
$\mathcal{I}_{1,2,3}$ are symmetric under $\grp{S}_3\times\grp{S}_2$.
The integrand $\mathcal{I}_{4}$ is
the only with full $\grp{S}_4$ permutation symmetry on four points. The integrands
$\mathcal{I}_{5}$
and $\mathcal{I}_{6}$ are symmetric under $\grp{S}_2\times\grp{S}_2$.
The last element $\mathcal{I}_{7}$ enjoys an $\grp{S}_2\ltimes\grp{S}_2^2$ symmetry.
By the $\grp{S}_5$ permutation symmetry of the generating function
$G_{5,2}$, the coefficients $f_k$ must obey the same symmetries as the
corresponding integrals $\mathcal{I}_k$.

Starting from~\eqref{eq:G42}, we can recast the two-loop four-point
integrand as a sum over the rational functions $\mathcal{I}_4$ and
$\mathcal{I}_5$ as:
\begin{equation}
\label{eq:G42intbasis}
G_{4,2} = \frac{1}{\Nc^2}\frac{2\,\mathcal{R}_{1234}}{\prod_{1\leq i<j\leq 4} w_{ij}}\left(\mathcal{I}_4\,\frac{x_{12}^2 x_{34}^2 w_{12}w_{34}}{8}+\mathcal{I}_5\,\frac{x_{12}^{2} w_{12}}{2}+(\grp{S}_4\text{ perms})\right)
\,,
\end{equation}
and its coefficient $C^{(5)}_{1234;5}$, entering~\eqref{eq:G52}, is defined in~\eqref{eq:defC5}.

With the abundant data of fixed-charge correlators we computed, we are
able to find all seven coefficients $f_k$ of the simple-pole part~as:
\begin{align}
f_1 &=-f_2=
-\sfrac{1}{6}
{x_{12}^2 x_{23}^2 x_{31}^2}\,P_1
\,,\nn\\
f_3 &=
\sfrac{1}{12}\,
{x_{23}^2} \brk*{
x_{14}^2 x_{15}^2 x_{23}^2\,P_{3,1}
+2x_{13}^2 x_{15}^2 x_{24}^2\,P_{3,2}
-\sfrac{1}{6}x_{12}^2 x_{13}^2 x_{45}^2\,P_1
}
+\brk{\grp{S}_3\times\grp{S}_2\text{ perms}}
\,,\nn\\
f_4 &=\sfrac{1}{96}\,\mathcal{R}_{1234}\,x_{12}^2x_{34}^2\,d_{15}d_{25}w_{34}(d_{35}d_{45}w_{12}-2w_{35}w_{45})
+\brk{\grp{S}_4\text{ perms}}
\,,\nn\\
f_5&=-f_{6}+
\sfrac{1}{2}\,x_{12}^2 w_{12}
\mathcal{R}_{1234,5}
\,,\nn\\
f_6&=\sfrac{1}{8}\,
x_{12}^2
\brk*{
x_{12}^2 x_{34}^2 P_{6,1}
+ 2 x_{13}^2 x_{24}^2 P_{6,2}
}
+\brk{\grp{S}_2\times\grp{S}_2\text{ perms}}
\,,\nn\\
f_{7}&=
\sfrac{1}{16}
\brk*{
x_{15}^2 x_{25}^2 x_{34}^2\,P_{7,1}
+2x_{15}^2 x_{24}^2 x_{35}^2\,P_{7,2}
}
+\brk{\grp{S}_2\ltimes\grp{S}_2\text{ perms}}
\,,
\label{eq:fk}
\end{align}
where the $P$ coefficients stand for polynomial functions of $d_{ij}$
given below, and the coefficients
$\mathcal{R}_{1234}$ and $\mathcal{R}_{1234,5}$ are defined
in~\eqref{eq:R1234first} and~\eqref{eq:R1234c5} above. The permutation groups
that are summed over in the definitions of $f_3$, $f_4$, $f_6$, and
$f_7$ are the symmetry groups that preserve the respective integrands
$\mathcal{I}_k$ in~\eqref{eq:5pt2LoopInts}.
For example, the permutation group that is summed over in the definition
of $f_7$ is generated by $(1234)\to(2134)$ and $(1234)\to(3412)$,
which is the symmetry group of $\mathcal{I}_7$. The fractional
numbers, such as $1/96$ in $f_4$, serve to avoid overcounting repeated
terms when performing the permutations, both in eq.~\eqref{eq:fk} and in eq.~\eqref{eq:G52}.

The coefficients $P_i$ and $P_{i,j}$ in~\eqref{eq:fk} are polynomials
in $d_{ij}$ given by (we use $V$ and $w$ defined in
eqs.~\eqref{eq:wDef} and~\eqref{eq:Vdet2}):
\begin{align}
P_1&=
\sfrac{1}{4} V^{12}_{34}V^{12}_{35}\,d_{45}\,w_{14}w_{15}w_{23}
+\sfrac{1}{2} d_{15} \brk{d_{14}-d_{24}} V^{12}_{35} V^{12}_{43} w_{45}
+\brk{\grp{S}_3
\times\grp{S}_2
\text{ perms}}\,,
\label{eq:mathF}
\\
P_{3,1}&=
\sfrac{1}{4}
V^{12}_{43}\,V^{13}_{52}\,
\brk!{
d_{45}\,w_{14}w_{15}w_{23}
+(d_{14}-d_{24})(d_{15}-d_{35})\,w_{45}
}
+\brk{\grp{S}_2
\times\grp{S}_2
\text{ perms}}
\,,\nn\\
P_{3,2}&=
-\sfrac{1}{2}
V^{12}_{34}\,V^{13}_{52}\,
\brk!{
d_{45}\,w_{14}w_{15}w_{23}
+(d_{14}-d_{24})(d_{15}-d_{35})\,w_{45}
}
+\brk{\grp{S}_2\text{ perms}}
\,,
\label{eq:P31and2}
\\
P_{6,1}&=
\sfrac{1}{2}
V^{12}_{34}
\brkleft!{
d_{12} d_{35} d_{45} w_{14} w_{23}
+2 d_{15} d_{45} d_{23} w_{12} w_{34}
+2 d_{15} d_{35} (d_{12}-d_{23}) (d_{14}-d_{34})
}
\nn\\ & \mspace{50mu}
+d_{15} d_{25} d_{35} d_{45} (1-d_{14} d_{23})
-2 d_{15} d_{25} d_{35} d_{14} w_{23}
\nn\\ & \mspace{50mu}
\brkright!{\mathord{}
-2 d_{15} d_{35} d_{45} (d_{23} w_{12} w_{14}+d_{12} w_{23} w_{34})
+2 d_{15} d_{25} d_{34} w_{14} w_{35}
}
+\brk{\grp{S}_2\times\grp{S}_2\text{ perms}}
\,,\nn\\
P_{6,2}&=
\sfrac{1}{2}
V^{12}_{34} V^{13}_{42} d_{15} d_{25} d_{35} (2-d_{45})
\nn\\ & \mspace{30mu}
+\sfrac{1}{2}V^{12}_{43} d_{35} \brk!{d_{12} (2 d_{15} (d_{24}+d_{45} w_{24})-d_{45} w_{13} w_{24})-2 d_{13} d_{15} d_{24} d_{45}}
\nn\\ & \mspace{30mu}
+\sfrac{1}{2}V^{12}_{34} \brkleft!{
    2 d_{13} d_{15} (d_{14}-d_{24}) d_{25}
    -d_{35} d_{12} (2 d_{14} d_{15}+d_{45} w_{14} (2 d_{15} d_{23}+w_{23}))
    }
\nn\\ & \mspace{80mu}
    \brkright!{\mathord{}
    +2 d_{35} (d_{15} d_{24} (d_{23} (-1+d_{45})-d_{25} w_{13} w_{14})+d_{14} (d_{15} d_{23}+d_{25} w_{13} w_{24}))
}
\nn\\ & \mspace{30mu}
+\sfrac{1}{2}V^{13}_{42} \brkleft!{
    -2 (d_{12}-d_{13}) d_{25} (d_{24}-d_{34}) d_{35}
    }
\nn\\ & \mspace{80mu}
    +d_{15} d_{24} d_{35} (d_{13} d_{25} (-2+d_{45})
    +2 (-1+d_{12} d_{34} d_{45}+d_{34} w_{12}+d_{45} w_{12} w_{13}))
\nn\\ & \mspace{80mu}
    \brkright!{\mathord{}
    -d_{15} d_{25} (2 d_{34} w_{13}+d_{35} (-2+d_{45}+2 w_{24} w_{34}))
}
+\brk{\grp{S}_2\text{ perms}}
\,.
\label{eq:P61andP62}
\end{align}
Again, the permutation groups that we sum over in the above
definitions are the symmetry groups that preserve the coefficients of
the respective polynomials $P_{k,i}$, \ie the corresponding integrand
$\mathcal{I}_k$, as well as the respective $x_{ij}^2$ prefactors
in~\eqref{eq:fk}. Explicitly, these symmetry groups are given in \tabref{tab:Psym}.
See \appref{app:graphical-analysis} for a discussion of the pole
structure of the various coefficients $f_k$.

\begin{table}[h]
\centering
\begin{tabular}{ccll}
\toprule
Polynomial & Symmetry Group & Generators $(12345)\to\#$ & Cycle notation \\
\midrule
$P_1$ & $\grp{S}_3\times\grp{S}_2$ & $(21345)$, $(23145)$, $(12354)$ & $(12)$, $(123)$, $(45)$ \\
$P_{3,1}$ & $\grp{S}_2\times\grp{S}_2$ & $(13245)$, $(12354)$ & $(23)$, $(45)$\\
$P_{3,2}$ & $\grp{S}_2$ & $(21354)$ & $(12)(45)$ \\
$P_{6,1}$ & $\grp{S}_2\times\grp{S}_2$ & $(21345)$, $(12435)$ & $(12)$, $(34)$ \\
$P_{6,2}$ & $\grp{S}_2$ & $(21435)$ & $(12)(34)$\\
$P_{7,1}$ & $\grp{S}_2\times\grp{S}_2$ & $(21345)$, $(12435)$ & $(12)$, $(34)$ \\
$P_{7,2}$ & $\grp{S}_2$ & $(34125)$ & $(13)(24)$ \\
\bottomrule
\end{tabular}
\caption{Symmetries of polynomial coefficients $P$ in eq.~\protect\eqref{eq:fk}.}
\label{tab:Psym}
\end{table}

Furthermore, we find that the polynomials $P_{7,1}$ and $P_{7,2}$ in $f_7$ can be
mostly expressed in terms of the other polynomials. Using the notation
\begin{equation}
P_{k,i}^{(12345)}\equiv P_{k,i}
\,,\qquad
P_{k,i}^{(abcde)}=
\eval*{P_{k,i}^{(12345)}}_{(12345)\to(abcde)}
\,,
\end{equation}
(and similarly for $P_1$), we observe
\begin{align}
P_{7,1}&=-P_{6,1}^{(52,34,1)}+(1 \leftrightarrow 2)
\,,\nn\\
P_{7,2}&=
-P_{6,2}^{(52,34,1)}+(1 \leftrightarrow 3, 2 \leftrightarrow 4)
-P_1^{(245,13)}
+p_7
\,.
\label{eq:P71andP72}
\end{align}
Here, the commas in the superscripts serve to illustrate the permutation
symmetries of the polynomials. The remaining polynomial $p_7$ is given by
\begin{multline}
p_{7}=
(d_{35} V^{12}_{54}-d_{15} V^{24}_{53})
\brk[s]2{
    (d_{25} d_{35}w_{14}-w_{25}w_{35}) V^{13}_{42}
    +\sfrac{1}{2}(1-d_{14} d_{23}) (d_{35} V^{14}_{52}-d_{25} V^{13}_{45})
}
\\
+(1 \leftrightarrow 3, 2 \leftrightarrow 4)
\,.
\label{eq:p7polynomial}
\end{multline}
This completes our exposition of the generating function in the
integrand/integral basis of~\eqref{eq:5pt2LoopInts}.
We include the representation~\eqref{eq:G52} of $G_{5,2}$ in terms of
conformal integrals in the attached \mathematica file
\ancfile{G52IntegralBasis.m}.

\paragraph{Gram Identity and a Two-Loop Pentagon Integral.}

In four-dimensional spacetime we have the following (polynomial)
seven-point Gram identity:
\begin{align}
0&=
\text{Gram}_7\equiv\det\brk[s]{x_{ij}^2}
\end{align}
Multiplying by the factor $60/\brk{x_{67}^2\prod_{i=1}^{5}x_{i6}^2 x_{i7}^2}$,
we obtain the following identity among the rational basis of functions $\mathcal{I}_{k}$~\eqref{eq:5pt2LoopInts}:
\begin{align}
0&=-20 x_{12}^2 x_{13}^2 x_{23}^2 \,\mathcal{I}_{1}
+20 x_{12}^2 x_{13}^2 x_{23}^2 \,\mathcal{I}_{2}
-20 x_{12}^2 (-6 x_{13}^2 x_{25}^2 x_{34}^2 + 3 x_{12}^2 x_{34}^2 x_{35}^2 + 2 x_{13}^2 x_{23}^2 x_{45}^2) \,\mathcal{I}_{3}
\nn\\ & \mspace{20mu}
+15 x_{14}^2 x_{23}^2 (x_{14}^2 x_{23}^2 - 2 x_{13}^2 x_{24}^2) \mathcal{I}_{4}
-60 x_{12}^2 (-2 x_{14}^2 x_{23}^2 + x_{12}^2 x_{34}^2) \mathcal{I}_{5}
\nn\\ & \mspace{20mu}
+60 x_{12}^2 (-2 x_{14}^2 x_{23}^2 + x_{12}^2 x_{34}^2) \mathcal{I}_{6}
-120 x_{15}^2 (x_{25}^2 x_{34}^2 - x_{24}^2 x_{35}^2) \mathcal{I}_{7}
\nn\\ & \mspace{20mu}
+2x_{15}^2x_{23}^2\brk{5x_{15}^2x_{24}^2-6x_{14}^2x_{25}^2}x_{34}^2\,\mathcal{I}_0
+(\grp{S}_5\text{ perms})
\,.
\label{eq:GramZero}
\end{align}
This rational function vanishes for any random configuration of
four-dimensional vectors~$\set{x_i}$.
The relation includes another five-point two-loop conformal integral
\begin{align}
\includegraphics[align=c,hsmash=c]{FigInt0}
\mspace{100mu} \mathcal{I}_{0}^{
 [\col1,\col2,\col3,\col4,\col5]}
&=\frac{x_{67}^2}{(x_{\col16}^2 x_{\col26}^2 x_{\col36}^2 x_{\col46}^2 x_{\col56}^2 ) (x_{\col17}^2 x_{\col27}^2 x_{\col37}^2 x_{\col47}^2 x_{\col57}^2 )}
\label{eq:newpenta2loop}
\end{align}
This integral is excluded from the expansion~\eqref{eq:G52} of the
generating function $G_{5,2}$, because it is linearly related to the
other integrals $\set{\mathcal{I}_1,\dots,\mathcal{I}_7}$ through
the Gram identity~\eqref{eq:GramZero}.
Conversely,
the Gram identity~\eqref{eq:GramZero} can be used to shift
the coefficients~$f_k$ as:
\begin{equation}
G_{5,2}^{\text{other}}
=G_{5,2}
+\text{Gram}_7\times(\text{factor with }\grp{S}_{5}\times\grp{S}_{2}\text{ symmetry})
\,,
\end{equation}
at the cost of inserting the two-loop pentagon
integral~$\mathcal{I}_{0}$.%
\footnote{Note that the Gram identity cannot be used to remove any
other of the basis integrands $\mathcal{I}_1,\dots,\mathcal{I}_7$ from
the expression for the generating function~$G_{5,2}$,
because~\eqref{eq:GramZero} only contains specific $\grp{S}_5$
symmetric combinations of each of those integrands.}
See \appref{appIntegralsDone} for more details on this integral. This ambiguity of the integrand $G_{5,2}$ arises due to the projection $y_{6},y_{7}\to 0$. On the other hand, the correlator $\tilde{G}_{5,2}$, which includes the KK partners of the Lagrangian, should not suffer from this ambiguity.
$\tilde{G}_{5,2}$ should only have ten-dimensional poles, such that
the four-dimensional Gram identity will not induce a linear relation
among the occurring rational functions $\mathcal{\tilde{I}}_k$
(ten-dimensional analogues of $\mathcal{I}_k$).
The ambiguity of the Gram identity therefore complicates the search for a
potential ten-dimensional symmetry based on the knowledge of only
$G_{5,2}$ (as opposed to the full $\tilde{G}_{5,2}$).

\subsubsection{The Double-Trace OPE}
\label{sec:OPEdoubletrace}

We consider an OPE limit of our generating function which results on a lower-point function involving a double-trace operator. We define this new correlator as:
\begin{equation}
\label{eq:dTcorrelator}
G_{n,\ell}^{\text{dT}} \equiv \left\langle \mathcal{O}^{2}(x_1,y_1) \prod_{j=2}^{n}\mathcal{O}(x_i,y_i)\prod_{j=n+1}^{n+\ell}L_{\text{int}}(x_i) \right\rangle_{\text{SDYM}}
\,.
\end{equation}
This can be obtained in the Euclidean OPE limit of the single-trace correlator:
\begin{equation}
\label{eq:dTope}
G_{n,\ell}^{\text{dT}} =
\left[\lim_{x_{2}\to x_1}
\lim_{y_{2}\to y_1}
\, G_{n+1,\ell}\right]\Big{|}_{
  \begin{subarray}{l}
    \text{for } i\geq 3:\\
    x_i,y_i\to x_{i-1},y_{i-1}
  \end{subarray}
  }
\,.
\end{equation}
The last replacement just enforces a relabeling of points to match the definition of the double-trace correlator~\eqref{eq:dTcorrelator}.

From the OPE limit of our five-point loop-integrands, \eqref{eq:G51} and~\eqref{eq:G52}, we obtain the  loop-integrands of a four-point function involving a double trace:
\begin{align}
G_{4,1}^{\text{dT}} &= \frac{1}{\Nc^3}\, 4\,\mathcal{R}_{1234}\prod\limits_{ i<j}^4 x_{ij}^2  \times \,\frac{C_{4,1}^{\text{dT}}}{\prod\limits_{1\leq i<j\leq 5} X_{ij}^2}\,\bigg{|}_{y_5\to0}\,,
\label{eq:GdT41}\\
G_{4,2}^{\text{dT}} &=\frac{1}{\Nc^3}\,4\,\mathcal{R}_{1234}\prod\limits_{ i<j}^4 x_{ij}^2  \times \, \frac{X_{12}^2 X_{34}^2 X_{56}^2\, C_{4,2}^{\text{dT}} \,+\, (14\text{ perm.})}{ \prod\limits_{1\leq i<j\leq 6} X_{ij}^2 }\bigg{|}_{y_5,y_6\to0}\label{eq:GdT42}
\,,
\end{align}
with the coefficients $C^{\text{dT}}$ given by polynomials in $D_{ij}$ as:
\begin{align}
C^{\text{dT}}_{4,1}& = \;\sum_{\mathclap{2\leq i<j\leq 5}}\;
D_{1i}D_{1j}(1+D_{ij})\,,
\label{eq:CdT41}
\\
C^{\text{dT}}_{4,2}& =\;\sum_{\mathclap{2\leq i<j\leq 6}}\; D_{1i}D_{1j}(1+D_{ij})+ D_{12}^2
+\sum_{\!\!k\in \{3,5\}\!\!}\left[D_{12}^2D_{k,k+1}-D_{12}(1+D_{k,k+1})(D_{1,k}+D_{1,k+1})\right]
\,.\nn
\end{align}
The resulting one- and two-loop integrands take a form similar to the single-trace four-point correlators in eqs.~\eqref{eq:G41}
and~\eqref{eq:G42}, but now the unit coefficients of $X_{ij}^2$ in the numerator get promoted from ``1'' to polynomials in $D_{ij}$. It would be nice to compare this weak-coupling structure of double-trace correlators with its strong-coupling counterpart, which could be obtained following the recent supergravity results in~\cite{Bissi:2024tqf,Aprile:2024lwy,Aprile:2025hlt}.

In terms of conformal integrals, the one-loop correlator can be
expanded in box diagrams, while the two-loop correlator can be written
in terms of the integrals~$\mathcal{I}_4$ and~$\mathcal{I}_5$ in~\eqref{eq:5pt2LoopInts}
(which are the squared box and two-loop
ladder integrals),
and both contain the polynomial
$\mathcal{R}_{1234}$, see~\eqref{eq:R1234first}, as an overall factor. This
latter fact is expected, since there is a single super-invariant at
four points. This is more generally true for the correlator of
determinants~\cite{Caron-Huot:2023wdh}.

The OPE limit~\eqref{eq:dTope} of the integrand $G_{5,1}$, in~\eqref{eq:G51}, only receives contributions from the double-pole part of the correlator. This is summarized in the following relation for the double-pole coefficient in $G_{5,1}$:
\beq
 \left[\lim_{x_{2}\to x_1}
\lim_{y_{2}\to y_1}
\, C^{(5)}_{2345;1}\,\frac{\mathcal{R}_{2345}}{x_{16}^2 x_{26}^2 x_{36}^2 x_{46}^2}\,\prod_{i<j}^{5}\frac{x_{ij}^2}{X_{ij}^2}\right]\Bigg{|}_{
  \begin{subarray}{l}
    \text{for } i\geq 3:\\
    x_i,y_i\to x_{i-1},y_{i-1}
  \end{subarray}
  } \mspace{-30mu}
  =-\frac{\Nc^3}{2}\,G_{4,1}^{\mathrm{dT}}
\,,
\eeq
and similarly for the permutation $C^{(5)}_{3451;2}\times \mathcal{R}_{3451}$, while the other permutations of the polynomial $\mathcal{R}_{1234}$ and the polynomial $\mathcal{R}_{1234,5}$ vanish individually in this OPE limit.
Hence also $G_{4,1}\suprm{dT}$ exclusively consists of double-pole
terms, as can also be seen, by combining the poles in $X_{ij}^2$ of~\eqref{eq:GdT41} with the poles in $D_{ij}$ of~\eqref{eq:CdT41}.
The same is true for $G_{4,2}\suprm{dT}$.

On the other hand, the structure of the two-loop integrand~\eqref{eq:GdT42} does not follow directly from our two-loop results when
organized in a basis of fixed-charge
correlators as in~\eqref{eq:G52splitcharges}, or in a basis of conformal
integrals as in~\eqref{eq:G52}.
Below, we review the conditions that are imposed by this OPE on the
polynomials~$P$ that appear in~\eqref{eq:fk}.

\paragraph{OPE Consistency and \texorpdfstring{$P$}{P}-Relations.}

In order to obtain a ``healthy'' OPE, some cancellations are required
between the terms $f_k\,\mathcal{I}_{k}$ in the two-loop correlator in eq.~\eqref{eq:G52}. We can make a list of the
necessary conditions on $f_k$ to cancel the contribution of unwanted
integrals with double poles of the type $(x_{i6}^2)^2$ or $(x_{i7}^2)^2$,
which appear in the OPE limit of various $\mathcal{I}_k$. Here, we only
write the necessary conditions that involve the coefficients $f_6$ and $f_7$:
\begin{align}
P_{7,1}^{(15342)}+P_{6,1}^{(12,34,5)}+P_{6,1}^{(52,34,1)} &
\xrightarrow{\;d_{i5}\to d_{i1}\;} 0
\label{eq:P71CombinationLimit}
\,,\\
P_{6,1}^{(12,35,4)}+P_{6,2}^{(12,35,4)}&
\xrightarrow{\;d_{i5}\to d_{i1}\;} 0
\label{eq:P62plusP61limit}
\,,\\
P_1^{(123,45)}+P_{6,2}^{(23,15,4)}&
\xrightarrow{\;d_{i5}\to d_{i1}\;} 0
\label{eq:P62plusFlimit}
\,,\\
P_{7,2}^{(15432)}+P_{6,2}^{(12,34,5)}&
\xrightarrow{\;d_{i5}\to d_{i1}\;} 0
\label{eq:P72CombinationLimit}
\,.
\end{align}
These relations guarantee that only the good integrals
$\mathcal{I}_{4}$ and $\mathcal{I}_5$ survive in the OPE limit.
Furthermore, the combinations~\eqref{eq:P71CombinationLimit}--\eqref{eq:P72CombinationLimit}
are nicer than their constituents even before taking the OPE limit. In particular,
the first relation in~\eqref{eq:P71CombinationLimit} already holds before taking the limit.
The relations above helped organizing our two-loop results, for
instance, it allowed us to discover the relations
in~\eqref{eq:P71andP72}, which define the polynomial coefficient
$P_{7,i}$ in terms of other, simpler polynomials.

\section{Six-Point Integrand}
\label{sec:six-point-integrand}

We present results on the six-point one-loop integrand extracted from the
supercorrelator $\mathbb{G}_{7}^{\NMHV}$. Besides simple 10d poles,
this correlator presents nested five-point integrands sitting at
double poles and four-point integrands sitting at third-order 10d
poles.

\subsection{One-Loop Integrand \texorpdfstring{$G_{6,1}$}{G61}}
\label{sec:Oneloop6point}

The six-point one-loop integrand, at leading planar order, is given by
\begin{multline}
G_{6,1}=
\prod_{i<j=1}^{6}\frac{1}{w_{ij}}\times
\Big{(}\brk[a]{222222}_1+\brk[a]{333333}^\star_1+\frac{1}{\Nc}\left[C^{(6,1)}_{12345,6}\,\brk[a]{22222}_1+\text{(5 perm.)}\right]
\\
+\frac{1}{\Nc^2}\left[C^{(6,2)}_{1234,56}\,\brk[a]{2222}_1+\text{(14 perm.)}\right]\Big{)}\,.
\label{eq:G61}
\end{multline}
Here, $\brk[a]{222222}_1$, $\brk[a]{22222}_1$, $\brk[a]{2222}_1$
denote the one-loop $\twentyprime$ integrands at six, five, and four
points. In each case, the operators are understood to be labeled by 1
through 6, 5, and 4, respectively. For convenience, we repeat the
expressions for the latter two integrands in terms of the invariants
discussed in \secref{sec:Oneloop5point}:
\begin{equation}
\brk[a]{22222}_1 = -2 R^{(6b)}_{12345;7}/\Nc^3 \quad\text{and}\quad \brk[a]{2222}_1 = -2 R^{(5)}_{1234;7}/\Nc^2
\,.
\end{equation}
$\brk[a]{333333}^\star_1$ appears as an additional structure, and has an
R-charge weight of~3 in each of the six points. As a result, in the charge-expansion~\eqref{eq:GnlFromFixedCharges}, the
six-point correlator that contains only operators of weight~3 is the
first correlator in which this structure appears. More explicitly we have:
\begin{align}
\brk[a]{333333}_1
&=\brk[a]{333333}^\star_1
+ \left(\frac{d_{12}d_{34}d_{56}}{48}\brk[a]{222222}_1
+ \frac{d_{12}d_{35}d_{46}d_{56}^2}{4\,\Nc^2}\brk[a]{2222}_1
+ (\grp{S}_6\text{ perms})\right)\,.
\end{align}
Correlators that contain this structure cannot be written as linear combinations of $\twentyprime$ correlators.
The new invariant $\brk[a]{333333}^\star_1$ and the $\twentyprime$ six-point correlator
$\brk[a]{222222}_1$ are given in terms of $V^{ij}_{kl}$~\eqref{eq:Vdet2} by
\begin{align}
\brk[a]{222222}_1&=\frac{1}{\Nc^4}
\frac{1}{4}\,d_{26}V^{14}_{32}\left(2d_{56}V^{14}_{53}+d_{45}V^{15}_{63}\right)\frac{x_{14}x_{23}}{x_{17}x_{27}x_{37}x_{47}}
+(\grp{S}_{6}\text{ perms})
\,,\\
\brk[a]{333333}^\star_1&=\frac{1}{\Nc^4}
\frac{1}{4}\,d_{26} d_{45} V^{14}_{32} \left(2 d_{16}d_{35}d_{56}V^{12}_{43}
+2 d_{16} d_{25} d_{36} V^{14}_{53}
+d_{12} d_{34} d_{56}
V^{15}_{63}\right)\frac{x_{14}x_{23}}{x_{17}x_{27}x_{37}x_{47}}\nn\\
&\quad
+(\grp{S}_{6}\text{ perms})
\,.
\end{align}
Finally, the coefficient of the five-point invariant in the generating function is:
\begin{align}
C^{(6,1)}_{12345,6}=&d_{16} d_{26}\left(
\frac{- d_{36} d_{46}}{24}
+\frac{ d_{36} d_{46} d_{56}}{40}  +\frac{ 1-3 d_{36}+3 d_{36} d_{46}- d_{36} d_{46} d_{56}}{12\,w_{12}}\right)+(\grp{S}_{5}\text{ perms.})
\,,
\end{align}
and presents a simple ten-dimensional pole on $w_{ij}=1-d_{ij}$ with
$i,j\neq 6$, which combined with the overall factor in~\eqref{eq:G61} gives a double pole.
On the other hand, the four-point coefficient $C^{(6,2)}_{1234,56}$ contains higher-order
10d poles, and can be most compactly written when using the variable
$D_{ij}\equiv{d_{ij}}/\brk{1-d_{ij}} ={-y_{ij}^2}/{X_{ij}^2}$. For comparison, here we write
both coefficients, $C^{(6,1)}$ and $C^{(6,2)}$, making use of $D_{ij}$:
\begin{align}
C^{(6,1)}_{12345,6}&=
\prod\limits_{i=1}^{5}w_{i6}\times
D_{16}D_{26}\left(1+D_{12}-\sfrac{1}{2} D_{36} D_{46}-\sfrac{1}{5} D_{36} D_{46} D_{56}\right)+(\grp{S}_{5}\text{ perms.})
\label{eq:C61coeff}
\\
C^{(6,2)}_{1234,56}&=
\frac{w_{56}}{24}\prod\limits_{i=1}^{4}w_{i5}w_{i6}\times D_{16}
\bigg{[} 3 D_{26}D_{35} D_{45} (D_{12}
   (D_{34}+2)+1) \nonumber\\
 &+4 D_{56}
   D_{15}^2 \big{(}3 D_{12} (D_{25}+1)+D_{25} (3-D_{35} D_{45})+2\big{)}  \nonumber \\
   &+D_{15} \bigg{(}4 (D_{12} (6 D_{25}+3)+D_{25} (6-2 D_{35}
   D_{45})+2) D_{56} \nonumber \\
   &\qquad+D_{26} (6 (2 (D_{12} (D_{13}+2)+1)
   D_{35} +(D_{12}+1) D_{25} (D_{12}-D_{35} D_{45}+1)) )  \nonumber
   \\
   &\qquad+2D_{26}
   D_{25}D_{56} (3 D_{12}+D_{36} D_{45} (D_{35}
   (D_{46}+4)+3)+3) \bigg{)} \nonumber  \\
 &   +2 D_{25}D_{56} (3 (D_{12}+1)-2
   D_{35} D_{45})  \bigg{]} + \left(\grp{S}_{4}\times \grp{S}_{2}\text{ perms.}\right)
\label{eq:C62coeff}
\end{align}
We further observe that the five-point coefficient~\eqref{eq:C61coeff} reduces to the four-point coefficient in~\eqref{eq:defC5} by taking the limit:
\beq
\lim_{y_{5}\to 0}C^{(6,1)}_{12345,6} = C^{(5)}_{1234,6}
\eeq
while the four-point coefficient in~\eqref{eq:C62coeff} vanishes in this same limit.
Moreover, we notice that also the four-point coefficient
$C^{(5)}_{1234,5}$ vanishes in the limit $y_5\to0$, which suggests an
iterative structure.

Finally, this one-loop integrand can be easily upgraded to the
integrated level by replacing the factors
${1}/\brk{x^2_{al}x^2_{bl}x^2_{cl}x^2_{dl}}$ by the one-loop scalar
box integrals in~\eqref{eq:scalarbox}. While the content on arbitrary
R-charge correlators remains on the rational coefficients with 10d
poles.

\section{Structure of Higher Poles}
\label{sec:higher-poles}

Beyond four points, the loop integrand $G_{n,\ell}$ acquires higher
ten-dimensional poles. This was anticipated in
\secref{sec:structure-integrand}, and in~\eqref{eq:G51-Rbasis},
\eqref{eq:G52splitcharges}, and~\eqref{eq:G61}, we observed that
double-poles and triple poles can be written in terms of lower-point
integrands $G_{m,\ell}$, $m<n$.

This feature can be understood by the ``saturation of bridges''
property discussed at the end of \secref{sec:G52Rchargebasis}: A
ten-dimensional pole (propagator $D_{ij}$) stands for a bundle
(``bridge'') of arbitrarily many parallel ordinary propagators
$d_{ij}$, and thus cannot be crossed by loop corrections in the planar
limit, due to supersymmetry and the BPS property of the operator
insertions. A double pole $D_{ij}\times{D_{ij}}$ between two operators
$i$ and $j$ therefore introduces a contour formed by the two
ten-dimensional propagators $D_{ij}$ together with the two operator
insertions, which splits the color sphere into two disc-like regions.
Interactions are confined to the two separate regions, and thus the loop
corrections factorize at the locus of the double pole.

We can illustrate this general principle at the example of the
five-point integrand $G_{n=5,\ell}$. The two propagators $D_{ij}$ must
be homotopically distinct, therefore each of the two disc-like regions
must contain at least one of the remaining $n-2$ operators. For the
five-point function, this means there must be a substructure of the form:
\begin{equation}
\includegraphics[align=c]{FigDoublePole}
\,.
\label{eq:DoublePoleStructure}
\end{equation}
It is clear that the sum of all terms that surround this substructure sum up to
a part of the generating function $G_{4,\ell}$ that has one less
operator, namely the part that contains a bridge $1/w_{ij}$, which gets replaced by the
structure~\eqref{eq:DoublePoleStructure}. We can therefore conclude:
\begin{equation}
G_{5,\ell}\brk[s]*{\includegraphics[align=c]{FigDoublePoleLabeled}}
=\frac{1}{\Nc}
\frac{D_{i5}D_{j5}}{w_{ij}}
\times\eval*{G_{4,\ell}}_{1/w_{ij}\text{
term}}
\,,
\label{eq:DoublePoleRelation}
\end{equation}
where the left-hand-side stands for all terms in $G_{5,\ell}$ that
contain the substructure inside the square brackets.
The parts of $G_{4,\ell}$ that
do not contain a $1/w_{ij}$ pole do not contribute to the double pole.
The right-hand-side of~\eqref{eq:DoublePoleRelation} is exactly equal
to the double-pole part that we observed at
one-loop~\eqref{eq:G51-Rbasis} and two-loop
order~\eqref{eq:G52splitcharges}, which is produced by
the second term in the coefficient $C^{(5)}_{1234;5}$~\eqref{eq:defC5}.

The substructure~\eqref{eq:DoublePoleStructure} is a three-point
function, in which the bridge $D_{ij}$ has split in two. Drawing the
three-point function inserted on a sphere, this can be pictured as a
cut on the sphere along the $D_{ij}$ bridge. Similarly, drawing also
the four-point generating function $G_{4,\ell}$ on a sphere, and
cutting it along its $D_{ij}$ bridge, the two spheres get glued
together to form the double-pole part of the five-point function
$G_{5,\ell}$:
\begin{align}
\includegraphics[align=c]{FigSphere}
\quad
\scalebox{-1}[1]{\includegraphics[align=c]{FigSphere}}
\quad \longrightarrow \quad
\includegraphics[align=c]{FigSphereJoined}
\end{align}
At higher points $G_{n,\ell}$, $n>5$, both spheres can contain
non-trivial lower-point functions $G_{m_1,\ell_1}$, $G_{m_2,\ell_2}$,
with $m_1+m_2=n+2$ and $\ell_1+\ell_2=\ell$.
Higher-point functions will also have poles $1/w_{ij}^p$ of higher orders $p$,
with $p\leq{n-3}$, which are obtained by gluing $p$ spheres, \ie
taking products of $p$ lower-point generating functions.
Hence we can conclude that all double- and higher-pole
terms of higher-point generating functions $G_{n,\ell}$ will be given
by products of lower-point generating functions.
An example of this structure can be seen in the six-point
integrand~\eqref{eq:G61}, where the double and triple
poles are given by products of five-point
and four-point correlators.

To be precise, we can deduce that the double poles of
$G_{n,\ell}$ can be completely reconstructed from products of
single-pole terms of lower-point (and lower-loop) functions $G_{m,k}$
as follows:
\begin{equation}
\hat{W}_{ij}^2\brk[s]{G_{n,\ell}}
=
\frac{1}{2}
\sideset{}{'}\sum_{\mathbold{m}}
\sum_{\mathbold{k}}
\hat{W}_{ij}^1\brk[s]{G_{\mathbold{m},\mathbold{k}}}
\times
\hat{W}_{ij}^1\brk[s]{G_{\mathbold{\bar{m}},\mathbold{\bar{k}}}}
\,.
\label{eq:DoublePoleDecomposition}
\end{equation}
The first sum runs over all
subsets $\mathbold{m}\subset\set{1,\dots,n}$
with $i,j\in\mathbold{m}$ and $3\leq\abs{\mathbold{m}}\leq{n-1}$, and
$\mathbold{\bar{m}}$ is the ``complement'', such that
$\mathbold{m}\cup\mathbold{\bar{m}}=\set{1,\dots,n}$
and $\mathbold{m}\cap\mathbold{\bar{m}}=\set{i,j}$, while the second
sum just runs over all bipartions
$\mathbold{k}\mathbin{\dot\cup}\mathbold{\bar{k}}=\set{n+1,\dots,n+\ell}$. We indicate
this difference in summing over partitions with a prime.
$G_{\mathbold{m}, \mathbold{k}}$ denotes $G_{\abs{\mathbold{m}},
\abs{\mathbold{k}}}$ with operators labeled according to the numbers
in the sets $\mathbold{m}$ and $\mathbold{k}$.
The operator $\hat{W}_{ij}^p$ extracts the $1/w_{ij}^p$ pole part of
the generating function:
\begin{equation}
\hat{W}_{ij}^p\brk[s]{f(w_{ij})}
\equiv
\frac{1}{w_{ij}^p}\res_{w_{ij}=0}\brk[s]*{w_{ij}^{p-1}f(w_{ij})}
\,,
\end{equation}
where we assume that all $G_{n,\ell}$
in~\eqref{eq:DoublePoleDecomposition} are written exclusively in
$w_{ij}$ variables, \ie all $d_{ij}$ and $D_{ij}$ are re-written in
terms of $w_{ij}$, see~\eqref{eq:wDef} for the respective relations.
The overall factor $1/2$ in~\eqref{eq:DoublePoleDecomposition}
is a symmetry factor that compensates a double-counting --
the corresponding symmetry is that of swapping the two spheres
in~\eqref{eq:DoublePoleStructure}, which leads to identical terms
after gluing.
The formula~\eqref{eq:DoublePoleDecomposition}
exactly matches with the double-pole parts of the
five-point functions~\eqref{eq:G51-Rbasis} and~\eqref{eq:G52splitcharges}.

Similarly, the triple poles of $G_{n,\ell}$ dissect the color sphere
into three discs, or equivalently three spheres with one cut each,
which means that the triple pole
can be written as a product of three single-pole functions:
\begin{equation}
\hat{W}_{ij}^3\brk[s]*{G_{n,\ell}}
=
\frac{1}{3}
\sideset{}{'}\sum_{\mathbold{m}_1,\mathbold{m}_2,\mathbold{m}_3}
\,\sum_{\mathbold{k}_1,\mathbold{k}_2,\mathbold{k}_3}
\!\hat{W}_{ij}^1\brk[s]{G_{\mathbold{m}_1,\mathbold{k}_1}}
\times
\hat{W}_{ij}^1\brk[s]{G_{\mathbold{m}_2,\mathbold{k}_2}}
\times
\hat{W}_{ij}^1\brk[s]{G_{\mathbold{m}_3,\mathbold{k}_3}}
\,,
\label{eq:TriplePoleDecomposition}
\end{equation}
where the first sum runs over $\mathbold{m}_r\subset\set{1,\dots,n}$ with
$\abs{\mathbold{m}_r}\geq3$ and $\bigcup_r
\mathbold{m}_r=\set{1,\dots,n}$ and
$\mathbold{m}_r\cap\mathbold{m}_s=\set{i,j}$,
and the symmetry factor $1/3$ compensates the symmetry of cyclically
rotating the three single-pole factors. The second sum runs over all
tripartitions $\mathop{\dot\bigcup}_r\mathbold{k_r}=\set{n+1,\dots,n+\ell}$.
Using~\eqref{eq:DoublePoleDecomposition}, the triple pole can also be
written as a product of a single-pole and a double-pole factor:
\begin{equation}
\hat{W}_{ij}^3\brk[s]*{G_{n,\ell}}
=
\frac{1}{3}
\sideset{}{'}\sum_{\mathbold{m}}
\sum_{\mathbold{k}}
\hat{W}_{ij}^1\brk[s]{G_{\mathbold{m},\mathbold{k}}}
\times
2\,\hat{W}_{ij}^2\brk[s]{G_{\mathbold{\bar{m}},\mathbold{\bar{k}}}}
\,.
\label{eq:TriplePole1times2}
\end{equation}
The double-pole~\eqref{eq:DoublePoleDecomposition} and
triple-pole~\eqref{eq:TriplePoleDecomposition} decompositions
straightforwardly generalize to poles of any order:
\begin{equation}
\hat{W}_{ij}^p\brk[s]*{G_{n,\ell}}
=
\frac{1}{p}\,
\sideset{}{'}\sum_{\mathbold{m}_1,\dots,\mathbold{m}_p}
\,\sum_{\mathbold{k}_1,\dots,\mathbold{k}_p}\,
\prod_{r=1}^{p}
\hat{W}_{ij}^1\brk[s]{G_{\mathbold{m}_r,\mathbold{k}_r}}
\,.
\end{equation}
The formula~\eqref{eq:TriplePole1times2} suggests that it could be
possible to write the complete higher-pole part of any $G_{n,\ell}$ as
a product of two lower-point $G_{m,k}$, dressed with suitable
operators that generate the appropriate symmetry factors for each pole.

As intuitive as these formulas are, there is an important caveat:
\emph{They only hold in the single-trace operator basis}, and not in
the single-particle operator basis that we work with everywhere else
in this paper, for the following reason: Also graphs that contain
operators of valency one can be glued to form graphs that have valency
two or more on all operators, \ie graphs that contribute to
single-particle generating functions. Therefore, when decomposing the
higher-order poles into lower-point generating functions, the
lower-point functions (on the right-hand-sides of the above formulas)
should be taken in the single-trace basis. This in turn will produce
the generating function in the single-trace basis also on the
left-hand side.%
\footnote{At leading order (with no Lagrangian insertions), the
equations should be correct when only the operators $i$ and $j$ on the
right-hand side are taken in the single-trace basis, and all other
operators are left in the single-particle basis (on both sides of the
equality). This however does not extend to loop level.}
We explicitly verified that the above formulas hold
for all higher-order poles of the tree-level functions
$G\suprm{st}_{n,0}$ in the single-trace basis up to $n=6$. Since loop
corrections will not affect the 10d pole structure, we can infer that
the formulas will also hold at loop level, in the single-trace basis.

That being said, we \emph{do} find that the above formulas \emph{are}
correct in the single-particle basis when we restrict ourselves to the
highest-order poles of order $p=n-3$. Further above, we already
verified this for the double-poles of the five-point function. At six
points, the highest poles are of order three. In~\eqref{eq:C62coeff},
they are encoded in the fourth line, in the term that is proportional
to $\propto D_{12}^2$. Extracting the third-order pole of
the six-point one-loop generating function explicitly yields
\begin{align}
\hat{W}_{12}^3\brk[s]*{G_{6,1}}
=&
\prod_{i<j=1}^{6}\frac{1}{w_{ij}}\times \frac{1}{24} D_{16}w_{56}\prod\limits_{i=1}^{4}w_{i5}w_{i6}
\times 48 \frac{D_{15}D_{25}D_{26}}{w_{12}^2} \frac{\brk[a]{2222}_1}{\Nc^2}+\text{(14 perm.)}
\nonumber \\
=\,& 2 \frac{D_{15}D_{16}D_{25}D_{26}}{w_{12}^3} \times\hat{W}_{12}^1\brk[s]*{G_{4,1}} +\text{(14 perm.)}
\,,
\end{align}
which is the expected product of the residue of the four-point
one-loop generating function and two tree-level triangles, in
agreement with~\eqref{eq:TriplePoleDecomposition}.

\section{Five-Point Correlator at Integrated Level and OPE}
\label{sec:OPEData}

\subsection{Correlators at Integrated Level}
\label{sec:integrated-correlators}

We define the notation for the super Yang--Mills correlator:
\begin{equation}
G_{n}^{\text{SYM}} \equiv   \left\langle \prod_{i=1}^{n}\mathcal{O}(x_i,y_i) \right\rangle_{\!\text{SYM}} \hspace{-1em} =\, G_{n,0} -g^{2}\int \frac{d^4x_{n+1}}{\pi^2}\,G_{n,1} +\frac{g^{4}}{2}\int\hspace{-0.5em}\int \frac{d^4x_{n+1}}{\pi^2} \frac{d^4x_{n+2}}{\pi^2}\,G_{n,2}+\order{g^6}
\label{eq:GnIntegrated}
\end{equation}
where $\mathcal{O}(x,y)$ was defined in~\eqref{eq:Oscalar} by resumming the tower of half-BPS single-trace scalar operators. The right-hand side is the perturbative expansion obtained via the Lagrangian insertion method; see eq.~\eqref{eq:integratedcorrelator}.
We use the label SYM to stress the difference between the $n$-point
correlator $G^{\text{SYM}}_{n}$ in the full Yang--Mills theory and the
loop integrands $G_{n,\ell}$, see~\eqref{eq:Gnl-integrand}, computed
as $(n+\ell)$-point correlators in the self-dual sector of the theory (SDYM). The
free-theory correlator $G_{n,0}$ is the same in both cases.

The integral over the odd part of the integrand evaluates to zero, so here we concentrate only on the even part when we refer to the integrand. Furthermore, by conformal symmetry, the complete five-point correlator and the five-point integrals are functions of five independent  conformal cross-ratios:
\begin{equation}
\label{eq:ufive}
u_{1}=\frac{x_{25}^2 x_{34}^2}{x_{24}^2 x_{35}^2}\,,\quad
u_{2}=\frac{x_{13}^2 x_{45}^2}{x_{14}^2 x_{35}^2}\,,\quad
u_{3}=\frac{x_{15}^2 x_{24}^2}{x_{14}^2 x_{25}^2}\,,\quad
u_{4}=\frac{x_{12}^2 x_{35}^2}{x_{13}^2 x_{25}^2}\,,\quad
u_{5}=\frac{x_{14}^2 x_{23}^2}{x_{13}^2 x_{24}^2}\,.
\end{equation}
We also introduce an overcomplete ten-element basis of light-cone cross-ratios as:
\begin{equation}\label{crossratiosZ}
    z_{5}\bar{z}_{5} = \frac{x_{12}^2x_{34}^2}{x_{13}^2x_{24}^2}
    \quad \text{and} \quad u_5=
    (1-z_{5})(1-\bar{z}_{5}) \equiv \frac{x_{14}^2x_{23}^2}{x_{13}^2x_{24}^2}
    \quad \text{and} \quad
    z_{i+1}\equiv z_{i}|_{x_i \to x_{i+1}}\,.
\end{equation}
where pairs of cross-ratios $z_i,\bar{z}_i$ are defined by cyclic
permutations of the four-point cross-ratios $z_5,\bar{z}_5$. These are
useful to write the four-point subcorrelators or integrals which
depend just on a pair of cross-ratios such as the ladder integrals~\eqref{BoxIntegrals}.

The one-loop integrands, \eqref{eq:G41}, \eqref{eq:G51}
and~\eqref{eq:G61}, depend on the Lagrangian position $x_{l}$ only
through factors of the form ${1}/\brk{x_{i l}^2\,x_{j l}^2\,x_{k l}^2\,x_{m
l}^2}$. At the integrated level, these get promoted to the one-loop
scalar box integral
\beq\label{eq:scalarbox}
\int \frac{d^4x_{l}}{\pi^2}\,\frac{1}{x_{1 l}^2\,x_{2 l}^2\,x_{3 l}^2\,x_{4 l}^2} \,=\, \frac{F_{1}(z_5,\bar{z}_5)}{x_{13}^2 x_{24}^2}
\,.
\eeq
This evaluates to the one-loop ladder function $F_{1}$, which is the first in a tower of loop integrals given by singled-valued polylogarithms~\cite{Usyukina:1993ch}:
\beq\label{BoxIntegrals}
F_{p}(z,\bar{z}) = \sum_{j=0}^{p}\frac{(-1)^{j}(2p-j)!}{p!(p-j)!j!}\,\log(z\bar{z})^{j}\times\frac{\Li_{2p-j}(z)-\Li_{2p-j}(\bar{z})}{z-\bar{z}}
\,.
\eeq
The two-loop integrands, \eqref{eq:G42intbasis} and~\eqref{eq:G52},  were written in the basis of rational functions
$\mathcal{I}_k$ defined in~\eqref{eq:5pt2LoopInts}. Now we promote these to integrals by integrating over
the points associated with the Lagrangian insertions, and introduce the notation:
\begin{equation}
    \mathbb{I}_{k}^{[...]} = \int\hspace{-0.5em}\int \frac{d^4 x_6}{\pi^2} \frac{d^4 x_7}{\pi^2} \mathcal{I}_k^{[...]}\,.
    \label{IntegrandToInt}
\end{equation}
In the four-point two-loop integrand~\eqref{eq:G42intbasis}, we get the one-loop scalar box squared and the scalar double-box integrals, which evaluate to:
\begin{align}
\mathbb{I}_{4}^{[\col1,\col2,\col3,\col4]}&= \int\hspace{-0.5em}\int \frac{d^4 x_6}{\pi^2} \frac{d^4 x_7}{\pi^2}\frac{1}{(x_{\col16}^2 x_{\col26}^2 x_{\col36}^2 x_{\col46}^2) (x_{\col17}^2 x_{\col27}^2 x_{\col37}^2 x_{\col47}^2)}= \frac{F_{1}(z_5,\bar{z}_5)^2}{x_{13}^4 x_{24}^4}\,,\\
\mathbb{I}_{5}^{[\col1,\col2|\col3,\col4]}
&=\int\hspace{-0.5em}\int \frac{d^4 x_6}{\pi^2} \frac{d^4 x_7}{\pi^2}\frac{1}{(x_{\col16}^2 x_{\col26}^2 x_{\col36}^2) x_{67}^2(x_{\col17}^2 x_{\col27}^2 x_{\col47}^2) } = \frac{F_{2}\big{(}\frac{1}{z_5},\frac{1}{\bar{z}_5}\big{)}}{x_{12}^2 x_{34}^4}
\,.
\end{align}
Then we can write the SYM four-point generating function up to two loops as~\cite{Caron-Huot:2021usw}:
\begin{align}\label{eq:G4SYM}
G^{\text{SYM}}_{4}&= G_{4,0} - \frac{2g^{2}}{\Nc^2}\,\frac{\mathcal{R}_{1234}}{\prod_{i<j}^4 w_{ij} }\frac{F_{1}(z_5,\bar{z}_5)}{x_{13}^2 x_{24}^2}\nonumber\\
&\quad+\frac{2g^{4}}{\Nc^2}\frac{ \mathcal{R}_{1234}}{\prod_{i<j}^4 w_{ij}} \left[\frac{F_{1}(z_5,\bar{z}_5)^2}{2x_{13}^4 x_{24}^4}\left(w_{12}w_{34}\,x_{12}^2x_{34}^2+w_{13}w_{24}\,x_{13}^2x_{24}^2+w_{14}w_{23}\,x_{14}^2x_{23}^2\right)\right.\nonumber\\
&\qquad\;\;+\left.
\frac{(w_{12}+w_{34})F_{2}\left(\frac{1}{z_5},\frac{1}{\bar{z}_5}\right)}{x_{12}^2 x_{34}^2}+
\frac{(w_{13}+w_{24})F_{2}(z_5,\bar{z}_5)}{x_{13}^2 x_{24}^2}+
\frac{(w_{14}+w_{23})F_{2}\left(\frac{z_5}{1-z_5},\frac{\bar{z}_5}{1-\bar{z}_5}\right)}{x_{14}^2 x_{23}^2}\right]\nonumber\\
&\quad+\order{g^6}
\,.
\end{align}
The factors of $\Nc$ are due to our normalization~\eqref{eq:dijdef}, under which we have
$G_{n,\ell}\sim G^{\text{SYM}}_n\sim \Nc^{2-n}$.
We use the notation
$w_{ij}\equiv\brk{x_{ij}^2+y_{ij}^2}/\brk{x_{ij}^2}$
of~\eqref{eq:wDef} and the polynomial $\mathcal{R}_{1234}$ defined
in~\eqref{eq:R1234first}. This latter vanishes in special kinematics
with enhanced supersymmetry: $\mathcal{R}_{1234}\sim
(z_5-\alpha_5)(z_5-\bar{\alpha}_5)(\bar{z}_5-\alpha_5)(\bar{z}_5-\bar{\alpha}_5)$,
where we have R-charge cross ratios defined as: $\alpha_i \equiv
z_{i}|_{x\to y}$. Furthermore, the ten-dimensional symmetry of the
integrand described in \secref{sec:4pointreview}
is now explicitly broken, since the integrals of~\eqref{IntegrandToInt} are
only performed on 4d spacetime, and the result only depends on 4d
cross-ratios. Nevertheless, these conformal integrals are still
dressed with rational coefficients that depend on 10d distances and
contain information on the whole tower of BPS single-trace operators
with arbitrary R-charge.

For the five-point generating function, we need a larger basis of conformal integrals, see~\eqref{eq:5pt2LoopInts} and \appref{appIntegralsDone}. Some of them evaluate to simple functions, such as:
\begin{align}
\mathbb{I}_{1}^{[\col1,\col2,\col3]}
&=\int\hspace{-0.5em}\int \frac{d^4 x_6}{\pi^2} \frac{d^4 x_7}{\pi^2}\frac{1}{(x_{\col16}^2 x_{\col26}^2 x_{\col36}^2)x_{67}^2(x_{\col17}^2 x_{\col27}^2 x_{\col37}^2)}= \frac{6 \zeta_3}{x_{12}^2 x_{23}^2 x_{13}^2}\,,\\
\mathbb{I}_{3}^{[\col1,\col2,\col3|\col4,\col5]}
&= \int\hspace{-0.5em}\int \frac{d^4 x_6}{\pi^2} \frac{d^4 x_7}{\pi^2}\frac{1}{(x_{\col16}^2 x_{\col26}^2 x_{\col36}^2 x_{\col46}^2)(x_{\col17}^2 x_{\col27}^2 x_{\col37}^2 x_{\col57}^2)} \,=\, \frac{F_{1}(z_5,\bar{z}_5)}{x_{13}^2 x_{24}^2}\frac{F_{1}(z_4,\bar{z}_4)}{x_{13}^2 x_{25}^2}
\,.
\end{align}
On the other hand, the genuine five-point integrals (the five-point
double-penta integral $\mathbb{I}_{2}$, the pentabox integral
$\mathbb{I}_{6}$, and the double-box integral $\mathbb{I}_7$) are not
known as closed-form functions in general kinematics. However, they
have been evaluated in some special kinematics, such as
multi-light-cone limits and 2d-plane
kinematics~\cite{Bork:2022vat,Fleury:2020ykw,Bercini:2024pya,Bork:2025ztu}. In the
following, we just leave them unevaluated when presenting the correlator.

Finally, at the integrated level, the SYM five-point generating function is given by:
\begin{align}\label{eq:G5SYM}
G_{5}^{\text{SYM}}&= G_{5,0} +\bigg{(}\frac{-C^{(5)}_{1234,5}}{2 \Nc\,w_{15}w_{25}w_{35}w_{45}}\times\left( G_{4}^{\text{SYM}} -G_{4,0}\right) + (\grp{C}_5\text{ perms.})\bigg{)}\nonumber \\
 &\qquad
 -\frac{g^{2}}{\Nc^3}\brk*{\frac{\mathcal{R}_{1234,5}}{\prod_{i<j}^5 w_{ij}}\,\frac{F_1(z_5,\bar{z}_5)}{x_{13}^2 x_{24}^2}+(\grp{C}_5\text{ perms.})}
\nonumber\\
&\qquad+\frac{g^{4}}{2\Nc^3}\frac{f_{1}\frac{6\zeta_3}{ x_{12}^2 x_{23}^2 x_{13}^2} +f_{3}\frac{F_{1}(z_5,\bar{z}_5)\,F_{1}(z_4,\bar{z}_4)}{x_{13}^4 x_{24}^2x_{25}^2}+ f_{4}\frac{F_{1}(z_5,\bar{z}_5)^2}{x_{13}^4 x_{24}^4}+ f_{5}\frac{F_{2}{(}{1}/{z_5},{1}/{\bar{z}_5}{)}}{x_{12}^2 x_{34}^4}+(\grp{S}_5\text{ perms.})}{\prod_{i<j}^5 w_{ij}}\nonumber \\
&\qquad+\frac{g^{4}}{2\Nc^3}\frac{f_{2}\,\mathbb{I}_2^{[1,2,3|4,5]} +f_{6}\,\mathbb{I}_6^{[1,2|3,4|5]}+ f_{7}\,\mathbb{I}_7^{[1,2|3,4|5]}+(\grp{S}_5\text{ perms.})}{\prod_{i<j}^5 w_{ij}} +\order{g^6}
\,.
\end{align}
On the first line, we have the free-theory correlator $G_{5,0}$, and the
four-point subcorrelator $G_{4}^{\text{SYM}}$ from
eq.~\eqref{eq:G4SYM}, as well as its other four cyclic permutations.
The coefficient of the latter contains 10d double poles, and is
composed of the coefficient~\eqref{eq:defC5} and the factors
$w_{ij}\equiv\brk{x_{ij}^2+y_{ij}^2}/\brk{x_{ij}^2}$.
We also need to
subtract $G_{4,0}$ to avoid overcounting of four-point tree level
graphs. The next lines are the loop corrections that come with 10d
simple poles in $w_{ij}$. The second line contains
the one-loop correction, given in terms of
scalar-box integrals and the polynomial $\mathcal{R}_{1234,5}$
of~\eqref{eq:R1234c5}. The third line contains the two-loop
contributions from the integrals that evaluate to ladder
functions~\eqref{BoxIntegrals}, and the last line has the more
complicated five-point integrals. The coefficients $f_i$ stand for polynomials in $x_{ij}^2$ and
$w_{ij}$. We
provide them explicitly
in eqs.~\eqref{eq:fk}, which require the polynomials given in eqs.~\eqref{eq:mathF}--\eqref{eq:P71andP72}.

From the generating functions~\eqref{eq:G4SYM} and~\eqref{eq:G5SYM},
we can access all four- and five-point planar correlators of half-BPS
single-trace operators with arbitrary R-charge up to two-loop order.
This is done by expanding the 10d poles $1/w_{ij}$ in a geometric
series in $y_{ij}^2/x_{ij}^2$, and selecting the desired R-charge
correlator by the exponents on the polarization vectors $y_i$, as
shown in~\eqref{eq:Rweightcomponent}. For instance, the four-point
correlator of the lightest operator $\mathcal{O}_2$ is obtained by
setting every factor of $w_{ij}$ in~\eqref{eq:G4SYM} to one, because
the polynomial $\mathcal{R}_{1234}$ already saturates the R-charge of
this correlator, see~\eqref{eq:R1234first}. Similarly, the five-point
correlator of $\mathcal{O}_{2}$ can be obtained from the small $y_i$
limit of~\eqref{eq:G5SYM}. This lightest correlator
does not receive contributions from the four-point subcorrelators on
the first line of~\eqref{eq:G5SYM}, and its loop corrections are
obtained by setting the denominator factors of $w_{ij}$ to one, and
projecting the $f_k$ coefficients to the appropriate R~charge (only
$f_4$ gives a zero contribution in this case).

The five-point generating function~\eqref{eq:G5SYM} is our main
result. At one-loop order, it repackages the results
of~\cite{Drukker:2008pi} for five-point correlators with arbitrary
R-charge. It generalizes the two-loop five-point correlator of the
lightest half-BPS operators of~\cite{Bargheer:2022sfd} by upgrading
the coefficients of the conformal integrals to rational functions with
10d poles, which contain the information on arbitrary half-BPS
single-trace operators. The evaluation of the two-loop five-point
conformal integrals in general kinematics still remains an open
problem. However, they were evaluated in~\cite{Bercini:2020msp} in
certain OPE limits. In the following section, we use these results to
obtain OPE structure constants of non-protected spinning operators. We
also find a perfect match for some of this OPE data with
integrability-based predictions.
This provides a non-trivial test of
our two-loop generating function.

\subsection{OPE Limit and Spinning Structure Constants}
\label{sec:OPELimit}

\begin{figure}[t]
    \centering
    \includegraphics[width=1\linewidth]{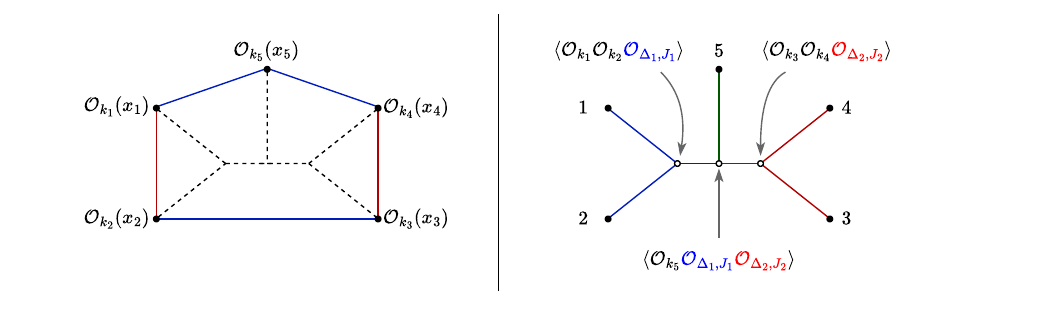}
    \caption{On the left, we depict the OPE
    limit~\protect\eqref{eq:DLC} that we consider for the five-point
    function: The light-like limits ($x_{12}^2,x_{34}^2 \to 0$) are
    displayed in red, the Euclidean limits ($x_{2} \to x_3$, $x_{1}
    \to x_{5}$ and $x_{4} \to x_{5}$) are shown in blue. On the right,
    we present the channel that we consider, together with the
    three-point functions involving one and two spinning operators
    that appear in this decomposition.}
    \label{figOPELimits}
\end{figure}

One of the several applications that
higher-point integrands have is extracting CFT data of many non-protected
operators. While scalar four-point functions encode the structure
constants of a single spinning operator in their OPE decomposition,
scalar five-point correlation functions encode richer structure
constants involving two spinning operators, as shown on the right
of \figref{figOPELimits}, and schematically written here:
\begin{equation}
    \langle k_1\,k_2\,k_3\,k_4\,k_5 \rangle = \frac{1}{N_c^3}\sum_{\textcolor{blue}{\tau_1},\textcolor{blue}{J_1}}\sum_{\textcolor{red1}{\tau_2},\textcolor{red1}{J_2}}\sum_{\textcolor{green1}{\ell}=0}^{\min(\textcolor{blue}{J_1},\textcolor{red1}{J_2})} C_{\textcolor{blue}{\tau_1}}^{\textcolor{blue}{J_1}}\,C_{\textcolor{red1}{\tau_2}}^{\textcolor{red1}{J_2}}\,C_{\textcolor{blue}{\tau_1},\textcolor{red1}{\tau_2}}^{\textcolor{blue}{J_1},\textcolor{red1}{J_2};\textcolor{green1}{\ell}} \times \mathcal{F}(\tau_i,J_i,\ell,x_{ij})\,,
    \label{schemeOPE}
\end{equation}
where $\mathcal{F}$ is a kinematical function completely fixed by
conformal symmetry~\cite{Goncalves:2019znr,Poland:2023vpn}, known as a
conformal block, $\tau_i$ are the twists ($\equiv\Delta_i-J_i$), $J_i$
are the spins, and $\ell$ is the
so-called spin-polarization, which accounts for the several tensor
structures in three-point functions with two spinning operators:
\begin{equation}\label{eq:CtwoSpins}
    \langle \mathcal{O}(\tau_1,J_1)\mathcal{O}(\tau_2,J_2)\mathcal{O}(\tau_3)\rangle =\frac{1}{N_c}  \sum_{\ell=0}^{\min(J_1,J_2)} \left(\frac{V_{1,23}^{J_1-\ell}V_{2,31}^{J_2-\ell} H_{23}^{\ell}}{x_{12}^{\kappa_1+\kappa_2-\kappa_3} x_{13}^{\kappa_1+\kappa_3-\kappa_2} x_{23}^{\kappa_2+\kappa_3-\kappa_1}}\right) C_{\tau_1,\tau_2}^{J_1,J_2;\ell}
    \,,
\end{equation}
being $\kappa_i = \tau_i+2J_i$ the conformal spin, $V$ and~$H$
the two conformally invariant tensors can be found on eqs.~(4.14)
and~(4.15) of~\cite{Costa:2011dw}, and finally
$C_{\tau_1,\tau_1}^{J_1,J_2;\ell}$ is the structure constant between a
scalar and two spinning operators, which reduces to the single
spinning operator~$C_{\tau}^{J}$ when one of the operators has zero
spin.

\paragraph{CFT Data Extraction.}

To extract the conformal data, we will follow the same logic
as~\cite{Bercini:2024pya}, which consists in considering two
expressions for the same five-point correlation function. One
expression is obtained by expanding the correlator~\eqref{eq:G5SYM} in
a particular kinematical limit, writing it as explicit functions of space-time coordinates.
The other expression is its equally explicit OPE decomposition in terms of structure
constants and conformal blocks~\eqref{schemeOPE}. By comparing these
two expressions, conformal-integral and conformal-block decompositions,
one can easily read off the structure constants.

We will focus on structure constants that involve one or two
non-protected operators lying in the $\alg{sl}(2)$ sector of the theory.
These are operators composed of a single type of complex scalar,
\eg~$Z$ (or its conjugate), and derivatives along a light-cone
direction: $\text{Tr}(D^{J}_{+} Z^{\tau^0})$, where the number of scalars
$\tau^0$ represents the tree-level twist, and $J$ is the Lorentz spin.
At non-zero coupling, the full twist is obtained by adding the
anomalous dimension: $\tau = \tau^{0}+ \gamma(g)$. In order to isolate
the contribution of this class of operators in the OPE of the
five-point correlator~\eqref{eq:G5SYM}, we need to perform some
specific R-charge projections of the external operators, and take light-cone
limits in the cross ratios~$u_i$.

\begin{figure}[t]
    \centering
    \includegraphics[width=0.75\linewidth]{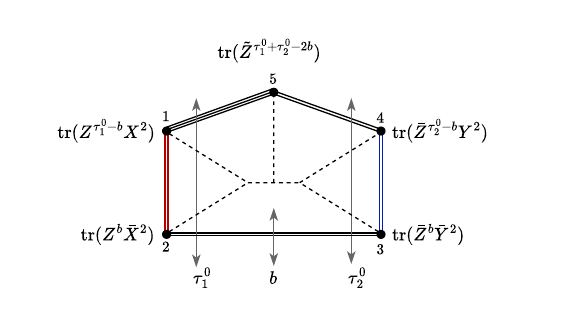}
    \vspace{-2em}
    \caption{The correlation functions that we will consider. Black
    lines are propagators of $Z$-$\bar{Z}$ scalars, blue lines are propagators
    of $Y$-$\bar{Y}$ scalars, and red lines are propagators between
    $X$-$\bar{X}$ scalars. The operator at the top is the rotated BPS
    operator made out of fields $\tilde{Z}=Z+\bar{Z}+X-\bar{X}$. We
    choose special values of R-charge for the external
    operators, such that the OPE decomposition starts with operators
    of leading twists $\tau_1$ and $\tau_2$, which lie in the $\alg{sl}(2)$
    sector of the spectrum. The value $b$ of the bottom bridge length
    controls the two-loop wrapping corrections to the structure
    constant  with two  non-protected spinning operators.}
    \label{fig5ptCorrelators}
\end{figure}

Namely, we consider five-point correlators of half-BPS operators
$\mathcal{O}_{k_i}$ with R-charges
\begin{equation}
\label{eq:kpolarized}
k_{1} = \tau_1^0-b+2\,,\quad k_{2}=k_3 = b+2\,,\quad k_4 = \tau_2^0-b+2\,,\quad k_5 = \tau_1^0+\tau_2^0 -2b
\end{equation}
with $\tau^{0}_1,\tau_{2}^{0}>b$, and perform specific R-charge
projections to obtain the scalar-field content shown in
\figref{fig5ptCorrelators}.
Specifically, we set the R-charge polarizations to (in $\grp{SO}(6)$
fundamental notation):
\begin{align}
y_1&=\frac{1}{{2}}\brk{t_1,+\ii t_1,0,0,1,\ii}\,,&
y_4&=\frac{1}{{2}}\brk{0,0,t_4,+\ii t_4,1,-\ii}\,,&
y_5&=\brk{0,\ii,0,0,1,0}\,,\nn\\
y_2&=\frac{1}{{2}}\brk{t_2,-\ii t_2,0,0,1,\ii}\,,&
y_3&=\frac{1}{{2}}\brk{0,0,t_3,-\ii t_3,1,-\ii}\,,&&
\label{eq:OPEPolarizations}
\end{align}
then apply
two derivatives in each of $t_1,\dots,t_4$, and finally set $t_i$ to
zero. For example, the top left operator in \figref{fig5ptCorrelators}
is obtained by:%
\footnote{In the single-particle basis, the right-hand side
of~\eqref{eq:Rpolarization} would also have double traces, however
these extra terms will have subleading $1/\Nc$ contribution.}
\begin{equation}
\label{eq:Rpolarization}
\frac{1}{2!}
\eval*{\frac{\partial^{2}}{\partial t_1^{2}}\,\mathcal{O}_{k_1}(x_1,y_1)}_{t_1\to 0} \,=\, \frac{1}{k_1} \tr\brk[s]*{Z^{\tau_1^0-b}X^2} + \brk{\text{permutations}}
\,,
\end{equation}
where
\begin{align}
X&=\frac{\phi_1+\ii\phi_2}{{2}}
\,,\qquad
Y=\frac{\phi_3+\ii\phi_4}{{2}}
\,,\qquad
Z=\frac{\phi_5+\ii\phi_6}{{2}}
\,,\\
\tilde{Z}&=Z+\bar{Z}+X-\bar{X}=\phi_5+\ii\phi_2
\,.
\end{align}
The permutations on the right-hand side of~\eqref{eq:Rpolarization} stand for different
inequivalent orderings of the complex scalars $Z$ and $X$ inside the
trace.
We perform similar R-charge projections on the other four operators,
such that, by conservation of R-charge, the number of free propagators
(``bridge length'') between the first and second operators is fixed
to~$2$, and their light-cone OPE starts with $\alg{sl}(2)$ operators with~$Z$
scalars and tree-level twist $\tau_1^{0}$. Similarly, the bridge length
between the third and fourth operators is also fixed to~$2$,%
\footnote{This choice of bridge lengths between operator pairs $1$-$2$ and $3$-$4$
only affects the structure constants with one single spinning
operator. The case with two spinning operators depends on the choice
of the bridge length $b$ between operators~$2$ and~$3$.}
and their light-cone OPE has $\alg{sl}(2)$ operators with $\bar{Z}$ scalars and
tree-level twist $\tau_2^{0}$.

At the level of the generating function~\eqref{eq:G5SYM}, the R-charge
projections $\mathcal{O}\to \mathcal{O}_{k_i}$,
see~\eqref{eq:Rweightcomponent}, and the
polarizations~\eqref{eq:Rpolarization}, only affect the coefficients
$f_i$ and can be easily performed. Furthermore,
following~\cite{Bercini:2024pya}, to make the $\alg{sl}(2)$ operators
dominant in the OPE decomposition, we consider the double light cone OPE
limits $x_{12}^2,x_{34}^2\to 0$, which focuses on the $\alg{sl}(2)$
towers of spinning operators with tree-level twists $\tau_1^0$ and
$\tau_2^0$, respectively. Subsequently, we take the Euclidean
coincident point OPE limits $x_{1}\to x_{5}$, $x_{2}\to x_{3}$, and
$x_{4}\to x_5$, which focus on the small-spin operators of the $\alg{sl}(2)$ towers.
In terms of the five independent conformal cross-ratios
of~\eqref{eq:ufive}, these limits are equivalent to:
\begin{equation}
\label{eq:DLC}
\text{OPE:}\quad u_1,\,u_4 \to 0\quad \text{and}\quad u_{2},\,u_{3},\,u_{5}\to 1
\,.
\end{equation}
This OPE limit also simplifies the problem of comparing the conformal-block expansion and the conformal-integral expressions, since now both turn into tractable objects.  First, the complicated functional form of the conformal blocks is reduced to
simple combinations of hypergeometric functions in the light-cone OPE
limits $u_1,u_4\to 0$, see~\cite{Bercini:2020msp}, and is further simplified to a series in powers and logarithms of the cross ratios in the
subsequent Euclidean OPE limits $u_2,u_3,u_5\to 1$, see~\appref{appConfBlock}. Second, the most complicated two-loop five-point conformal integrals, $\mathbb{I}_{2}$, $\mathbb{I}_{6}$ and $\mathbb{I}_7$, can
all be evaluated as expansions around this limit~\cite{Bercini:2024pya}. For example:
\begin{equation}
    \mathbb{I}_{6}^{[1,2|3,4|5]} \overset{\text{OPE}}{=}  \frac{1}{x_{12}^2x_{13}^2x_{24}^2}\left(6\zeta_3 + \frac{v_2^2}{4} + 3\zeta_3 v_1 v_2 v_3 +\frac{v_1 v_2^2 v_3}{4} +\left(1+24\zeta_3\right)\frac{v_1^2 v_2^2 v_3^2}{12}+\dots\right)
    \,.
\end{equation}
Similar expansions for the other integrals are presented in \appref{appIntegralsDone}. Here, all the cross-ratios $v_i$ are going to zero in the OPE limit~\eqref{eq:DLC} and are related to $u_i$ in~\eqref{eq:ufive} as:
\begin{equation}\label{eq:vcross}
    v_1 = 1- u_2
    \,, \quad v_2 = 1- u_3
    \quad \text{and} \quad v_1 v_2 v_3 = 1- u_5
    \,.
\end{equation}
This is exactly the configuration studied in~\cite{Bercini:2024pya},
where, by considering the correlator $\langle 22222 \rangle$,
structure constants for two spinning operators of twist two were
extracted at two loops. Now, equipped with the generating
function~\eqref{eq:G5SYM}, we can consider correlation functions with
arbitrary external dimensions, $k_i$ in~\eqref{eq:kpolarized}, and consequently extract two-loop CFT data with arbitrary values of leading twists ($\tau_1^{0}$ and $\tau_2^{0}$ in \figref{fig5ptCorrelators}), as exemplified in~\appref{appOPE}. The resulting data extracted
from these correlators is presented in \tabref{tab:twists2and3}, \tabref{tab:twists3and3} and \tabref{tabData}.

\begin{table}[t]
    \centering
    \begin{tabular}{c|ccccc}
    \diagbox[width=1.30cm,height=.75cm]{$\mspace{-8mu}\tau^0$}{$J$}
      & 2 & 3 & 4 & 5 & 6 \\
    \hline
    2 & 1 & 0 & 1 & 0 & 1 \rule{0pt}{2.7ex} \\
    3 & 1 & 2 & 1 & 2 & 3 \\
    4 & 2 & 2 & 5 & 4 & 8 \\
    5 & 2 & 4 & 7 & 12 & 16 \\
    \end{tabular}
    \caption{Degeneracies $\text{deg}(\tau^0,J)$ of
    $\alg{sl}(2)$ multiplets with tree-level twists $\tau^0=2,\dots,5$ and spins
    $J=2,\dots,6$.}
    \label{tab:degs}
\end{table}

Due to the presence of nearly degenerate operators at weak coupling,
it is generally not possible to determine all individual OPE
coefficients. These degeneracies grow with the values of tree-level
twist~$\tau^0$ and spin~$J$~\cite{Marboe:2017dmb}, as shown in
\tabref{tab:degs}. Beyond twist-two, the only non-degenerate cases are
twist-three operators
with spins $J=2$ and $J=4$. However, their structure constants with twist-two
operators still form an infinite set of novel two-loop OPE data,
extractable from our five-point generating function. Examples are
given in \tabref{tab:twists2and3}, while \tabref{tab:twists3and3}
lists structure constants of pairs of twist-three operators with spins $J=2$
or $J=4$.
In order to clarify our notation for these structure constants,
we represent it graphically as:
\begin{equation}\label{eq:3pointgraph}
C^{J_1,J_2;\,\ell}_{\tau_1,\tau_2;\,b}
\sim \mspace{-10mu}
\includegraphics[align=c,scale=.8]{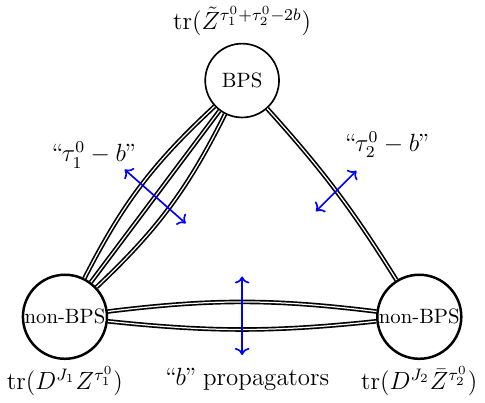}
\,.
\end{equation}
Besides the twists $\tau_i\equiv\tau_{i}^0+\gamma_i(g)$ and spins $J_i$ of the unprotected operators, these structure constants also depend on the spin polarization $\ell$, as defined in~\eqref{eq:CtwoSpins}. In addition, they depend on the parameter $b$, which represents the number of tree-level propagators between the two spinning operators (equivalently, between the BPS operators~2 and~3 in the five-point function of \figref{fig5ptCorrelators}). This dependence is evident in the contrast between the top and bottom tables of \tabref{tab:twists3and3}. As we explain below, the difference can be traced to distinct types of ``wrapping'' corrections in the hexagon formalism.

For operators with twist $\tau^0=4$ or higher, degeneracies occur for every spin (see \tabref{tab:degs}).
In perturbation theory, such states appear with nearly identical conformal blocks in the OPE
expansion~\eqref{schemeOPE}, and therefore only their averaged
contributions can be extracted in the form of sum rules:%
\footnote{All degeneracies eventually lift at higher orders in
perturbation theory. However, to fully resolve the averages
$P_{\tau_1,\tau_2;b}^{J_1,J_2;\ell}$ of $n$ degenerate states at a given loop order $\ell$, one
would need to compute the OPE expansion to a very large loop order
$\sim n\times\ell$.}
\begin{equation}
P^{J_1,J_2;\ell}_{\tau_1,\tau_2;b} = \sum_{n_1=1}^{\text{deg}(\tau_1,J_1)} \,\sum_{n_2=1}^{\text{deg}(\tau_2,J_2)} C_{\tau_1,n_1}^{J_1} \, C_{\tau_2,n_2}^{J_2}\, C_{\tau_1,n_1;\tau_2,n_2;b}^{J_1,J_2;\ell}
\,.
\label{combData}
\end{equation}
Here $n_i$ labels the nearly degenerate states that share the same tree-level twist~$\tau_i^0$ and
spin~$J_i$, while~$b$ denotes the bridge length between the two spinning
operators. The number of degenerate supermultiplets can be determined
using the Bethe Ansatz for any values of twist
and spin~\cite{Marboe:2017dmb}.
For instance, there are two non-protected operators
with tree-level twist $\tau^0=4$ and spin $J=2$, \ie $\text{deg}(4,2) = 2$. Using
integrability, one can easily compute the anomalous dimensions of these two states in perturbation theory
(\eg by solving the Bethe equations%
\footnote{To two-loop order accuracy, there are no wrapping corrections and it suffices to consider the Asymptotic Bethe equations.}):
\begin{align}
    \tau(1) &= 4 + (10-2\sqrt{5}) g^2 -(34-10\sqrt{5}) g^4 +\order{g^6} \,,\\
    \tau(2) &= 4 + (10+2\sqrt{5}) g^2 -(34+10\sqrt{5}) g^4 +\order{g^6} \,.
\end{align}
This spectral information serves as input for the OPE conformal block
decomposition. By comparing the latter with the OPE limit of the
five-point correlator in the integral basis,
we obtain the
two-loop sum rules $P^{J_1,J_2;\ell}_{\tau_1,\tau_2;b}$
of~\eqref{combData}, which are presented in \tabref{tabData}. This OPE
extraction procedure is illustrated in \appref{appOPE} for the
simplest case of twist-two spinning operators.
Besides the data points
shown in the tables, we provide higher-spin data for $P$'s
in the \mathematica file \ancfile{Psums.m}.
The OPE data is subject to comparison with hexagon predictions as described below.

\newcolumntype{C}{>{$}c<{$}} 
\newcolumntype{L}{>{$}l<{$}} 
{\renewcommand{\arraystretch}{1.7}
\begin{table}[t]
\centering
\begin{tabular}{l|>{$}l<{$}|>{$}l<{$}|>{$}l<{$}}
$J$& \multicolumn{1}{C|}{\ell=0} & \multicolumn{1}{C|}{\ell=1} & \multicolumn{1}{C}{\ell=2}\\
\hline
2& 1-7 g^2+60 g^4 & 1-5 g^2+44 g^4 & 1+2 g^2-26 g^4 \\
4&1-\frac{502 g^2}{63}+\frac{2191895 g^4}{31752} & 1-\frac{1861 g^2}{252}+\frac{614087 g^4}{7938} & 1-\frac{85 g^2}{42}+\frac{530515 g^4}{31752} \\
6&1-\frac{13999 g^2}{1650}+\frac{487292813 g^4}{6534000} & 1-\frac{7027g^2}{825}+\frac{212614463 g^4}{2178000} & 1-\frac{993 g^2}{275}+\frac{1298711449g^4}{32670000} \\
\end{tabular}
\captionsetup{singlelinecheck=off}
\caption{Two-loop OPE coefficients squared
$\brk!{C^{2,J;\ell}_{3,2}}^2$ of two spinning operators with one
protected scalar operator, normalized by their tree-level values
\setlength{\abovedisplayskip}{0.3ex}
\setlength{\belowdisplayskip}{0.3ex}
\begin{displaymath}
\left(C^{2,J;\ell}_{3,2}\right)^2\overset{\text{tree}}{=}
\frac{1}{6}\binom{2}{\ell}^2 \binom{J}{\ell}^2\frac{J!^2}{(2J)!}
\,.
\end{displaymath}
The first
spinning operator is fixed to have tree-level twist $\tau_1^0=3$ and
spin $J_1=2$. The second operator belongs to the twist-two family with
arbitrary even spin $J$. For illustration, results are restricted to
$J=2,4,6$ but include all spin polarizations $\ell=0,1,2$.}
\label{tab:twists2and3}
\end{table}}

{\renewcommand{\arraystretch}{1.7}
\begin{table}[t]
\begin{tabular}{l|L|L|L}
 \multicolumn{1}{c}{} &\multicolumn{1}{l}{$\boxed{\; b=1 \;}$}&\multicolumn{1}{c}{}&\multicolumn{1}{c}{} \\
 $\ell$& \multicolumn{1}{C|}{J_1=J_2=2} & \multicolumn{1}{C|}{J_1=2,\,J_2=4} & \multicolumn{1}{C}{J_1=J_2=4}\\
 \hline
$0$ & \textcolor{blue}{1-8 g^2+56 g^4} & \textcolor{blue}{1-9 g^2+\frac{1115 g^4}{18}} & \textcolor{blue}{1-\frac{28
   g^2}{3}+\frac{319 g^4}{6}} \\
$1$ & \textcolor{blue}{1-8 g^2+72 g^4} & \textcolor{blue}{1-10 g^2+\frac{1793 g^4}{18}} & \textcolor{blue}{1-\frac{34
   g^2}{3}+\frac{629 g^4}{6}} \\
$2$ & \textcolor{blue}{1} & \textcolor{blue}{1-\frac{13 g^2}{3}+\frac{113 g^4}{3}} & \textcolor{blue}{1-\frac{28
   g^2}{3}+\frac{262 g^4}{3}} \\
$3$& \multicolumn{1}{C}{}& \multicolumn{1}{C|}{} & \textcolor{blue}{1-\frac{23 g^2}{6}+\frac{1573 g^4}{48}} \\
$4$& \multicolumn{1}{C}{}& \multicolumn{1}{C|}{} & \textcolor{blue}{1+\frac{62 g^2}{3}-\frac{109 g^4}{3}} \\
\end{tabular}
\vspace{.5cm}

\begin{tabular}{c|@{}L@{}|@{}L@{}|@{}L@{}}
 \multicolumn{1}{c}{} &\multicolumn{1}{l}{$\boxed{\;b=2\;}$}&\multicolumn{1}{c}{}&\multicolumn{1}{c}{} \\
 $\ell$& \multicolumn{1}{C|}{J_1=J_2=2} & \multicolumn{1}{C|}{J_1=2,\,J_2=4} & \multicolumn{1}{C}{J_1=J_2=4}\\
 \hline
$0$ &\,1-24 g^2+g^4 \left(32 \zeta_3+\frac{1216}{3}\right)\, & \,1-\frac{91 g^2}{3}+g^4
   \left(\frac{216 \zeta_3}{5}+\frac{18113}{30}\right)\, & \,1-\frac{112
   g^2}{3}+g^4 \left(\frac{1296 \zeta_3}{35}+\frac{30883}{35}\right)\, \\
$1$ &\,1+g^4 \left(-32 \zeta_3-\frac{136}{3}\right)\, & \,1-\frac{31 g^2}{3}+g^4
   \left(\frac{72 \zeta_3}{5}+\frac{637}{10}\right)\, & \,1-\frac{67 g^2}{3}+g^4
   \left(\frac{324 \zeta_3}{35}+\frac{12297}{35}\right)\, \\
$2$ &\,1+g^4 \left(\frac{32 \zeta_3}{3}-\frac{56}{9}\right)\, & \,1-\frac{5 g^2}{3}+g^4
   \left(\frac{136 \zeta_3}{5}+\frac{422}{45}\right)\, & \,1-\frac{116 g^2}{9}+g^4
   \left(\frac{408 \zeta_3}{35}+\frac{860263}{5670}\right)\, \\
$3$& \multicolumn{1}{C}{}& \multicolumn{1}{C|}{} & \,1+\frac{80 g^2}{3}+g^4 \left(-\frac{3456 \zeta_3}{35}-\frac{68141}{420}\right)\, \\
$4$& \multicolumn{1}{C}{}& \multicolumn{1}{C|}{} & \,1-\frac{106 g^2}{57}+g^4 \left(\frac{864 \zeta_3}{35}+\frac{1067407}{75810}\right)\, \\
\end{tabular}
\caption{Two-loop OPE coefficients squared $\brk!{C^{J_1,J_2;\ell}_{3,3;b}}^2$ for two twist-three operators of spins $J_1$ and $J_2$ with one protected scalar operator. Here, $b$ denotes the number of tree-level propagators exchanged between the two spinning operators. The coefficients are normalized by their tree-level values and reported for all possible spin polarizations $\ell\leq\text{min}(J_1,J_2)$.}
\label{tab:twists3and3}
\end{table}}

{\renewcommand{\arraystretch}{1.7}
\begin{table}[t]
\begin{tabular}{@{\;}C@{\;\;}C|L|L|L@{\;}}
\multicolumn{1}{c}{} &\multicolumn{1}{c}{} & \multicolumn{1}{l}{$\boxed{\; b=1 \;}$}&\multicolumn{1}{c}{}&\multicolumn{1}{c}{}\\
\tau_1^0 & \tau_2^0 & \multicolumn{1}{C|}{\ell=0} & \multicolumn{1}{C|}{\ell=1} & \multicolumn{1}{C}{\ell=2}\\
\hline
2 & 4 & \frac{1}{40} -\frac{11}{30}g^2+\frac{217}{40}g^4 & \frac{1}{10} -\frac{7}{5}g^2+20g^4 & \frac{1}{40} -\frac{13}{6}g^2+\frac{137}{60}g^4 \\
3 & 4 & \color{blue} \frac{1}{40} -\frac{31}{120}g^2+ \frac{341}{120}g^4
  & \color{blue}\frac{1}{10} -\frac{11}{10}g^2 + \frac{379}{30}g^4
  & \color{blue} \frac{1}{40} -\frac{1}{5}g^2+ 2g^4
\\
4 & 4 & \color{blue} \frac{9}{400} -\frac{1}{5}g^2+\frac{11}{5}g^4 & \color{blue} \frac{9}{100} -\frac{9}{10}g^2+\frac{51}{5}g^4 & \color{blue} \frac{9}{400} -\frac{7}{40}g^2+\frac{7}{4}g^4 \\
\end{tabular}

\vspace{.5cm}

\begin{tabular}{@{\;}C@{\;\;}C|L|L|L@{\;}}
\multicolumn{1}{c}{} &\multicolumn{1}{c}{} & \multicolumn{1}{l}{$\boxed{\; b=2 \;}$}&\multicolumn{1}{c}{}&\multicolumn{1}{c}{}\\
   \tau_1^0 & \tau_2^0 & \multicolumn{1}{C|}{\ell=0} & \multicolumn{1}{C|}{\ell=1} & \multicolumn{1}{C}{\ell=2}\\
 \hline
3 & 4 &\frac{7}{120}-\frac{53}{60}g^2+\left(\frac{1963}{180}+\frac{\zeta_3}{3}\right) g^4  & \frac{7}{60}-g^2+\left(\frac{791}{90}-\frac{2\zeta_3}{3}\right)
   g^4 & \frac{3}{40}-\frac{13}{30}g^2+\left(\frac{119}{36}+\frac{\zeta_3}{3}\right)
   g^4 \\
3 & 5 &\frac{1}{15}-\frac{4}{5}g^2+ \left(\frac{769}{90}+\frac{4 \zeta_3}{15}\right)g^4  & \frac{2}{15}-\frac{16}{15}g^2+ \left(\frac{422}{45}-\frac{8 \zeta_3}{15}\right)g^4 & \frac{1}{15}-\frac{2}{5}g^2+ \left(\frac{164}{45}+\frac{4\zeta_3}{15}\right)g^4 \\
4 & 4 & \textcolor{red1}{\frac{49}{400}-\frac{7}{5}g^2+\frac{531}{40}g^4} & \textcolor{red1}{\frac{49}{200}-2
   g^2+\frac{319}{20}g^4} & \textcolor{red1}{\frac{33}{400}-\frac{3}{10}g^2+ \frac{17}{20}g^4}
   \\
4 & 5 & \textcolor{red1}{\frac{7}{50}-\frac{63}{50}g^2+\frac{991}{100}g^4} & \textcolor{red1}{\frac{7}{25}-2
   g^2+\frac{707}{50}g^4} & \textcolor{red1}{\frac{2}{25}-\frac{8}{25}g^2+\frac{9}{5}g^4} \\
5 & 5 & \textcolor{red1}{\frac{4}{25}-\frac{28}{25}g^2+7 g^4} & \textcolor{red1}{\frac{8}{25}-\frac{48}{25}g^2+\frac{278}{25}g^4} & \textcolor{red1}{\frac{2}{25}-\frac{8}{25}g^2+\frac{53}{25}g^4}\\
\end{tabular}
\caption{Two-loop OPE sum rules $P^{2,2;\ell}_{\tau_1,\tau_2;b}$ of nearly-degenerate
operators with tree-level twist $\tau_i^0$ and spins $J_1=J_2=2$, see
eq.~\protect\eqref{combData} for the definition. Here, $b$ denotes the
number of tree-level propagators exchanged between the pair of
spinning operators, and~$\ell$ is the spin polarization. The
degeneracies are shown in \protect\tabref{tab:degs}. We omit the
non-degenerate twist-two and twist-three cases, already presented in
previous tables. These as well as $P$'s for other spins are included
in the \mathematica file \ancfile{Psums.m}.}
\label{tabData}
\end{table}}

\paragraph{Comparison with Hexagons.}

In planar $\superN=4$ SYM, we can compute structure
constants via the integrability-based formalism known as
\emph{hexagons}~\cite{Basso:2015zoa}. In this formalism, the
three-point functions (``pair of pants'') are broken down into two
hexagon form factors. These are constrained by symmetries, and can be
bootstrapped at finite coupling. Gluing them back together yields the
structure constants. When more than one non-protected spinning
operator is present, the structure constants depend on the spin
polarization $\ell$, which can also be accommodated in the hexagon
formalism, see eq.~(31) in~\cite{Bercini:2022gvs}.%
\footnote{Our notation differs slightly
from~\cite{Bercini:2022gvs}. For example,  the two-spinning
structure constant is written here as:
$\brk!{C^{J_1,J_2;\ell}_{\tau_1,\tau_2}}^{\text{here}} \equiv
\brk!{C^{J_1,J_2,0}_{0,0,\ell}}^{\text{there}}$. In the latter,
the twists were not labeled explicitly.}

The gluing of hexagons in perturbation theory depends on the bridge
lengths, \ie the number of tree-level propagators exchanged between
pairs of operators. If the minimum bridge length is $\ell$, then
hexagons are glued together trivially up to order $g^{2\ell}$ (the
``asymptotic hexagon'' regime). Beyond this order, mirror particles
propagate across the bridges, producing wrapping
corrections~\cite{Basso:2015eqa}. For the present case, the onset of
wrapping was predicted as follows~\cite{Bercini:2022gvs}:
\begin{itemize}
\item The two bridges connecting to the BPS operator, of lengths
$\tau_1^0-b$ and $\tau_2^0-b$, develop wrapping corrections at loop
orders $\tau_1^{0}-b+1$ and $\tau_2^{0}-b+1$, respectively. At
two-loop order, such corrections appear in the twist-three structure
constants of \tabref{tab:twists2and3}, and the $b=2$ entries of
\tabref{tab:twists3and3}. In addition, they are also present in the
$P$ sum rules with $(\tau_1^0,b)=(2,1)$ and $(\tau_1^0,b)=(3,2)$ in
\tabref{tabData}. Most of these OPE results are identifiable by the
presence of $\zeta_3$ at order $g^4$.
\item The bridge connecting the two spinning operators, of length $b$,
develops wrapping corrections only at loop order $b+2$. For
non-extremal structure constants ($b\geq 1$), this implies that such
corrections are delayed at least to three loops, $\mathcal{O}(g^6)$.
\end{itemize}
Based on these predictions, structure constants with bridges
satisfying $\tau_1^0-b\geq 2,\,\tau_2^0-b\geq 2$, and $b\geq 1$ should
be free of wrapping corrections up to two loops, \ie order $g^4$. In
the following, we focus precisely on testing our OPE results in this
regime, where asymptotic hexagons are expected to be sufficient to
two-loop order. The corresponding data points are highlighted in blue
and red in \tabref{tab:twists3and3} and \tabref{tabData}. Our findings
are:
\begin{itemize}
\item At one loop, $\mathcal{O}(g^2)$, asymptotic hexagons match all OPE data extracted from our five-point correlators.
\item At two loops, $\mathcal{O}(g^4)$, we find perfect agreement between the red OPE data with $b=2$ in \tabref{tabData} and the asymptotic hexagon prediction.
\item At the same order, however, we find a mismatch for the blue OPE
data in \tabref{tab:twists3and3} and \tabref{tabData}. These
correspond to structure constants with bridge length $b=1$, twists $\tau_1^0,\tau_2^0\geq 3$ and arbitrary spins $J_i$.
\end{itemize}
The agreement in the $b=2$ case confirms that for sufficiently large
bridges, asymptotic hexagons suffice up to two loops. However, the
mismatch in the case $b=1$ indicates that the wrapping corrections on
this bridge appear earlier than predicted by~\cite{Bercini:2022gvs},
already at $\mathcal{O}(g^4)$.

Interestingly, this tension may be reconciled when summing over spin
polarizations. In~\cite{Bercini:2022gvs}, the prediction was supported
by matching the sum
$\sum_{\ell=0}^{\min(J_1,J_2)}C^{J_1,J_2;\,\ell}_{2,2;\,b=1}$ with
so-called ``abelian'' hexagons, corrected only by wrappings from the
BPS-connected bridges. However, no comparisons were made at the level of  individual spin
polarizations $\ell$. Extending this analysis with our twist-three OPE
data in \tabref{tab:twists3and3}, we find that also the corresponding twist-three sum $\sum_{\ell=0}^{\min(J_1,J_2)}C^{J_1,J_2;\,\ell}_{3,3;\,b=1}$ matches
the abelian hexagon computation, this time without requiring any
wrapping corrections. This suggests that while wrapping effects on the
$b=1$ bridge may indeed be present at $\mathcal{O}(g^4)$, they can
cancel in the polarization sum. A full resolution of this puzzle is
left for future work.

Finally, let us remark that to systematically delay the wrapping corrections at all orders in
perturbation theory, we need to consider the limit where all bridge
lengths go to infinity, giving rise to the so-called ``asymptotic''
correlation functions~\cite{Basso:2017khq}. Under this condition,
these correlators must factorize into the square of more fundamental
objects with disk topology, as was first observed in the case of four
points~\cite{Coronado:2018ypq}. This factorizes into ``octagons'' that
can be bootstrapped at all orders in perturbation
theory~\cite{Coronado:2018cxj} and computed at finite 't~Hooft coupling by leveraging integrability~\cite{Kostov:2019stn, Belitsky:2020qrm, Bajnok:2024epf}.
Although similar simplifications
should occur for the asymptotic five-point function, its factorized
decagon is still only known at two loops in perturbation
theory~\cite{Fleury:2020ykw}.

\section{Conclusions and Outlook}
\label{sec:conclusions-outlook}

We computed five- and six-point loop integrands of half-BPS
single-trace operators with arbitrary R-charge in the planar 't~Hooft
limit. These are presented in the form of compact generating functions,
which produce any desired R-charge correlator by expanding their
ten-dimensional poles in a geometric series in the R-charge
polarization vectors.

At one-loop order, in the weak 't~Hooft coupling limit, we obtained
generating functions for five and six arbitrary R-charged half-BPS
operators. These results can be contrasted with the one-loop
correlators of~\cite{Drukker:2008pi}, where the five-point function of
arbitrary single-trace operators was computed recursively, starting with the
five-point and four-point correlators of the smallest half-BPS
operator $\mathcal{O}_2$ of dimension two, given by a sum of scalar
box conformal integrals. Likewise, our one-loop generating function
can be expressed as the five-point function of $\mathcal{O}_2$ sitting
on 10d simple poles, plus the four-point function of $\mathcal{O}_2$
sitting on 10d double poles. These are poles in the
ten-dimensional distance $X_{ij}^2\equiv x_{ij}^2+y_{ij}^2$ that
combines spacetime and R-charge distances. Furthermore, we find a
similar structure for the six-point generating function, including
higher-order poles and more nested lower-point functions.

At two-loop order, we obtained state-of-the-art results on the
five-point generating function of half-BPS correlators.
These extend
previously known five-point correlators of light~\cite{Bargheer:2022sfd} and
heavier single-trace operators~\cite{Fleury:2020ykw,Bercini:2024pya}.
As in the one-loop
case, our generating function has 10d double poles that are controlled by the
four-point generating function. The numerator of the 10d-simple-pole contribution
is expressed as a linear combination of the lightest
five-point correlator plus other fixed-R-charge correlators with
maximum dimension four at each point. In fact, based on planarity, we
can argue that the numerator of the $\ell$-loop generating function
requires a basis of R-charge correlators with maximum dimension $\sim
2\ell$ at each point.

We also present our two-loop results in a basis of
conformal integrals. This basis of integrals is the same that already
appears in the correlator of the lightest
operator~\cite{Bargheer:2022sfd}. It includes the double-box four-point
integral and the more complicated penta-box and double-box five-point
integrals. When passing from the correlator of lightest operators to
the generating function, the coefficients of these
integrals are replaced by rational functions with 10d poles.
In this representation, it suffices to evaluate the basis of conformal
integrals to obtain all R-charge correlators at the integrated level.
However, some of these integrals have only been evaluated in special
kinematics~\cite{Bork:2022vat, Bercini:2024pya}. This includes the
light-cone OPE limit, which we used to compare our results with integrability-based predictions
of OPE structure constants~\cite{Basso:2015zoa, Bercini:2022gvs},
serving as a very non-trivial check of our generating function.

Ultimately, sharper control of conformal integrals in general
kinematics is essential for studying higher-point correlators with
arbitrary R-charge, particularly in the
multi-Regge~\cite{Costa:2023wfz} and multi-light-cone
limits~\cite{Alday:2013cwa,Alday:2016mxe,Bercini:2020msp,Bercini:2021jti}.
Furthermore, via light-ray transforms, these correlators of local
operators can be recast as correlators of detectors, such as the
three-point energy correlator
(EEEC)~\cite{Chen:2022jhb,Chang:2022ryc,Moult:2025nhu}. Since our
generating function gives access to arbitrary R-charges,  it offers a natural way to extend EEEC analyses to heavy states produced by operators with large or unrestricted R-charge~\cite{Chicherin:2023gxt}.

\paragraph{Higher Points and Loops.}

In principle, our correlator computation could be pushed to higher
points and/or higher loops. However, the computation quickly becomes
very demanding, mainly due to the following three bottlenecks: The rapidly
increasing number of contributing graphs, the difficulty of
performing algebra with many Grassmann numbers efficently on a
computer, and the increasing size of the ansätze for correlator
compontents, which make the corresponding dense linear systems hard or
impossible to solve. Still, with enough determination and computing
power, it might be possible to obtain the six-point two-loop or
five-point three-loop generating functions.
Another route to higher
points and/or loops is to consider ten-dimensional null limits, which
we comment on further below.

\paragraph{Higher-Order Poles and Nesting.}

Our five- and six-point loop integrands present a nested structure of
lower-point correlators sitting at double- and higher-order poles.  In
fact, this pattern is already present in the free-theory  correlator
of scalars and, at this level, it can be understood by analyzing the
graphs that represent the Wick contractions with effective 10d
propagators. This pole structure is inherited by the SDYM
supercorrelator, since the latter is obtained by dressing the free
graphs with $\Delta$-propagators that carry the superspace coordinate
dependence. And likewise for our loop integrands, which are obtained
as specific Grassmann components of the supercorrelator.%
\footnote{One might be worried that the ten-dimensional pole structure
of the free-theory correlators gets modified when passing to
the loop integrands $G_{n,\ell}$, due to the $y_i\to0$ projections of
the Lagrangian points, which reduce 10d~poles to 4d~poles. But in fact
the loop integrand $G_{n,\ell}$ inherits the relevant higher-order
pole structure from its free-theory counterpart $G_{n,0}$.}
However, supersymmetry does play a role on reducing the highest-order pole of
the skeleton graphs. More specifically, for the $n$-point loop integrand,
we expect the highest order to be $n-3$ for a 10d pole, and it should be
controlled by a four-point subintegrand.
Three-point subintegrands are excluded because they vanish due to supersymmetry.

\paragraph{Ten-Dimensional Symmetry and SUSY Ward Identities.}

The ten-dimensional symmetry is a $\grp{SO}(10,2)$ conformal symmetry
of the four-point half-BPS generating function at the loop-integrand
level, and it is conjectured to hold at all loop orders. It combines
the spacetime conformal symmetry $\grp{SO}(4,2)$ and R-symmetry
$\grp{SO}(6)$.%
\footnote{This is a symmetry only at the integrand level. It is explicitly broken at the integrated level,
when the Lagrangian insertions are integrated over 4d spacetime.}
Whether this 10d conformal symmetry generalizes to higher-point
correlators (at the integrand level) still remains an open question.
To address this question, it is important to highlight the role played
by supersymmetry. In the well-established four-point case, the 10d
symmetry is a property of the reduced correlator, which is defined by
stripping off an overall factor whose presence is a consequence of
superconformal Ward identities, also known as  partial
non-renormalization theorem~\cite{Eden:2000bk}. In contrast, the
consequences of supersymmetry for higher-point correlators are not yet
fully understood, nor is the generalization of the `reduced'
correlator(s). Nevertheless, based on a counting of superconformal
invariants at higher points,%
\footnote{See \eg~\cite{Chicherin:2015bza,Heslop:2022xgp} for a
construction of higher-point superinvariants.}
we could speculate on a possible
decomposition of the form: $G_{n,\ell}=\sum_{a}R_{a}(x,y)\times
\mathcal{H}_{a,\ell}(X_{ij}^2)$. Here, $R_a$ would be a basis of ``susy
factors'' that contain zeros where $x$ and $y$ kinematics are aligned to
enhanced the supersymmetry of the correlator. The $\mathcal{H}_a$
would be the corresponding basis of `reduced' correlators subject to 10d conformal
symmetry. If such a decomposition exists, it will likely mirror a
natural decomposition of the dual closed-string amplitudes into
different (super)polarization terms. We hope that the five-point generating function worked out
in this paper as well as the recent supergravity result
of~\cite{Fernandes:2025eqe} will help to further elucidate this susy/10d
structure in the future.%
\footnote{Another notable holographic five-point correlator with
hidden symmetry can be found in the $\grp{AdS}_5\times\grp{S}^{3}$
model of~\cite{Huang:2024dxr}. Also see~\cite{Du:2024xbd} for its weak
coupling counterpart.}

\paragraph{Ten-Dimensional Null Limits.}
Our generating functions simplify in ten-dimensional null limits.
$X_{ij}^2\to0$.
A particularly interesting combination of such limits is the
ten-dimensional null polygonal limit $X_{i,i+1}^2\to0$, $i=1,\dots,n$.
This is a generalization of the four-dimensional null-polygonal limit of correlators
of the lightest operator $\mathcal{O}_2$. The 4d case results in a
triality between 4d null $\mathcal{O}_2$-correlators, null polygonal
Wilson loops and massless scattering amplitudes~\cite{Alday:2007hr,
Bern:2008ap, Drummond:2008aq, Alday:2010zy, Caron-Huot:2010ryg,
Mason:2010yk}, with the latter relation
explained by a fermionic T-duality~\cite{Berkovits:2008ic,
Beisert:2008iq}. Our generating functions ($\mathcal{O}$-correlators)
now carry 10d pole singularities at the locus of the ten-dimensional
null limits,
and their 10d null polygon limit is dominated by
$\mathcal{O}_k$-correlators with large R-charge~$k$. This
correspondence was studied in~\cite{Caron-Huot:2021usw} for the
four-point generating function, relating its 10d null limit to the
large-R-charge ``octagon'' correlator~\cite{Coronado:2018cxj}.
In that work, the 10d null correlator was conjectured to be
dual to a massive scattering amplitude on the Coulomb branch of the
theory for a specific choice of scalar vacuum expectation value. The same limit can be
performed straightforwardly on our five and six-point correlators.
These results, as well as a generalization to higher-point 10d-null
polygons, will be the subject of a separate publication.

\subsection*{Acknowledgments}

We thank Francesco Aprile, Simon Caron-Huot, Bruno Fernandes, Paul Heslop, Zhongjie
Huang, Ian Moult and Ellis Ye Yuan for useful discussions.
The work of T.\,B., A.\,B., and C.\,B. was funded by the Deutsche
Forschungsgemeinschaft (DFG, German Research Foundation) Grant No.
460391856.
T.\,B., A.\,B., and C.\,B. acknowledge support from DESY
(Hamburg, Germany), a member of the Helmholtz Association HGF,
and by the Deutsche
Forschungsgemeinschaft (DFG, German Research Foundation) under
Germany's Excellence Strategy -- EXC 2121 ``Quantum Universe'' --
390833306.
A.\,B. is further supported by the Studienstiftung des Deutschen Volkes. The work of C.\,B. was partially supported by the European Research
Council (ERC) under the European Union's Horizon 2020 research and
innovation program - 60 - (grant agreement No.~865075) EXACTC.
The work of F.C. is supported in part by the Simons Foundation
grant 994306 (Simons Collaboration on Confinement and QCD Strings), as well the NCCR
SwissMAP that is also funded by the Swiss National Science Foundation.

\appendix

\section{Data Files}
\label{sec:datafiles}

The following ancillary files are provided at \url{https://arxiv.org/src/2509.14332/anc}:
\begin{description}
\item[\ancfile{G52CorrelatorBasis.m}:]
The file constructs the parity-even two-loop
five-point generating function $G_{5,2}$ from the basis of (modified) R-charge
correlators given in the first line of~\eqref{eq:G52splitcharges}
The R-charge correlators are given as explicit expressions of $d_{ij}$
and $x_{ij}^2$, which are fully symmetrized over the
permutation group that respects the charge configuration
of the respective R-charge correlator.
\item[\ancfile{G52IntegralBasis.m}:]
The file produces the expression for the two-loop five-point
generating function $G_{5,2}$, given in terms of a basis of seven rational
functions $\mathcal{I}_k$ that are integrands of conformally invariant integrals.
It contains explicit expressions for
$\mathcal{R}_{1234}$~\eqref{eq:R1234first},
$\mathcal{R}_{1234,5}$~\eqref{eq:R1234c5},
$C^{(5)}_{1234,5}$~\eqref{eq:defC5},
the polynomials $f_k$~\eqref{eq:fk}
and $P_{i,j}$~\eqref{eq:mathF}--\eqref{eq:p7polynomial},
the basis of conformal integrals $\mathcal{I}_k$~\eqref{eq:5pt2LoopInts},
the four-point and five-point generating functions at one and two loops
$G_{4,1}$~\eqref{eq:G41},
$G_{5,1}$~\eqref{eq:G51-Rbasis},
$G_{4,2}$~\eqref{eq:G42intbasis},
$G_{5,2}$~\eqref{eq:G52},
and the full five-point correlator up to two loops $G_5^{\mathrm{SYM}}$~\eqref{eq:GnIntegrated}.
\item[\ancfile{G60.m}:]
The file generates the expression for the leading-order six-point generating function $G_{6,0}$.
It is computed from a minimal expression $G_{6,0}\suprm{seed}$ by summing over $\grp{S}_6$ permutations.
\item[\ancfile{G70.m}:]
The file generates the expression for the leading-order seven-point generating function $G_{7,0}$.
It is computed from a minimal expression $G_{7,0}\suprm{seed}$ by summing over $\grp{S}_7$ permutations.
\item[\ancfile{Psums.m}:]
The file contains the OPE sum rules $P_{\tau_1,\tau_2;b}^{J_1,J_2;\ell}$~\eqref{combData}.
Some of these are displayed in \tabref{tabData} in the paper.
\item[\ancfile{checks.m}:]
The file verifies some properties of the generating functions,
and reproduces the OPE expansion.
In particular, it verifies equation~\eqref{example1}.
The file also verifies the consistency of
\ancfile{G52CorrelatorBasis.m} and \ancfile{G52IntegralBasis.m}, \ie
the equality of the two expressions~\eqref{eq:G52splitcharges}
and~\eqref{eq:G52} for the two-loop generating function $G_{5,2}$.
\item[\ancfile{intOPE.m}:]
The file is adapted from the original \filename{intsOPE.wl}, which is part
of~\cite{Bercini:2024pya}.
It only differs from the original by re-labeling
of symbols and different normalizations, as well as the expansions
of the one-loop box integrals.
The file contains the expansions of all integrals that appear in
the OPE limit described in \secref{sec:OPELimit}. See the code
at the end of the file \ancfile{checks.m}, which reproduces the OPE
expansion.
\end{description}
%

\section{Details on Fixed-Charge Correlator Construction}
\label{app:fixedChargeAnsatz}

We demonstrate the explicit construction of the fixed-charge
ansatz~\eqref{eq:ansatz} by considering integrand of the
five-point correlator $\brk[a]*{33222}$ at one and two loops. This
component is $\grp{S}_2\times\grp{S}_3$ symmetric, which leaves seven different
$y$-structures that needs to be summed over in~\eqref{eq:ansatz}, up to
those permutations (see \tabref{tab:33222}).

The maximal number of independent terms of each polynomial
$P_\mathbf{a}^\ell(x_{ij}^2)$ does not depend on the specific charges of
the operators, but on the number of points~$n$ and the loop
order~$\ell$. At one-loop order, $P_{\mathbf{a}}^1$ has maximally 15
different terms. At two loops, the number of
terms can be reduced by identifying terms that are related under
the permutations of the two internal Lagrangian insertions, which
leaves 442 different terms for each polynomial $P_{\mathbf{a}}^2$.
In general, the $\ell$-loop polynomials exhibit an
$\grp{S}_\ell$ symmetry in the internal points, which reflects the permutation
symmetry of the integration points.

This number can be lowered further by using the seven-point conformal
Gram identity
\begin{equation}
	0=\det \{x_{ij}^2\}_{i,j=1,\dots,7}\,,
\end{equation}
which relates products of the squared
distances $x_{ij}^2$. It can be used to exclude the integrands of
certain conformal integrals from the $\brk[a]{33222}_2$ ansatz, see~\eqref{eq:GramZero}. We
choose to remove the integrand~\eqref{eq:newpenta2loop} that has
a factor $x_{67}^2$ in the numerator, or equivalently to
remove terms in the polynomials $P_{\mathbf{a}}^2$ that have factors
$(x_{67}^2)^2$. This reduces the number of different terms to~420.

Each polynomial $P_\mathbf{a}^\ell(x_{ij}^2)$ does not exhibit the full
permutation symmetry of the correlator, but rather a subgroup thereof
that is leaving the respective $y$-structure invariant. By demanding
this invariance for each polynomial, the number of independent
coefficients can be further reduced.

Finally, certain terms in the polynomials can be excluded by considering
the Euclidean OPE limit of two operators. At leading order,
it holds for $x_i\rightarrow x_j$ that
\begin{align}
	\mathcal{O}_{k_i}(x_i)\mathcal{O}_{k_j}(x_j)\,\longrightarrow\, & \left(d_{ij}\right)^{\min\{k_i,k_j\}} \mathcal{O}_{|k_i-k_j|}(x_j)\nonumber\\
	&+\left(\text{terms in $(d_{ij})^k$ with $k<\min\{k_i,k_j\}$}\right)\,,
\end{align}
thus reducing an $n$-point correlator to an $(n-1)$-point correlator. Applying
this relation to two operators with identical charge yields the
identity operator at leading order. For five points, this implies that
we are left with a three-point function that does not obtain loop
corrections. Hence, terms in the ansatz contributing to that limit
cannot appear in the ultimate expression of the loop integrand.
Specifically, for $y$-structures that contain two operators $i$ and $j$ whose
polarization vectors only contract with each other (this can happen
\eg~for the first two
lines in \tabref{tab:33222}), one can infer that every term in the
corresponding polynomials must contain a factor $x_{ij}^2$. This
further reduces the amount of independent coefficients in these polynomials.

\begin{table}
	\centering
	\begin{tabular}{c c c c}
    	\toprule
		$y$-structure & \# of coeff.~in $P_\mathbf{a}^1$ & \# of coeff.~in $P_\mathbf{a}^2$  \\
		\midrule
		$d_{12}^3 d_{34} d_{35} d_{45}$ & 1 & 22 \\
		$d_{12}^2 d_{13} d_{23} d_{45}^2$ & 2 & 48 \\
		$d_{13} d_{14}d_{15} d_{23} d_{24} d_{25}$& 3 & 59 \\
		$d_{12}^2 d_{13} d_{24} d_{35} d_{45}$ & 9 & 228 \\
		$d_{12} d_{13} d_{14} d_{23} d_{25} d_{45}$ & 9 & 228 \\
		$d_{13}^2 d_{14} d_{24} d_{25}^2$ & 9 & 228 \\
		$d_{12} d_{13}^2 d_{24} d_{25} d_{45}$ & 9 & 231  \\
		\midrule
		\# of coeff.~in the ansatz & 42 & 1044\\
		\bottomrule
	\end{tabular}
	\caption{Numbers of coefficients that enter the one- and two-loop
    integrands of the $\brk[a]*{33222}$ component via the polynomials
    $P_\mathbf{a}^\ell$. The number of coefficients is reduced by the
    respective permutation symmetry of the corresponding
    $y$-structure. Also, the leading OPE behavior of disconnected
    $y$-structures is used to lower the number of independent
    coefficients.}
	\label{tab:33222}
\end{table}

After taking all these reductions into account,
the remaining number of coefficients
for the different $y$-structures of the $\brk[a]*{33222}$ ansatz at
one and two loops is listed in \tabref{tab:33222}.

At one loop, the total number of coefficients that needs to be fixed
via solving a linear equation system by matching against the numerical
twistor answer is~42. In fact, for a large class
of one-loop integrands this number of coefficients
stays reasonably small, such that fixing those coefficients is
possible and even sufficiently fast. At two loops, however, that
number of independent coefficient is much larger, as can already be
seen from this example. Ansätze of specific R-charge components with charges that are
invariant under fewer permutations or even exhibit no permutation symmetry are
much less constrained, so that the number of independent coefficients
is too large to be solved in a linear equation system. For example,
the ansatz of the two-loop integrand of the component $\brk[a]*{65432}_2$
has 28980 coefficients.

One way  to still obtain a final expression for the components is to
sequentially probe those ansätze by choosing different polarizations
such that some $d_{ij}=0$. This effectively sets a part of the
$y$-structures to zero, thus reducing the amount of coefficients. In
that way, the coefficients can be fixed step by step, such that the
equation systems can be solved numerically. This works very
effectively for specific R-charge components of $G_{5,2}$.

\section{Constraints on Poles and Numerators from Graphs}
\label{app:graphical-analysis}

Planar five-point single-particle graphs have maximally
three propagators connecting the same pair of operators. In fact,
there is exactly one such graph, which is displayed in~\eqref{eq:maxPoleGraph}.%
\footnote{This graph produces the first term in the tree-level
generating function~\eqref{eq:G50}.}
As explained there, this graph receives no loop corrections. Loop
corrections (correlators with Lagrangian insertions) are only
non-trivial if the graph of $D_{ij}$ propagators has at least one
square (or larger) face. At five points, such graphs have
at most two $D_{ij}$ propagators that connect the same pair of points.
Hence in the loop generating function $G_{5,\ell}$, $\ell>0$, the
poles at $D_{ij}\to\infty$ ($w_{ij}\to0$) have maximally degree two.

By a similar analysis, we can understand the pole as well as numerator
structures of the
coefficients $f_i$ that accompany the various integrands
$\mathcal{I}_i$ of conformal integrals in the
decomposition~\eqref{eq:G52}.
For example, consider the coefficient $f_7$ that multiplies the
integrand $\mathcal{I}_7$~\eqref{eq:5pt2LoopInts}. At
leading order in $1/\Nc$, the integrand $\mathcal{I}_7$ can only be
embedded into an ambient graph of $D_{ij}$ propagators if that graph
has at least one face that involves all five points, and therefore is
at least pentagonal. An example graph is:
\begin{equation}
\includegraphics[align=c]{FigF52I7MaxGraph}
\,.
\end{equation}
One can easily convince oneself that such graphs do not admit multiple
identical $D_{ij}$ propagators. Hence the series $f_7$ must become
finite after pulling out the product $1/\prod_{1\leq{i<j}\leq5}w_{ij}$ of single
poles. Moreover, we can see that all planar graphs that admit an
embedding of $\mathcal{I}_7$ have at most seven propagators $D_{ij}$,
hence all terms in the finite numerator of $f_7$ must be at least
cubic in $w_{ij}$ (after converting all $d_{ij}$ and $D_{ij}$ to
$w_{ij}$). In contrast, the two-loop ladder integrand
$\mathcal{I}_5$ only requires a square face for a planar embedding,
for example:
\begin{equation}
\includegraphics[align=c]{FigF52I5MaxGraphNoDouble}
\qquad
\includegraphics[align=c]{FigF52I5MaxGraphWithDouble}
\end{equation}
One can see that such graphs admit at most one pair of identical
propagators $D_{ij}$, and have at most eight propagators $D_{ij}$ in
total. Hence we have to pull out at least one extra factor
$1/w_{ij}$ (besides the overall $1/\prod_{1\leq{i<j}\leq5}w_{ij}$) to
obtain a finite numerator for $f_5$, and all terms in the numerator must be at
least of quadratic order in~$w_{ij}$.
By this type of analysis, we can see that $\mathcal{I}_1$
admits triple poles $1/w_{ij}^3$,
$\mathcal{I}_{4}$ and $\mathcal{I}_5$ admit double
poles, and $\mathcal{I}_{2}$, $\mathcal{I}_3$, $\mathcal{I}_6$,
and $\mathcal{I}_7$ only admit simple poles. The maximal number of
concurrent poles (including double/triple poles) is
eight for $f_1$ and $f_5$,
six for $f_2$, and
seven for $f_3$, $f_4$, $f_6$, and $f_7$ (seven concurrent poles for
$f_4$ can only occur in double-pole terms).
Therefore, after
factoring out the ten-fold product $\prod_{i<j=1}^51/w_{ij}$,
the remaining numerators of the
various terms in each~$f_i$ must be of the following orders in~$w_{ij}$:
$f_2,f_4\sim\order{w_{ij}^4}$,
$f_3,f_6,f_7\sim\order{w_{ij}^3}$
and $f_1,f_5\sim\order{w_{ij}^2}$.%
\footnote{The integrand $\mathcal{I}_1$ behaves differently than
expected, see the discussion around~\eqref{eq:I1graphs} below: It
comes with only single-pole contributions, and has at most six
concurrent poles, thus the numerators of $f_1$ are at least $\order{w_{ij}^4}$.}

We can verify the pole structure of the various coefficients $f_k$
anticipated above. For example, consider $f_6$ and $f_7$. Even though
not apparent from the expressions~\eqref{eq:P61andP62},
after expanding $V^{ij}_{kl}$ and $d_{ij}$ in terms of $w_{ij}$ and
summing over permutations, all terms in the polynomials~$P_6$ are at
least of degree four in $w_{ij}$. The same is true for the
polynomials~$P_7$. However, some of the $w^4$ terms contain a factor
$w_{ij}^2$, hence only three of the $1/w_{ij}$ factors of the overall
prefactor in $f_6$ and $f_7$ get canceled. In other words, both
$\mathcal{I}_6$ and $\mathcal{I}_7$ get multiplied by at most seven
factors of $1/w_{ij}$ (all of them distinct), as expected.

For $f_3$, one can see that all terms in the polynomials $P_3$
in~\eqref{eq:P31and2} are manifestly of degree at least three in
$w_{ij}$. In fact, expanding everything in terms of $w_{ij}$ and
summing over permutations, all terms in $f_3$ are at least of order
five in $w_{ij}$. However, some of the $w^5$ terms contain two factors
of $w_{ij}^2$, hence $\mathcal{I}_3$ is accompanied by at most
seven factors $1/w_{ij}$, as anticipated.

Finally, we find that $P_1$ cancels four or more factors
$1/w_{ij}$ of the overall prefactor, hence $\mathcal{I}_1$ and
$\mathcal{I}_2$ are multiplied by at most six pole factors $1/w_{ij}$.
For $f_2$, this is expected based on the graphical analysis above. Considering
$f_1$, the graphical analysis indicated that $\mathcal{I}_1$ could be
accompanied by up to eight pole factors $w_{ij}$. However, the graphs
with seven or eight $D_{ij}$ propagators apparently give a zero
contribution. In particular, no double (or triple) poles appear.
At the integrated level, this is expected, as such terms would
constitute loop corrections to three-point functions of BPS operators,
which would be in conflict with supersymmetry.
The graphs that contribute to $f_1$ with a maximal number of $D_{ij}$
propagators are (up to permutations):
\begin{equation}
\includegraphics[align=t,vshift=1.6cm]{FigF52I1Graph1}
\qquad
\includegraphics[align=t,vshift=1.6cm]{FigF52I1Graph2}
\qquad
\includegraphics[align=t,vshift=1.6cm]{FigF52I1Graph3}
\,.
\label{eq:I1graphs}
\end{equation}
One can see that the integrand $\mathcal{I}_1$ only gives a non-zero
contribution when it is embedded in a (degenerate) hexagon formed by
$D_{ij}$ propagators.

This knowledge of the ten-dimensional pole structure can be employed to
determine the generating functions $G_{n,\ell}$ from a finite list of
fixed-charge correlators. For example, we know that the full two-loop
five-point generating function $G_{5,2}$ has simple and double poles.
We can remove all these poles by pulling out an appropriate factor
$\prod_{ij}1/w_{ij}^{n_{ij}}$, with $n_{ij}\in\set{1,2}$.
The full generating function $G_{5,2}$ expands to an
infinite tower of fixed-charge correlators. Removing the
prefactor renders this series finite, which means that it can be
reconstructed from a finite set of fixed-charge correlators.
This is exactly the strategy that we used to find the
expressions~\eqref{eq:G52splitcharges} and~\eqref{eq:G52}.

Since the conformal integrals are linearly independent,
we can apply the same strategy for each of the
coefficients $f_k$ separately. Some of them truncate after a prefactor
of single poles, others only truncate with a prefactor of double poles.
By the analysis above,
we can understand the prefactor that needs to be pulled out of each
coefficient $f_k$ to render its series finite.

\section{Five-Point Conformal Block}
\label{appConfBlock}

The five-point conformal block in general kinematics is a complicated
object that is not known in closed form~\cite{Goncalves:2019znr}. However, when
considering the light-cone limit ($u_1,u_4 \to 0$,
see~\eqref{eq:ufive} for the definition of the cross ratios), it is possible to
write this object as the following integral representation~\cite{Bercini:2020msp}
\begin{align}
    \mathcal{F}_{J_1,J_2,\ell}(u_i) &= \frac{u_4^{\brk{\tau_1-k_1-k_2}/{2}}u_1^{\brk{\tau_2-k_3-k_4}/{2}}(1-u_5)^\ell\,u_2^{\brk{\tau_1+k_3-k_4-k_5}/{2}+\ell}u_3^{\brk{\tau_2+k_2-k_1-k_5}/{2}+\ell}}{x_{13}^{k_1+k_2+k_3-k_4-k_5}x_{14}^{k_4+k_5+k_1-k_2-k_3}x_{24}^{k_2+k_3+k_4-k_5-k_1}x_{25}^{k_5+k_1+k_2-k_3-k_4}x_{35}^{k_3+k_4+k_5-k_1-k_2}}  \nonumber \\
    &\hspace{-5em}\times\frac{\Gamma\left( 2J_1+\tau_1\right)}{\Gamma\left(J_1+\frac{\tau_1}{2}+\frac{k_1}{2}-\frac{k_2}{2}\right)\Gamma\left(J_1+\frac{\tau_1}{2}-\frac{k_1}{2}+\frac{k_2}{2}\right)}\frac{\Gamma\left( 2J_2+\tau_2\right)}{\Gamma\left(J_2+\frac{\tau_2}{2}+\frac{k_3}{2}-\frac{k_4}{2}\right)\Gamma\left(J_2+\frac{\tau_2}{2}-\frac{k_3}{2}+\frac{k_4}{2}\right)} \nonumber \\
    &\hspace{-5em}\times\!\int_{0}^1\hspace{-0.4em}\int_{0}^{1}\mspace{-6mu}dt_1 dt_2\, t_1^{J_1-1+\brk{k_1-k_2+\tau_1}/{2}}(1\!-\mspace{-1mu}t_1)_{\phantom{1}}^{J_1-1+\brk{k_2-k_1+\tau_1}/{2}}t_2^{J_2-1+\brk{k_3-k_4+\tau_2}/{2}}(1\!-\mspace{-1mu}t_2)_{\phantom{2}}^{J_2-1+\brk{k_4-k_3+\tau_2}/{2}}\nonumber \\
    &\hspace{-5em}\times \frac{(1-u_3-t_2(1-u_2-u_3+u_5u_2u_3))^{J_1-\ell}(1-u_2-t_1(1-u_2-u_3+u_5u_2u_3))^{J_2-\ell}}{\brk!{1-t_1(1-u_3)-t_2(1-u_2)+t_1t_2(1-u_2-u_3+u_5u_2u_3)}^{J_1+J_2+\brk{\tau_1+\tau_2-k_5}/{2}}}\,.\label{5ptBlock}
\end{align}
For any values of spins $J_i$ and spin-polarization $\ell$, it is trivial
to further expand the block in the OPE limit ($u_2,u_3,u_5 \to 1$) up
to arbitrary orders. In that limit, it is also equally easy to perform the
integrations over the auxiliary variables $t_i$ to obtain the explicit
dependence of the cross-ratio shown in~\eqref{example2}.

By changing to the cross-ratios $v_i$ defined in~\eqref{eq:vcross}, we
make the OPE expansion very transparent. At leading order in the OPE
series, each average $P_{\tau_1,\tau_2;b}^{J_1,J_2;\ell}$ of OPE
coefficients~\eqref{combData} is multiplied by a different monomial
$v_1^{J_1} v_2^{J_2} v_3^{\ell}$, as seen
in~\eqref{example2}, making it easier to perform the data extraction
procedure.

\section{Five-Point Conformal Integrals}
\label{appIntegralsDone}

The evaluation of the five-point conformal integrals in particular
kinematical limits (such as the OPE limit of~\eqref{eq:DLC}) was done
in~\cite{Bercini:2024pya}. The relation between the notation used here
for the integrals~\eqref{eq:5pt2LoopInts} and~\eqref{IntegrandToInt}
and the notation used there is
\begin{align}
    \mathbb{I}_{1}^{[1,2,3]} & = \int\hspace{-0.5em}\int \frac{d^4x_6d^4x_7}{\pi^4} \frac{1}{(x_{\col16}^2 x_{\col26}^2 x_{\col36}^2)x_{67}^2(x_{\col17}^2 x_{\col27}^2 x_{\col37}^2)}=  \frac{1}{x_{12}^2 x_{13}^2 x_{23}^2} \mathbb{E}_{123} \,,\\
    \mathbb{I}_{2}^{[1,2,3|4,5]} & = \int\hspace{-0.5em}\int \frac{d^4x_6d^4x_7}{\pi^4} \frac{x_{\col56}^2\,x_{\col47}^2}{(x_{\col16}^2 x_{\col26}^2 x_{\col36}^2 x_{\col46}^2 )x_{67}^2(x_{\col17}^2 x_{\col27}^2 x_{\col37}^2 x_{\col57}^2 )}=  \frac{1}{x_{12}^2 x_{13}^2 x_{23}^2} \mathbb{D}_{45;123} \,,\\
    \mathbb{I}_{3}^{[1,2,3|4,5]} & = \int\hspace{-0.5em}\int \frac{d^4x_6d^4x_7}{\pi^4} \frac{1}{(x_{\col16}^2 x_{\col26}^2 x_{\col36}^2 x_{\col46}^2)(x_{\col17}^2 x_{\col27}^2 x_{\col37}^2 x_{\col57}^2)} = \frac{1}{x_{13}^2 x_{24}^2}\frac{1}{x_{13}^2 x_{25}^2}\Phi^{(1)}_{12;34}\Phi^{(1)}_{12;35} \,,\\
    \mathbb{I}_{4}^{[1,2,3,4]} & =\int\hspace{-0.5em}\int \frac{d^4x_6d^4x_7}{\pi^4} \frac{1}{(x_{\col16}^2 x_{\col26}^2 x_{\col36}^2 x_{\col46}^2) (x_{\col17}^2 x_{\col27}^2 x_{\col37}^2 x_{\col47}^2)} =  \frac{1}{x_{13}^2 x_{24}^2}\frac{1}{x_{13}^2 x_{24}^2}\Phi^{(1)}_{12;34}\Phi^{(1)}_{12;34} \,,\\
    \mathbb{I}_{5}^{[1,2|3,4]} & = \int\hspace{-0.5em}\int \frac{d^4x_6d^4x_7}{\pi^4}\frac{1}{(x_{\col16}^2 x_{\col26}^2 x_{\col36}^2) x_{67}^2(x_{\col17}^2 x_{\col27}^2 x_{\col47}^2) }
 =\frac{1}{x_{12}^2 x_{13}^2 x_{24}^2}\Phi^{(2)}_{12;34} \,,\\
    \mathbb{I}_{6}^{[1,2|3,4|5]} & = \int\hspace{-0.5em}\int \frac{d^4x_6d^4x_7}{\pi^4} \frac{x_{\col56}^2}{(x_{\col16}^2 x_{\col26}^2 x_{\col36}^2 x_{\col46}^2 )x_{67}^2 (x_{\col17}^2 x_{\col27}^2 x_{\col57}^2 ) }=\frac{1}{x_{12}^2 x_{13}^2 x_{24}^2} \mathbb{S}_{5;12;34} \,,\\
    \mathbb{I}_{7}^{[1,2|3,4|5]} & =\int\hspace{-0.5em}\int \frac{d^4x_6d^4x_7}{\pi^4}\frac{1}{(x_{\col36}^2 x_{\col46}^2 x_{\col56}^2 ) x_{67}^2 (x_{\col17}^2 x_{\col27}^2 x_{\col57}^2   ) } = \frac{1}{x_{12}^2 x_{24}^2 x_{35}^2} \mathbb{L}_{5;12;34}\,.
\end{align}
Each of these integrals was evaluated in the OPE limit~\eqref{eq:DLC}. Taking the map~\eqref{eq:vcross} from $u_i$ to $v_i$ into account,
the first few terms of their expansions are:
\begin{align}
    \mathbb{E}_{123} &= 6\zeta_3 \,,\\
    \mathbb{D}_{45;123} &= 6\zeta_3 + \frac{1}{4}v_2^2 + \dots \,,\\
    \Phi^{(1)}_{12;34} &= \frac{1}{3}\left(\frac{2}{3}-\log u_4 -\log u_1 \right)v_1^2v_2^2v_3^2 + \dots \,,\\
    \Phi^{(1)}_{12;35} &= \frac{1}{2}\left(1-\log u_4\right)v_1v_2v_3 - \frac{1}{6}\left(\frac{1}{3}+\log u_4\right)v_1v_2^2v_3 +\dots \,,\\
    \Phi^{(2)}_{12;34} &= 6\zeta_3 +3\zeta_3 v_1 v_2 v_3 +\frac{1}{12}\left(1+24\zeta_3\right)v_1^2v_2^2v_3^2+\dots \,,\\
    \mathbb{S}_{5;12;34} & = 6\zeta_3 +3\zeta_3 v_1 v_2 v_3 + \frac{1}{4}v_2^2 +\frac{1}{4}v_1v_3 + \frac{1}{12}\left(1+24\zeta_3\right)v_1^2v_3^2 +\dots\,,\\
    \mathbb{L}_{5;12;34} & = (4-2\log u_4 - 2\log u_1 +\log u_4 \log u_1) + \Big(2 -\frac{1}{2}\log u_4 -\frac{3}{2}\log u_1 \\
    &+\frac{1}{2}\log u_4 \log u_1\Big) v_1 + \Big(\frac{43}{36} -\frac{2}{9}\log u_4 -\frac{7}{6}\log u_1 + \frac{1}{3}\log u_4 \log u_1\Big)v_1^2 \nonumber \\
    & -\Big(2 -\frac{1}{2}\log u_4 -\frac{3}{2}\log u_1 + \frac{1}{2}\log u_4 \log u_1\Big)v_2 - \Big(\frac{1}{2}-\log u_4 \nonumber \\
    & +\frac{1}{4}\log u_4 \log u_1\Big)v_1v_2 - \Big(\frac{1}{8} +\frac{1}{18}\log u_4 -\frac{3}{4}\log u_1 +\frac{1}{6}\log u_4 \log u_1\Big)v_1^2v_2\nonumber \\
    & + \Big(\frac{29}{36} -\frac{1}{3}\log u_4 -\frac{5}{18}\log u_1 + \frac{1}{6}\log u_4 \log u_1\Big)v_2^2 - \Big(\frac{31}{72}-\frac{1}{12}\log u_4 \nonumber \\
    & - \frac{2}{9}\log u_1 +\frac{1}{12}\log u_4 \log u_1\Big)v_1v_2^2 - \Big( \frac{175}{648} - \frac{1}{27}\log u_4 - \frac{19}{108}\log u_1 \nonumber \\
    & +\frac{1}{18}\log u_4 \log u_1 \Big)v_1^2v_2^2 - \frac{1}{4}\Big(\log u_4 +\log u_1 -\log u_4 \log u_1\Big)v_1v_2v_3\nonumber \\
    & - \Big(\frac{1}{18} -\frac{1}{36}\log u_4 +\frac{1}{6}\log u_1 - \frac{1}{12}\log u_4 \log u_1\Big)v_1^2v_2v_3 - \Big(\frac{1}{18}-\frac{1}{12}\log u_4 \nonumber\\
    & +\frac{1}{6}\log u_1 -\frac{1}{12}\log u_4 \log u_1 \Big)v_1^2v_2v_3 - \Big(\frac{1}{18}-\frac{1}{12}\log u_4 - \frac{5}{18}\log u_1 \nonumber \\
    & +\frac{1}{6}\log u_4 \log u_1 \Big)v_1v_2^2v_3 + \Big(\frac{31}{324} -\frac{5}{108}\log u_4 +\frac{4}{27}\log u_1 - \frac{1}{18}\log u_4 \log u_1\Big)v_1^2v_2^2v_3 \nonumber \\
    & -\Big(\frac{29}{324} +\frac{2}{27}\log u_4 +\frac{2}{27}\log u_1 - \frac{1}{9}\log u_4 \log u_1\Big)v_1^2v_2^2v_3^2 +\dots \,.\nonumber
\end{align}
where the ellipses in the expansions stand for terms that have higher
powers of the cross-ratios.

The sub-leading terms in powers of~$u_1$
and~$u_4$ have not been computed, whereas the
expansion in $v_1$, $v_2$, and $v_3$ can be computed to arbitrarily high
orders. Note that the OPE limit~\eqref{eq:DLC} and the permutation symmetry of the
integrals do not commute. Taking this limit breaks the
permutation symmetry, and forces us to compute conformal integrals
with permuted indices independently. These results are in the
\mathematica file of~\cite{Bercini:2024pya}.
We also attach a version of that file, which is adapted to our notation in \ancfile{intOPE.m}.

The only integral we have not commented on yet is $\mathcal{I}_0$,
which is linearly dependent on the others via the Gram identity~\eqref{eq:GramZero}. But it turns out that this integral can be explicitly computed in terms of box integrals, for any value of the cross-ratios. We start by stripping out a simple kinematical factor
\begin{equation}
\mathbb{I}_0(z_i,\bar{z}_i)=x_{13}^2x_{14}^2x_{24}^2x_{25}^2x_{35}^2 \int \frac{d^4x_6}{\pi^2} \int \frac{d^4x_7}{\pi^2}\,\mathcal{I}_0
\,.
\end{equation}

In the cross-ratios $z_i,\bar{z}_i$ of~\eqref{crossratiosZ}, the integral above is given by
\begin{multline}
    \mathbb{I}_0(z_i,\bar{z}_i) = \frac{1}{d}\Bigg(-\sum_{i<j=1}^{5}(z_i+\bar{z}_i)(z_j+\bar{z}_j)F_1(z_i,\bar{z}_i)F_1(z_j,\bar{z}_j)+\\
    +\sum_{i=1}^{5}F_1(z_i,\bar{z}_i)\brk!{a_i F_1(z_i,\bar{z}_i) + b_iF_1(z_{i+1},\bar{z}_{i+1}) +c_i F_1(z_{i+2},\bar{z}_{i+2})}\Bigg)
    \,,
    \label{intI0}
\end{multline}
where the functions $F_1(z_i,\bar{z}_i)$ are the well-known one-loop box integrals of~\eqref{BoxIntegrals}, and the first coefficients are
\begin{align}
    a_1 &= -4z_1\bar{z}_1 \,,\\
    b_1 &= 2-2(1-z_{2})(1-\bar{z}_{2})-2(1-z_{1})(1-\bar{z}_{1})(1+z_3\bar{z}_3)\nonumber \,,\\
    c_1 & =2(1-z_2)(1-\bar{z}_2)\Big(1-z_5\bar{z}_5-z_4\bar{z}_4+(1-z_4)(1-\bar{z}_4)\frac{z_1\bar{z}_1}{2}+(1-z_5)(1-\bar{z}_5)\frac{z_3\bar{z}_3}{2}\Big) \,.\nonumber
\end{align}
The other coefficients $a_i$, $b_i$, $c_i$ are obtained via cyclic permutations, \ie
$a_{i+1}=a_i|_{z_i,\bar{z}_i\to z_{i+1},\bar{z}_{i+1}}$. Finally, the denominator~$d$ is given by
\begin{multline}
    d = \sum_{i=1}^{5}\bigg((1-z_i)(1-\bar{z}_i)(1-(1-z_{i+2})(1-\bar{z}_{i+2}))-\frac{z_i\bar{z}_i z_{i+1}\bar{z}_{i+1}}{5}(1-z_{i+1})(1-\bar{z}_{i+3})+\mathord{}\\
    -\frac{1}{5}-\frac{z_i\bar{z}_i}{2}((1-z_{i-2})(1-\bar{z}_{i-2})+(1-z_{i+2})(1-\bar{z}_{i+2})\bigg)\,.
\end{multline}
One can use this result as a consistency check by comparing this
expression with the linear combination of conformal integrals arising
from the Gram determinant~\eqref{eq:GramZero}.

\section{CFT Data Extraction}
\label{appOPE}

The correlation function described in \figref{fig5ptCorrelators} is
given by the following fixed-weight correlator:
\begin{equation}
\hat{G}_{\tau_1^0,\tau_2^0}=
x_{12}^4
x_{34}^4
x_{23}^{2b}
x_{15}^{2\brk{\tau_1^0-b}}
x_{45}^{2\brk{\tau_2^0-b}}
\eval*{
\brk*{
\prod_{i=1}^{4}\frac{1}{2!}\frac{\partial^2}{\partial t_i^2}
}
\brk*{
\prod_{i=1}^{5}\frac{1}{k_i!}\frac{\partial^{k_i}}{\partial r_i^{k_i}}
}
\,
G_5\suprm{SYM}(x_i,r_iy_i)
}_{t_i,r_i=0}
\,,
\end{equation}
where the polarizations $y_i$ are given
in~\eqref{eq:OPEPolarizations}, and we have rescaled them with $r_i$
in order to extract the fixed-charge operators as
in~\eqref{eq:Rweightcomponent}, with weights $k_i$ given
in~\eqref{eq:kpolarized}. The derivatives with respect to $t_i$
extract the specific operator of \figref{fig5ptCorrelators}, as
described in~\eqref{eq:Rpolarization}. The overall product of
$x_{ij}^2$ ensures that $\hat{G}$ is a function of conformally
invariant cross ratios only.

Specializing to $\tau_1^0=\tau_2^0=2$,
the OPE limit~\eqref{eq:DLC} for the two expressions (integrals
and conformal block expansions) of this correlator are given by%
\footnote{See the attached \mathematica file \ancfile{checks.m} for
an explicit derivation of this expansion, starting from the generating
function~\eqref{eq:GnIntegrated}.}
\begin{align}
\frac{\Nc^3}{\tau_1^0\tau_2^0}\,
    \hat{G}_{2,2} &\overset{\text{OPE}}{=} 1 - \frac{1}{2}v_1 - \frac{1}{2}v_2 +\frac{1}{4}v_1v_2 - \frac{3}{4}v_1v_2v_3+\frac{1}{2}v_1^2v_2v_3 +\frac{1}{2}v_1v_2^2v_3-\frac{1}{4}v_1^2v_2^2v_3 + \label{example1} \\
    &+\textcolor{red1}{v_1^2 v_2^2 v_3^2} \Bigg(0-\frac{g^2}{2}\left(3-\log u_1 -\log u_4\right)+g^4\Big(\frac{227}{12}-7\log u_1+\frac{3}{2}\log^2 u_1\nonumber\\
    & - 7\log u_4+\log u_1 \log u_4+\frac{3}{2}\log^2 u_4 +4\zeta_3\Big)   \Bigg)+\dots \nonumber \\
\frac{\Nc^3}{\tau_1^0\tau_2^0}\,
    \hat{G}_{2,2} & \overset{\text{OPE}}{=} P^{0,0,0}_0 - \frac{P^{0,0,0}_0}{2}(v_1+v_2)+\frac{P^{0,0,0}_0}{4}v_1v_2 -\frac{3P^{0,0,0}_0}{4}v_1v_2v_3 + \label{example2} \\
    &+\Big(P^{2,0,0}_0-\frac{P^{0,0,0}_0}{6}\Big)(v_1^2+v_2^2) +\Big(\frac{P^{0,0,0}_0}{12}-\frac{P^{0,0,0}_0}{2}\Big)v_1v_2(v_1+v_2)+ \nonumber \\
    &+\Big(\frac{P^{0,0,0}_0}{3}+P^{2,0,0}_0\Big)v_1v_2(v_1+v_2)v_3+\Big(\frac{P^{0,0,0}_0}{36}-\frac{P^{2,0,0}_0}{3}+P_0^{2,2,0}\Big)v_1^2v_2^2+\nonumber\\
    &-\Big(\frac{5P^{0,0,0}_0}{36}+\frac{4P^{2,0,0}_0}{3}-P_0^{2,2,1}\Big)v_1^2v_2^2v_3+\textcolor{red1}{v_1^2 v_2^2 v_3^2}\Bigg(\Big(\frac{2P^{2,0,0}_0}{3}+P^{2,2,2}_0-\frac{5P^{0,0,0}_0}{36}\Big)+\nonumber \\
    &+g^2\Big(\frac{2P^{2,0,0}_1}{3}+P^{2,2,2}_1-\frac{5P^{0,0,0}_1}{36}+2\log (u_1 u_4)\left(P^{2,0,0}_0+3P^{2,2,2}_0\right)\Big)+g^4\Big(\frac{2P^{2,0,0}_2}{3}+\nonumber \\
    &+P^{2,2,2}_2-\frac{5P^{0,0,0}_2}{36}+6\log^2 (u_1u_4)\left(P^{2,0,0}_0+3P^{2,2,2}_0\right) - 12 \log u_1 \log u_4 P^{2,0,0}_0 +\nonumber\\
    &-2\log(u_1 u_4)\left(4P^{0,0,0}_0+12P^{2,2,2}_0-P^{2,0,0}_1-3P^{2,2,2}_1\right)\Bigg)+\dots \nonumber
\end{align}
In the OPE limit, both expressions are given by powers of the
cross-ratios $v_i$ and logarithms $\log u_1$ and $\log u_4$. The
ellipses stand for higher orders in the OPE expansion, namely higher
powers of the cross-ratios and higher loop orders. For simplicity, we
only presented the perturbative expansion for the $v_1^2v_2^2v_3^2$ term.

The first expression~\eqref{example1} is obtained by explicitly evaluating the
conformal integrals $\mathbb{I}_i$ defined in~\eqref{IntegrandToInt}, and written in \appref{appIntegralsDone}. The second expression~\eqref{example2} is the equally explicit OPE
expansion~\eqref{schemeOPE} in terms of the structure constants~\eqref{combData}
expanded in perturbation theory:
\begin{equation}
P^{J_1,J_2;\ell}_{\tau_1=2,\tau_2=2;b=1} = \sum_{k=0}^\infty g^{2k} P^{J_1,J_2,\ell}_{k}\,.
\end{equation}
For this twist-two example, there are no degeneracies, so $P$ only stands for the product of three structure constants. In this situation, and inputting the spectrum of anomalous dimensions in the conformal blocks, we are only left with the products $P$ as the fundamental unknowns that we need to solve for.

Comparing both Taylor series, we get a linear system of equations for
the products $P$. As one can easily see at tree-level, by considering
all equations from these two series up to a cutoff in the exponents of
$v_i$, we can solve for all OPE coefficients up to a certain spin. In
the example above, this procedure fixes the twist-two structure
constants of spin-two operators at tree level.

Furthermore, up to two loops we obtain the twist-two $P$-sums with $J_1=J_2=2$:
\begin{align}
P^{2,2;\,0}_{2,2}&=\frac{1}{36}-\frac{2 g^2}{3}+\frac{51 g^4}{4}+\mathcal{O}(g^6)\,,
\nn\\
P^{2,2;\,1}_{2,2}&=\frac{1}{9}-2 g^2+\frac{197 g^4}{6}+\mathcal{O}(g^6)\,,
\nn\\
P^{2,2;\,2}_{2,2}&=\frac{1}{36}-\frac{g^2}{6}+\frac{g^4}{4}+\mathcal{O}(g^6)\,,
\label{eq:P22}
\end{align}
where we omit the label $b=1$ since this is the only possibility for
twist-two operators. Since this case is non-degenerate, we can isolate the two-spin data by diving by the single-spin structure constant:
\beq\label{eq:C2}
\left(C^{J=2}_{\tau=2}\right)^2 =\frac{1}{3}-4 g^2+56 g^4+\mathcal{O}(g^6)
\eeq
This latter is just given by asymptotic hexagons due to our choice that makes its ``bottom'' bridge equal  to 2, preventing wrapping corrections at order $\mathcal{O}(g^4)$. Dividing~\eqref{eq:P22} by~\eqref{eq:C2}, we isolate the two-spin structure constants~\cite{Bianchi:2019jpy}:
\begin{align}
C^{2,2;\,0}_{2,2}&=\frac{1}{12}-g^2+\frac{49 g^4}{4}+\mathcal{O}(g^6) \,,
\nn\\
C^{2,2;\,1}_{2,2}&=\frac{1}{3}-2 g^2+\frac{37 g^4}{2}+\mathcal{O}(g^6) \,,
\nn\\
C^{2,2;\,2}_{2,2}&=\frac{1}{12}+\frac{g^2}{2}-\frac{29 g^4}{4}+\mathcal{O}(g^6) \,.
\end{align}
We repeated these same steps to isolate the non-degenerate twist-three
data in \tabref{tab:twists2and3} and \tabref{tab:twists3and3}.

Performing the same procedure at
two loops for correlators with different weights fixes all OPE sum
rules $P^{J_1,J_2,\ell}_{\tau_1,\tau_2,b}$ to two loops, and results
in the data presented in~\tabref{tabData}.

There is one small caveat when extracting OPE data. In perturbation theory, not only the sum rules $P_{\tau_1,\tau_2;b}^{J_1,J_2;\ell}$ of~\eqref{combData} appear. Similar sums, depending on the anomalous dimensions are also present as coefficients of the logarithms of cross ratios. For example, at two-loop order we have
\begin{equation}
\sum_{n_1=1}^{\text{deg}(\tau_1,J_1)} \,\sum_{n_2=1}^{\text{deg}(\tau_2,J_2)} \gamma(J_1,n_1)\gamma(J_2,n_2)C_{\tau_1,n_1}^{J_1} \, C_{\tau_2,n_2}^{J_2}\, C_{\tau_1,n_1;\tau_2,n_2;b}^{J_1,J_2;\ell}\,.
\label{combData1}
\end{equation}
Both $P$-sums \eqref{combData} and $\gamma$-weighted sums \eqref{combData1} are the fundamental independent variables we should solve for when extracting OPE data. However, instead, we take an equivalent approach where we input the spectral information of $\gamma$ from integrability, and treat the summand $P_{n_1,n_2}\equiv C_{n_1}C_{n_2}C_{n_1,n_2}$ as the fundamental variables. Solving for these latter variables only gives partial solutions, \ie relations among them; nevertheless, these are sufficient to reconstruct the OPE sums $P_{\tau_1,\tau_2;b}^{J_1,J_2;\ell}$ considered in the main text.

\bibliographystyle{nb}
\bibliography{references}

\begin{thebibliography}{10}
\addcontentsline{toc}{section}{\refname}
\providecommand{\href}[2]{#2}
\providecommand{\arxivref}[2]{\href{http://arxiv.org/abs/#1}{#2}}
\providecommand{\doiref}[2]{\href{http://dx.doi.org/#1}{#2}}
\providecommand{\nbbstauthor}[1]{#1}
\providecommand{\nbbstjournal}[1]{\textsf{#1}}
\providecommand{\nbbsttitle}[1]{\textit{#1}}
\providecommand{\nbbsturl}[1]{\texttt{#1}}
\providecommand{\nbbsteprint}[1]{\texttt{#1}}
\providecommand{\nbbststyle}{\raggedright\small\parskip0pt}
\nbbststyle

\bibitem{Eden:2011we}
\nbbstauthor{B.~Eden, P.~Heslop, G.~P.~Korchemsky and E.~Sokatchev},
\nbbsttitle{Hidden symmetry of four-point correlation functions and amplitudes
  in {$\mathcal{N}=\mathord{}$4} {SYM}},
\nbbstjournal{\doiref{10.1016/j.nuclphysb.2012.04.007}{Nucl.~Phys.~B~862,~193~(2012)}},
\nbbsteprint{\arxivref{1108.3557}{arxiv:1108.3557}}.

\bibitem{Drummond:2013nda}
\nbbstauthor{J.~Drummond, C.~Duhr, B.~Eden, P.~Heslop, J.~Pennington and
  V.~A.~Smirnov},
\nbbsttitle{Leading singularities and off-shell conformal integrals},
\nbbstjournal{\doiref{10.1007/JHEP08(2013)133}{JHEP~1308,~133~(2013)}},
\nbbsteprint{\arxivref{1303.6909}{arxiv:1303.6909}}.

\bibitem{Arutyunov:2000py}
\nbbstauthor{G.~Arutyunov and S.~Frolov},
\nbbsttitle{Four point functions of lowest weight CPOs in
  {$\mathcal{N}=\mathord{}$4} {SYM}(4) in supergravity approximation},
\nbbstjournal{\doiref{10.1103/PhysRevD.62.064016}{Phys.~Rev.~D~62,~064016~(2000)}},
\nbbsteprint{\arxivref{hep-th/0002170}{hep-th/0002170}}.

\bibitem{Goncalves:2014ffa}
\nbbstauthor{V.~Gon\c{c}alves},
\nbbsttitle{Four point function of {$\mathcal{N}=\mathord{}$4} stress-tensor
  multiplet at strong coupling},
\nbbstjournal{\doiref{10.1007/JHEP04(2015)150}{JHEP~1504,~150~(2015)}},
\nbbsteprint{\arxivref{1411.1675}{arxiv:1411.1675}}.

\bibitem{Alday:2023mvu}
\nbbstauthor{L.~F.~Alday and T.~Hansen},
\nbbsttitle{The {AdS} {Virasoro}-{Shapiro} amplitude},
\nbbstjournal{\doiref{10.1007/JHEP10(2023)023}{JHEP~2310,~023~(2023)}},
\nbbsteprint{\arxivref{2306.12786}{arxiv:2306.12786}}.

\bibitem{Beem:2016wfs}
\nbbstauthor{C.~Beem, L.~Rastelli and B.~C.~van~Rees},
\nbbsttitle{More {$\mathcal{N}=\mathord{}$4} superconformal bootstrap},
\nbbstjournal{\doiref{10.1103/PhysRevD.96.046014}{Phys.~Rev.~D~96,~046014~(2017)}},
\nbbsteprint{\arxivref{1612.02363}{arxiv:1612.02363}}.

\bibitem{Chester:2023ehi}
\nbbstauthor{S.~M.~Chester, R.~Dempsey and S.~S.~Pufu},
\nbbsttitle{Level repulsion in {$\mathcal{N}=\mathord{}$4}
  super-{Yang}--{Mills} via integrability, holography, and the bootstrap},
\nbbstjournal{\doiref{10.1007/JHEP07(2024)059}{JHEP~2407,~059~(2024)}},
\nbbsteprint{\arxivref{2312.12576}{arxiv:2312.12576}}.

\bibitem{Caron-Huot:2022sdy}
\nbbstauthor{S.~Caron-Huot, F.~Coronado, A.-K.~Trinh and Z.~Zahraee},
\nbbsttitle{Bootstrapping {$\mathcal{N}=\mathord{}$4} sYM correlators using
  integrability},
\nbbstjournal{\doiref{10.1007/JHEP02(2023)083}{JHEP~2302,~083~(2023)}},
\nbbsteprint{\arxivref{2207.01615}{arxiv:2207.01615}}.

\bibitem{Caron-Huot:2024tzr}
\nbbstauthor{S.~Caron-Huot, F.~Coronado and Z.~Zahraee},
\nbbsttitle{Bootstrapping {$\mathcal{N}=\mathord{}$4} sYM correlators using
  integrability and localization},
\nbbstjournal{\doiref{10.1007/JHEP05(2025)220}{JHEP~2505,~220~(2025)}},
\nbbsteprint{\arxivref{2412.00249}{arxiv:2412.00249}}.

\bibitem{Bargheer:2022sfd}
\nbbstauthor{T.~Bargheer, T.~Fleury and V.~Gon\c{c}alves},
\nbbsttitle{Higher-Point Integrands in {$\mathcal{N}=\mathord{}$4} super
  {Yang}--{Mills} Theory},
\nbbstjournal{\doiref{10.21468/SciPostPhys.15.2.059}{SciPost~Phys.~15,~059~(2023)}},
\nbbsteprint{\arxivref{2212.03773}{arxiv:2212.03773}}.

\bibitem{Goncalves:2019znr}
\nbbstauthor{V.~Gon\c{c}alves, R.~Pereira and X.~Zhou},
\nbbsttitle{$20'$ Five-Point Function from $AdS_5\times S^5$ Supergravity},
\nbbstjournal{\doiref{10.1007/JHEP10(2019)247}{JHEP~1910,~247~(2019)}},
\nbbsteprint{\arxivref{1906.05305}{arxiv:1906.05305}}.

\bibitem{Goncalves:2025jcg}
\nbbstauthor{V.~Gon\c{c}alves, M.~Nocchi and X.~Zhou},
\nbbsttitle{Dissecting supergraviton six-point function with lightcone limits
  and chiral algebra},
\nbbstjournal{\doiref{10.1007/JHEP06(2025)173}{JHEP~2506,~173~(2025)}},
\nbbsteprint{\arxivref{2502.10269}{arxiv:2502.10269}}.

\bibitem{Alday:2010zy}
\nbbstauthor{L.~F.~Alday, B.~Eden, G.~P.~Korchemsky, J.~Maldacena and
  E.~Sokatchev},
\nbbsttitle{From correlation functions to {Wilson} loops},
\nbbstjournal{\doiref{10.1007/JHEP09(2011)123}{JHEP~1109,~123~(2011)}},
\nbbsteprint{\arxivref{1007.3243}{arxiv:1007.3243}}.

\bibitem{Alday:2007hr}
\nbbstauthor{L.~F.~Alday and J.~M.~Maldacena},
\nbbsttitle{Gluon scattering amplitudes at strong coupling},
\nbbstjournal{\doiref{10.1088/1126-6708/2007/06/064}{JHEP~0706,~064~(2007)}},
\nbbsteprint{\arxivref{0705.0303}{arxiv:0705.0303}}.

\bibitem{Berkovits:2008ic}
\nbbstauthor{N.~Berkovits and J.~Maldacena},
\nbbsttitle{Fermionic {T}-Duality, Dual Superconformal Symmetry, and the
  Amplitude/{Wilson} Loop Connection},
\nbbstjournal{\doiref{10.1088/1126-6708/2008/09/062}{JHEP~0809,~062~(2008)}},
\nbbsteprint{\arxivref{0807.3196}{arxiv:0807.3196}}.

\bibitem{Beisert:2008iq}
\nbbstauthor{N.~Beisert, R.~Ricci, A.~A.~Tseytlin and M.~Wolf},
\nbbsttitle{Dual Superconformal Symmetry from {AdS$_5$ $\times$ S$^5$}
  Superstring Integrability},
\nbbstjournal{\doiref{10.1103/PhysRevD.78.126004}{Phys.~Rev.~D78,~126004~(2008)}},
\nbbsteprint{\arxivref{0807.3228}{arxiv:0807.3228}}.

\bibitem{Bern:2008ap}
\nbbstauthor{Z.~Bern, L.~J.~Dixon, D.~A.~Kosower, R.~Roiban, M.~Spradlin,
  C.~Vergu and A.~Volovich},
\nbbsttitle{The Two-Loop Six-Gluon {MHV} Amplitude in Maximally Supersymmetric
  {Yang}--{Mills} Theory},
\nbbstjournal{\doiref{10.1103/PhysRevD.78.045007}{Phys.~Rev.~D78,~045007~(2008)}},
\nbbsteprint{\arxivref{0803.1465}{arxiv:0803.1465}}.

\bibitem{Drummond:2008aq}
\nbbstauthor{J.~M.~Drummond, J.~Henn, G.~P.~Korchemsky and E.~Sokatchev},
\nbbsttitle{Hexagon {Wilson} loop = six-gluon {MHV} amplitude},
\nbbstjournal{\doiref{10.1016/j.nuclphysb.2009.02.015}{Nucl.~Phys.~B815,~142~(2009)}},
\nbbsteprint{\arxivref{0803.1466}{arxiv:0803.1466}}.

\bibitem{Mason:2010yk}
\nbbstauthor{L.~Mason and D.~Skinner},
\nbbsttitle{The Complete Planar {S}-matrix of {$\mathcal{N}=\mathord{}$4} {SYM}
  as a {Wilson} Loop in Twistor Space},
\nbbstjournal{\doiref{10.1007/JHEP12(2010)018}{JHEP~1012,~018~(2010)}},
\nbbsteprint{\arxivref{1009.2225}{arxiv:1009.2225}}.

\bibitem{Caron-Huot:2010ryg}
\nbbstauthor{S.~Caron-Huot},
\nbbsttitle{Notes on the scattering amplitude / {Wilson} loop duality},
\nbbstjournal{\doiref{10.1007/JHEP07(2011)058}{JHEP~1107,~058~(2011)}},
\nbbsteprint{\arxivref{1010.1167}{arxiv:1010.1167}}.

\bibitem{Eden:2010zz}
\nbbstauthor{B.~Eden, G.~P.~Korchemsky and E.~Sokatchev},
\nbbsttitle{From correlation functions to scattering amplitudes},
\nbbstjournal{\doiref{10.1007/JHEP12(2011)002}{JHEP~1112,~002~(2011)}},
\nbbsteprint{\arxivref{1007.3246}{arxiv:1007.3246}}.

\bibitem{Belitsky:2013ofa}
\nbbstauthor{A.~V.~Belitsky, S.~Hohenegger, G.~P.~Korchemsky, E.~Sokatchev and
  A.~Zhiboedov},
\nbbsttitle{Energy-Energy Correlations in {$\mathcal{N}=\mathord{}$4}
  Supersymmetric {Yang}--{Mills} Theory},
\nbbstjournal{\doiref{10.1103/PhysRevLett.112.071601}{Phys.~Rev.~Lett.~112,~071601~(2014)}},
\nbbsteprint{\arxivref{1311.6800}{arxiv:1311.6800}}.

\bibitem{Moult:2025nhu}
\nbbstauthor{I.~Moult and H.~X.~Zhu},
\nbbsttitle{Energy Correlators: A Journey From Theory to Experiment},
\nbbsteprint{\arxivref{2506.09119}{arxiv:2506.09119}}.

\bibitem{Basso:2015zoa}
\nbbstauthor{B.~Basso, S.~Komatsu and P.~Vieira},
\nbbsttitle{Structure Constants and Integrable Bootstrap in Planar
  {$\mathcal{N}=\mathord{}$4} {SYM} Theory},
\nbbsteprint{\arxivref{1505.06745}{arxiv:1505.06745}}.

\bibitem{Fleury:2016ykk}
\nbbstauthor{T.~Fleury and S.~Komatsu},
\nbbsttitle{Hexagonalization of Correlation Functions},
\nbbstjournal{\doiref{10.1007/JHEP01(2017)130}{JHEP~1701,~130~(2017)}},
\nbbsteprint{\arxivref{1611.05577}{arxiv:1611.05577}}.

\bibitem{Coronado:2018cxj}
\nbbstauthor{F.~Coronado},
\nbbsttitle{Bootstrapping the simplest correlator in planar
  {$\mathcal{N}=\mathord{}$4} {SYM} at all loops},
\nbbstjournal{\doiref{10.1103/PhysRevLett.124.171601}{Phys.~Rev.~Lett.~124,~171601~(2020)}},
\nbbsteprint{\arxivref{1811.03282}{arxiv:1811.03282}}.

\bibitem{Kostov:2019stn}
\nbbstauthor{I.~Kostov, V.~B.~Petkova and D.~Serban},
\nbbsttitle{Determinant formula for the octagon form factor in
  {$\mathcal{N}=\mathord{}$4} {SYM}},
\nbbstjournal{\doiref{10.1103/PhysRevLett.122.231601}{Phys.~Rev.~Lett.~122,~231601~(2019)}},
\nbbsteprint{\arxivref{1903.05038}{arxiv:1903.05038}}.

\bibitem{Belitsky:2020qrm}
\nbbstauthor{A.~V.~Belitsky and G.~P.~Korchemsky},
\nbbsttitle{Octagon at finite coupling},
\nbbstjournal{\doiref{10.1007/JHEP07(2020)219}{JHEP~2007,~219~(2020)}},
\nbbsteprint{\arxivref{2003.01121}{arxiv:2003.01121}}.

\bibitem{Bajnok:2024epf}
\nbbstauthor{Z.~Bajnok, B.~Boldis and G.~P.~Korchemsky},
\nbbsttitle{{Tracy}--{Widom} Distribution in Four-Dimensional Supersymmetric
  {Yang}--{Mills} Theories},
\nbbstjournal{\doiref{10.1103/PhysRevLett.133.031601}{Phys.~Rev.~Lett.~133,~031601~(2024)}},
\nbbsteprint{\arxivref{2403.13050}{arxiv:2403.13050}}.

\bibitem{Caron-Huot:2018kta}
\nbbstauthor{S.~Caron-Huot and A.-K.~Trinh},
\nbbsttitle{All tree-level correlators in {AdS}$_{5}$\texttimes{}S$_{5}$
  supergravity: hidden ten-dimensional conformal symmetry},
\nbbstjournal{\doiref{10.1007/JHEP01(2019)196}{JHEP~1901,~196~(2019)}},
\nbbsteprint{\arxivref{1809.09173}{arxiv:1809.09173}}.

\bibitem{Caron-Huot:2021usw}
\nbbstauthor{S.~Caron-Huot and F.~Coronado},
\nbbsttitle{Ten dimensional symmetry of {$\mathcal{N}=\mathord{4}$} {SYM}
  correlators},
\nbbstjournal{\doiref{10.1007/JHEP03(2022)151}{JHEP~2203,~151~(2022)}},
\nbbsteprint{\arxivref{2106.03892}{arxiv:2106.03892}}.

\bibitem{Intriligator:1998ig}
\nbbstauthor{K.~A.~Intriligator},
\nbbsttitle{Bonus Symmetries of {$\mathcal{N}=\mathord{}$4}
  Super-{Yang}--{Mills} Correlation Functions via {AdS} Duality},
\nbbstjournal{\doiref{10.1016/S0550-3213(99)00242-4}{Nucl.~Phys.~B551,~575~(1999)}},
\nbbsteprint{\arxivref{hep-th/9811047}{hep-th/9811047}}.

\bibitem{Caron-Huot:2023wdh}
\nbbstauthor{S.~Caron-Huot, F.~Coronado and B.~M{\"u}hlmann},
\nbbsttitle{Determinants in self-dual {$\mathcal{N}=\mathord{}$4} {SYM} and
  twistor space},
\nbbstjournal{\doiref{10.1007/JHEP08(2023)008}{JHEP~2308,~008~(2023)}},
\nbbsteprint{\arxivref{2304.12341}{arxiv:2304.12341}}.

\bibitem{Eden:2000bk}
\nbbstauthor{B.~Eden, A.~C.~Petkou, C.~Schubert and E.~Sokatchev},
\nbbsttitle{Partial non-renormalisation of the stress-tensor four-point
  function in {$\mathcal{N}=\mathord{}$4} {SYM}$_4$ and {AdS}/{CFT}},
\nbbstjournal{Nucl.~Phys.~B607,~191~(2001)},
\nbbsteprint{\arxivref{hep-th/0009106}{hep-th/0009106}}.

\bibitem{Fernandes:2025eqe}
\nbbstauthor{B.~Fernandes, V.~Gon\c{c}alves, Z.~Huang, Y.~Tang, J.~Vilas~Boas
  and E.~Y.~Yuan},
\nbbsttitle{{AdS}{\texttimes}S Mellin Bootstrap, Hidden 10D Symmetry and
  Five-Point {Kaluza}--{Klein} Functions in {$\mathcal{N}=\mathord{}$4}
  Supersymmetric {Yang}--{Mills} Theory},
\nbbstjournal{\doiref{10.1103/3qyd-n621}{Phys.~Rev.~Lett.~136,~081602~(2026)}},
\nbbsteprint{\arxivref{2507.14124}{arxiv:2507.14124}}.

\bibitem{Beisert:2010jr}
\nbbstauthor{N.~Beisert et~al.},
\nbbsttitle{Review of {AdS/CFT} Integrability: An Overview},
\nbbstjournal{\doiref{10.1007/s11005-011-0529-2}{Lett.~Math.~Phys.~99,~3~(2012)}},
\nbbsteprint{\arxivref{1012.3982}{arxiv:1012.3982}}.

\bibitem{Gromov:2013pga}
\nbbstauthor{N.~Gromov, V.~Kazakov, S.~Leurent and D.~Volin},
\nbbsttitle{Quantum Spectral Curve for Planar {$\mathcal{N}=\mathord{}$4}
  Super-{Yang}--{Mills} Theory},
\nbbstjournal{\doiref{10.1103/PhysRevLett.112.011602}{Phys.~Rev.~Lett.~112,~011602~(2014)}},
\nbbsteprint{\arxivref{1305.1939}{arxiv:1305.1939}}.

\bibitem{Eden:2016xvg}
\nbbstauthor{B.~Eden and A.~Sfondrini},
\nbbsttitle{Tessellating cushions: four-point functions in
  {$\mathcal{N}=\mathord{}$4} {SYM}},
\nbbstjournal{\doiref{10.1007/JHEP10(2017)098}{JHEP~1710,~098~(2017)}},
\nbbsteprint{\arxivref{1611.05436}{arxiv:1611.05436}}.

\bibitem{Bargheer:2017nne}
\nbbstauthor{T.~Bargheer, J.~Caetano, T.~Fleury, S.~Komatsu and P.~Vieira},
\nbbsttitle{Handling Handles: Nonplanar Integrability in $\mathcal{N}=4$
  Supersymmetric {Yang}--{Mills} Theory},
\nbbstjournal{\doiref{10.1103/PhysRevLett.121.231602}{Phys.~Rev.~Lett.~121,~231602~(2018)}},
\nbbsteprint{\arxivref{1711.05326}{arxiv:1711.05326}}.

\bibitem{Bargheer:2018jvq}
\nbbstauthor{T.~Bargheer, J.~Caetano, T.~Fleury, S.~Komatsu and P.~Vieira},
\nbbsttitle{Handling handles. Part {II}. Stratification and Data Analysis},
\nbbstjournal{\doiref{10.1007/JHEP11(2018)095}{JHEP~1811,~095~(2018)}},
\nbbsteprint{\arxivref{1809.09145}{arxiv:1809.09145}}.

\bibitem{Chicherin:2014uca}
\nbbstauthor{D.~Chicherin, R.~Doobary, B.~Eden, P.~Heslop, G.~P.~Korchemsky,
  L.~Mason and E.~Sokatchev},
\nbbsttitle{Correlation functions of the chiral stress-tensor multiplet in
  {$\mathcal{N}=\mathord{}$4} {SYM}},
\nbbstjournal{\doiref{10.1007/JHEP06(2015)198}{JHEP~1506,~198~(2015)}},
\nbbsteprint{\arxivref{1412.8718}{arxiv:1412.8718}}.

\bibitem{Fleury:2019ydf}
\nbbstauthor{T.~Fleury and R.~Pereira},
\nbbsttitle{Non-planar data of {$\mathcal{N}=\mathord{}4$} {SYM}},
\nbbstjournal{\doiref{10.1007/JHEP03(2020)003}{JHEP~2003,~003~(2020)}},
\nbbsteprint{\arxivref{1910.09428}{arxiv:1910.09428}}.

\bibitem{Aprile:2020uxk}
\nbbstauthor{F.~Aprile, J.~M.~Drummond, P.~Heslop, H.~Paul, F.~Sanfilippo,
  M.~Santagata and A.~Stewart},
\nbbsttitle{Single Particle Operators and their Correlators in Free
  {$\mathcal{N}=\mathord{}$4} {SYM}},
\nbbstjournal{\doiref{10.1007/JHEP11(2020)072}{JHEP~2011,~072~(2020)}},
\nbbsteprint{\arxivref{2007.09395}{arxiv:2007.09395}}.

\bibitem{Arutyunov:1999en}
\nbbstauthor{G.~Arutyunov and S.~Frolov},
\nbbsttitle{Some cubic couplings in type IIB supergravity on {AdS}(5) x S**5
  and three point functions in {SYM}(4) at large N},
\nbbstjournal{\doiref{10.1103/PhysRevD.61.064009}{Phys.~Rev.~D~61,~064009~(2000)}},
\nbbsteprint{\arxivref{hep-th/9907085}{hep-th/9907085}}.

\bibitem{Intriligator:1999ff}
\nbbstauthor{K.~A.~Intriligator and W.~Skiba},
\nbbsttitle{Bonus Symmetry and the Operator Product Expansion of
  {$\mathcal{N}=\mathord{}$4} Super-{Yang}--{Mills}},
\nbbstjournal{\doiref{10.1016/S0550-3213(99)00430-7}{Nucl.~Phys.~B559,~165~(1999)}},
\nbbsteprint{\arxivref{hep-th/9905020}{hep-th/9905020}}.

\bibitem{Chalmers:1996rq}
\nbbstauthor{G.~Chalmers and W.~Siegel},
\nbbsttitle{Self-dual sector of QCD amplitudes},
\nbbstjournal{\doiref{10.1103/PhysRevD.54.7628}{Phys.~Rev.~D~54,~7628~(1996)}},
\nbbsteprint{\arxivref{hep-th/9606061}{hep-th/9606061}}.

\bibitem{Drukker:2009sf}
\nbbstauthor{N.~Drukker and J.~Plefka},
\nbbsttitle{Superprotected n-point correlation functions of local operators in
  {$\mathcal{N}=\mathord{}$4} super {Yang}--{Mills}},
\nbbstjournal{\doiref{10.1088/1126-6708/2009/04/052}{JHEP~0904,~052~(2009)}},
\nbbsteprint{\arxivref{0901.3653}{arxiv:0901.3653}}.

\bibitem{Chicherin:2015edu}
\nbbstauthor{D.~Chicherin, J.~Drummond, P.~Heslop and E.~Sokatchev},
\nbbsttitle{All three-loop four-point correlators of half-{BPS} operators in
  planar {$\mathcal{N}=\mathord{}$4} {SYM}},
\nbbstjournal{\doiref{10.1007/JHEP08(2016)053}{JHEP~1608,~053~(2016)}},
\nbbsteprint{\arxivref{1512.02926}{arxiv:1512.02926}}.

\bibitem{Chicherin:2018avq}
\nbbstauthor{D.~Chicherin, A.~Georgoudis, V.~Gon\c{c}alves and R.~Pereira},
\nbbsttitle{All five-loop planar four-point functions of half-{BPS} operators
  in {$\mathcal{N}=\mathord{}$4} {SYM}},
\nbbstjournal{\doiref{10.1007/JHEP11(2018)069}{JHEP~1811,~069~(2018)}},
\nbbsteprint{\arxivref{1809.00551}{arxiv:1809.00551}}.

\bibitem{Drummond:2008vq}
\nbbstauthor{J.~M.~Drummond, J.~Henn, G.~P.~Korchemsky and E.~Sokatchev},
\nbbsttitle{Dual superconformal symmetry of scattering amplitudes in
  {$\mathcal{N}=\mathord{}$4} super-{Yang}--{Mills} theory},
\nbbstjournal{\doiref{10.1016/j.nuclphysb.2009.11.022}{Nucl.~Phys.~B828,~317~(2010)}},
\nbbsteprint{\arxivref{0807.1095}{arxiv:0807.1095}}.

\bibitem{Eden:2011yp}
\nbbstauthor{B.~Eden, P.~Heslop, G.~P.~Korchemsky and E.~Sokatchev},
\nbbsttitle{The super-correlator/super-amplitude duality: Part I},
\nbbstjournal{\doiref{10.1016/j.nuclphysb.2012.12.015}{Nucl.~Phys.~B869,~329~(2013)}},
\nbbsteprint{\arxivref{1103.3714}{arxiv:1103.3714}}.

\bibitem{Eden:2011ku}
\nbbstauthor{B.~Eden, P.~Heslop, G.~P.~Korchemsky and E.~Sokatchev},
\nbbsttitle{The super-correlator/super-amplitude duality: Part {II}},
\nbbstjournal{\doiref{10.1016/j.nuclphysb.2012.12.014}{Nucl.~Phys.~B869,~378~(2013)}},
\nbbsteprint{\arxivref{1103.4353}{arxiv:1103.4353}}.

\bibitem{Basso:2023bwv}
\nbbstauthor{B.~Basso and A.~G.~Tumanov},
\nbbsttitle{{Wilson} loop duality and {OPE} for super form factors of
  half-{BPS} operators},
\nbbstjournal{\doiref{10.1007/JHEP02(2024)022}{JHEP~2402,~022~(2024)}},
\nbbsteprint{\arxivref{2308.08432}{arxiv:2308.08432}}.

\bibitem{tHooft:1974jz}
\nbbstauthor{G.~'t~Hooft},
\nbbsttitle{A Planar Diagram Theory for Strong Interactions},
\nbbstjournal{\doiref{10.1016/0550-3213(74)90154-0}{Nucl.~Phys.~B72,~461~(1974)}}.

\bibitem{Ambrosio:2013pba}
\nbbstauthor{R.~G.~Ambrosio, B.~Eden, T.~Goddard, P.~Heslop and C.~Taylor},
\nbbsttitle{Local integrands for the five-point amplitude in planar
  {$\mathcal{N}=\mathord{}$4} {SYM} up to five loops},
\nbbstjournal{\doiref{10.1007/JHEP01(2015)116}{JHEP~1501,~116~(2015)}},
\nbbsteprint{\arxivref{1312.1163}{arxiv:1312.1163}}.

\bibitem{He:2024cej}
\nbbstauthor{S.~He, C.~Shi, Y.~Tang and Y.-Q.~Zhang},
\nbbsttitle{The cusp limit of correlators and a new graphical bootstrap for
  correlators/amplitudes to eleven loops},
\nbbstjournal{\doiref{10.1007/JHEP03(2025)192}{JHEP~2503,~192~(2025)}},
\nbbsteprint{\arxivref{2410.09859}{arxiv:2410.09859}}.

\bibitem{Bourjaily:2025iad}
\nbbstauthor{J.~L.~Bourjaily, S.~He, C.~Shi and Y.~Tang},
\nbbsttitle{The Four-Point Correlator of Planar sYM at Twelve Loops},
\nbbsteprint{\arxivref{2503.15593}{arxiv:2503.15593}}.

\bibitem{Eden:2017fow}
\nbbstauthor{B.~Eden, P.~Heslop and L.~Mason},
\nbbsttitle{The Correlahedron},
\nbbstjournal{\doiref{10.1007/JHEP09(2017)156}{JHEP~1709,~156~(2017)}},
\nbbsteprint{\arxivref{1701.00453}{arxiv:1701.00453}}.

\bibitem{He:2024xed}
\nbbstauthor{S.~He, Y.-t.~Huang and C.-K.~Kuo},
\nbbsttitle{All-loop geometry for four-point correlation functions},
\nbbstjournal{\doiref{10.1103/PhysRevD.110.L081701}{Phys.~Rev.~D~110,~L081701~(2024)}},
\nbbsteprint{\arxivref{2405.20292}{arxiv:2405.20292}}.

\bibitem{He:2025rza}
\nbbstauthor{S.~He, Y.-t.~Huang and C.-K.~Kuo},
\nbbsttitle{Leading singularities and chambers of Correlahedron},
\nbbsteprint{\arxivref{2505.09808}{arxiv:2505.09808}}.

\bibitem{He:2025vqt}
\nbbstauthor{S.~He, X.~Jiang, J.~Liu and Y.-Q.~Zhang},
\nbbsttitle{Notes on conformal integrals: {Coulomb} branch amplitudes, magic
  identities, and bootstrap},
\nbbstjournal{\doiref{10.1103/gmp7-r9dz}{Phys.~Rev.~D~112,~076012~(2025)}},
\nbbsteprint{\arxivref{2502.08871}{arxiv:2502.08871}}.

\bibitem{He:2025lzd}
\nbbstauthor{S.~He and X.~Jiang},
\nbbsttitle{Solving Infinite Families of Dual Conformal Integrals and Periods},
\nbbsteprint{\arxivref{2506.20095}{arxiv:2506.20095}}.

\bibitem{Chicherin:2015bza}
\nbbstauthor{D.~Chicherin, R.~Doobary, B.~Eden, P.~Heslop, G.~P.~Korchemsky and
  E.~Sokatchev},
\nbbsttitle{Bootstrapping correlation functions in {$\mathcal{N}=\mathord{}$4}
  {SYM}},
\nbbstjournal{\doiref{10.1007/JHEP03(2016)031}{JHEP~1603,~031~(2016)}},
\nbbsteprint{\arxivref{1506.04983}{arxiv:1506.04983}}.

\bibitem{Fleury:2020ykw}
\nbbstauthor{T.~Fleury and V.~Gon\c{c}alves},
\nbbsttitle{Decagon at Two Loops},
\nbbstjournal{\doiref{10.1007/JHEP07(2020)030}{JHEP~2007,~030~(2020)}},
\nbbsteprint{\arxivref{2004.10867}{arxiv:2004.10867}}.

\bibitem{Bercini:2024pya}
\nbbstauthor{C.~Bercini, B.~Fernandes and V.~Gon\c{c}alves},
\nbbsttitle{Two-loop five-point integrals: light, heavy and large-spin
  correlators},
\nbbstjournal{\doiref{10.1007/JHEP10(2024)242}{JHEP~2410,~242~(2024)}},
\nbbsteprint{\arxivref{2401.06099}{arxiv:2401.06099}}.

\bibitem{Bargheer:2019kxb}
\nbbstauthor{T.~Bargheer, F.~Coronado and P.~Vieira},
\nbbsttitle{Octagons I: Combinatorics and Non-Planar Resummations},
\nbbstjournal{\doiref{10.1007/JHEP08(2019)162}{JHEP~1908,~162~(2019)}},
\nbbsteprint{\arxivref{1904.00965}{arxiv:1904.00965}}.

\bibitem{Bissi:2024tqf}
\nbbstauthor{A.~Bissi, G.~Fardelli and A.~Manenti},
\nbbsttitle{Composite operators in $\mathcal{N} = 4$ Super {Yang}--{Mills}},
\nbbstjournal{\doiref{10.1007/JHEP07(2025)074}{JHEP~2507,~074~(2025)}},
\nbbsteprint{\arxivref{2412.19788}{arxiv:2412.19788}}.

\bibitem{Aprile:2024lwy}
\nbbstauthor{F.~Aprile, S.~Giusto and R.~Russo},
\nbbsttitle{Holographic correlators with {BPS} bound states in $\mathcal{N} =
  4$ {SYM}},
\nbbstjournal{\doiref{10.1103/PhysRevLett.134.091602}{Phys.~Rev.~Lett.~134,~091602~(2025)}},
\nbbsteprint{\arxivref{2409.12911}{arxiv:2409.12911}}.

\bibitem{Aprile:2025hlt}
\nbbstauthor{F.~Aprile, S.~Giusto and R.~Russo},
\nbbsttitle{Four-point correlators with {BPS} bound states in {AdS}$_{3}$ and
  {AdS}$_{5}$},
\nbbstjournal{\doiref{10.1007/JHEP08(2025)193}{JHEP~2508,~193~(2025)}},
\nbbsteprint{\arxivref{2503.02855}{arxiv:2503.02855}}.

\bibitem{Usyukina:1993ch}
\nbbstauthor{N.~I.~Ussyukina and A.~I.~Davydychev},
\nbbsttitle{Exact results for three- and four-point ladder diagrams with an
  arbitrary number of rungs},
\nbbstjournal{\doiref{10.1016/0370-2693(93)91118-7}{Phys.~Lett.~B305,~136~(1993)}}.

\bibitem{Bork:2022vat}
\nbbstauthor{L.~V.~Bork, N.~B.~Muzhichkov and E.~S.~Sozinov},
\nbbsttitle{Infrared properties of five-point massive amplitudes in
  {$\mathcal{N}=\mathord{}$4} {SYM} on the Coulomb branch},
\nbbstjournal{\doiref{10.1007/JHEP08(2022)173}{JHEP~2208,~173~(2022)}},
\nbbsteprint{\arxivref{2201.08762}{arxiv:2201.08762}}.

\bibitem{Bork:2025ztu}
\nbbstauthor{L.~V.~Bork, R.~N.~Lee and A.~I.~Onishchenko},
\nbbsttitle{Method of regions for dual conformal integrals},
\nbbstjournal{\doiref{10.1007/JHEP12(2025)107}{JHEP~2512,~107~(2025)}},
\nbbsteprint{\arxivref{2509.12056}{arxiv:2509.12056}}.

\bibitem{Drukker:2008pi}
\nbbstauthor{N.~Drukker and J.~Plefka},
\nbbsttitle{The Structure of n-point functions of chiral primary operators in
  {$\mathcal{N}=\mathord{}$4} super {Yang}--{Mills} at one-loop},
\nbbstjournal{\doiref{10.1088/1126-6708/2009/04/001}{JHEP~0904,~001~(2009)}},
\nbbsteprint{\arxivref{0812.3341}{arxiv:0812.3341}}.

\bibitem{Bercini:2020msp}
\nbbstauthor{C.~Bercini, V.~Gon\c{c}alves and P.~Vieira},
\nbbsttitle{Light-Cone Bootstrap of Higher Point Functions and {Wilson} Loop
  Duality},
\nbbstjournal{\doiref{10.1103/PhysRevLett.126.121603}{Phys.~Rev.~Lett.~126,~121603~(2021)}},
\nbbsteprint{\arxivref{2008.10407}{arxiv:2008.10407}}.

\bibitem{Poland:2023vpn}
\nbbstauthor{D.~Poland, V.~Prilepina and P.~Tadi\'c},
\nbbsttitle{The five-point bootstrap},
\nbbstjournal{\doiref{10.1007/JHEP10(2023)153}{JHEP~2310,~153~(2023)}},
\nbbsteprint{\arxivref{2305.08914}{arxiv:2305.08914}}.

\bibitem{Costa:2011dw}
\nbbstauthor{M.~S.~Costa, J.~Penedones, D.~Poland and S.~Rychkov},
\nbbsttitle{Spinning Conformal Blocks},
\nbbstjournal{\doiref{10.1007/JHEP11(2011)154}{JHEP~1111,~154~(2011)}},
\nbbsteprint{\arxivref{1109.6321}{arxiv:1109.6321}}.

\bibitem{Marboe:2017dmb}
\nbbstauthor{C.~Marboe and D.~Volin},
\nbbsttitle{The full spectrum of {AdS}$_5$/{CFT}$_4$ I: Representation theory
  and one-loop Q-system},
\nbbstjournal{\doiref{10.1088/1751-8121/aab34a}{J.~Phys.~A~51,~165401~(2018)}},
\nbbsteprint{\arxivref{1701.03704}{arxiv:1701.03704}}.

\bibitem{Bercini:2022gvs}
\nbbstauthor{C.~Bercini, V.~Gon\c{c}alves, A.~Homrich and P.~Vieira},
\nbbsttitle{Spinning hexagons},
\nbbstjournal{\doiref{10.1007/JHEP09(2022)228}{JHEP~2209,~228~(2022)}},
\nbbsteprint{\arxivref{2207.08931}{arxiv:2207.08931}}.

\bibitem{Basso:2015eqa}
\nbbstauthor{B.~Basso, V.~Gon\c{c}alves, S.~Komatsu and P.~Vieira},
\nbbsttitle{Gluing Hexagons at Three Loops},
\nbbstjournal{\doiref{10.1016/j.nuclphysb.2016.04.020}{Nucl.~Phys.~B907,~695~(2016)}},
\nbbsteprint{\arxivref{1510.01683}{arxiv:1510.01683}}.

\bibitem{Basso:2017khq}
\nbbstauthor{B.~Basso, F.~Coronado, S.~Komatsu, H.~T.~Lam, P.~Vieira and
  D.-l.~Zhong},
\nbbsttitle{Asymptotic Four Point Functions},
\nbbstjournal{\doiref{10.1007/JHEP07(2019)082}{JHEP~1907,~082~(2019)}},
\nbbsteprint{\arxivref{1701.04462}{arxiv:1701.04462}}.

\bibitem{Coronado:2018ypq}
\nbbstauthor{F.~Coronado},
\nbbsttitle{Perturbative Four-Point Functions in Planar
  {$\mathcal{N}=\mathord{}$4} {SYM} from Hexagonalization},
\nbbstjournal{\doiref{10.1007/JHEP01(2019)056}{JHEP~1901,~056~(2019)}},
\nbbsteprint{\arxivref{1811.00467}{arxiv:1811.00467}}.

\bibitem{Costa:2023wfz}
\nbbstauthor{M.~S.~Costa, V.~Gon\c{c}alves, A.~Salgarkar and J.~Vilas~Boas},
\nbbsttitle{Conformal multi-{Regge} theory},
\nbbstjournal{\doiref{10.1007/JHEP09(2023)155}{JHEP~2309,~155~(2023)}},
\nbbsteprint{\arxivref{2305.10394}{arxiv:2305.10394}}.

\bibitem{Alday:2013cwa}
\nbbstauthor{L.~F.~Alday and A.~Bissi},
\nbbsttitle{Higher-spin correlators},
\nbbstjournal{\doiref{10.1007/JHEP10(2013)202}{JHEP~1310,~202~(2013)}},
\nbbsteprint{\arxivref{1305.4604}{arxiv:1305.4604}}.

\bibitem{Alday:2016mxe}
\nbbstauthor{L.~F.~Alday and A.~Bissi},
\nbbsttitle{Crossing symmetry and Higher spin towers},
\nbbstjournal{\doiref{10.1007/JHEP12(2017)118}{JHEP~1712,~118~(2017)}},
\nbbsteprint{\arxivref{1603.05150}{arxiv:1603.05150}}.

\bibitem{Bercini:2021jti}
\nbbstauthor{C.~Bercini, V.~Gon\c{c}alves, A.~Homrich and P.~Vieira},
\nbbsttitle{The {Wilson} loop \textemdash{} large spin OPE dictionary},
\nbbstjournal{\doiref{10.1007/JHEP07(2022)079}{JHEP~2207,~079~(2022)}},
\nbbsteprint{\arxivref{2110.04364}{arxiv:2110.04364}}.

\bibitem{Chen:2022jhb}
\nbbstauthor{H.~Chen, I.~Moult, J.~Sandor and H.~X.~Zhu},
\nbbsttitle{Celestial blocks and transverse spin in the three-point energy
  correlator},
\nbbstjournal{\doiref{10.1007/JHEP09(2022)199}{JHEP~2209,~199~(2022)}},
\nbbsteprint{\arxivref{2202.04085}{arxiv:2202.04085}}.

\bibitem{Chang:2022ryc}
\nbbstauthor{C.-H.~Chang and D.~Simmons-Duffin},
\nbbsttitle{Three-point energy correlators and the celestial block expansion},
\nbbstjournal{\doiref{10.1007/JHEP02(2023)126}{JHEP~2302,~126~(2023)}},
\nbbsteprint{\arxivref{2202.04090}{arxiv:2202.04090}}.

\bibitem{Chicherin:2023gxt}
\nbbstauthor{D.~Chicherin, G.~P.~Korchemsky, E.~Sokatchev and A.~Zhiboedov},
\nbbsttitle{Energy correlations in heavy states},
\nbbstjournal{\doiref{10.1007/JHEP11(2023)134}{JHEP~2311,~134~(2023)}},
\nbbsteprint{\arxivref{2306.14330}{arxiv:2306.14330}}.

\bibitem{Heslop:2022xgp}
\nbbstauthor{P.~Heslop},
\nbbsttitle{The {SAGEX} Review on Scattering Amplitudes, Chapter 8: Half {BPS}
  correlators},
\nbbstjournal{\doiref{10.1088/1751-8121/ac8c71}{J.~Phys.~A~55,~443009~(2022)}},
\nbbsteprint{\arxivref{2203.13019}{arxiv:2203.13019}}.

\bibitem{Huang:2024dxr}
\nbbstauthor{Z.~Huang, B.~Wang, E.~Y.~Yuan and J.~Zhang},
\nbbsttitle{All Five-Point {Kaluza}--{Klein} Correlators and Hidden 8D Symmetry
  in {AdS}5{\texttimes}S3},
\nbbstjournal{\doiref{10.1103/PhysRevLett.134.161601}{Phys.~Rev.~Lett.~134,~161601~(2025)}},
\nbbsteprint{\arxivref{2408.12260}{arxiv:2408.12260}}.

\bibitem{Du:2024xbd}
\nbbstauthor{X.-E.~Du, Z.~Huang, B.~Wang, E.~Y.~Yuan and X.~Zhou},
\nbbsttitle{Meson correlators in 4d $\mathcal{N}=2$ {SCFT}s and hints for 8d
  structures at weak coupling},
\nbbstjournal{\doiref{10.1007/JHEP04(2025)128}{JHEP~2504,~128~(2025)}},
\nbbsteprint{\arxivref{2412.17260}{arxiv:2412.17260}}.

\bibitem{Bianchi:2019jpy}
\nbbstauthor{M.~S.~Bianchi},
\nbbsttitle{On structure constants with two spinning twist-two operators},
\nbbstjournal{\doiref{10.1007/JHEP04(2019)059}{JHEP~1904,~059~(2019)}},
\nbbsteprint{\arxivref{1901.00679}{arxiv:1901.00679}}.

\end{thebibliography}

\end{document}